\documentclass{aa}  

\usepackage{graphicx}
\usepackage{txfonts}
\usepackage[colorlinks=true, allcolors = blue]{hyperref}
\usepackage{subfigure}

\usepackage{xcolor}
\usepackage{pdflscape}
\usepackage{todonotes}
\usepackage{multirow}
\usepackage{threeparttable}
\usepackage{capt-of}
\usepackage{longtable}

\newcommand{\blue}[1]{\textcolor{black}{#1}}

\newcommand{\grey}[1]{\textcolor{gray}{#1}}

\begin{document}

   \title{X-Shooting ULLYSES: Massive Stars at low metallicity IX:\\ 
   Empirical constraints on mass-loss rates and clumping parameters for OB supergiants in the Large Magellanic Cloud}

   \subtitle{}

   \author{O.\ Verhamme
          \inst{\inst{\ref{inst:KUL}}},
          J.\ Sundqvist\inst{\ref{inst:KUL}},
           A.\ de Koter \inst{\ref{inst:KUL}, \ref{inst:UAmst}},
          H.\ Sana\inst{\ref{inst:KUL}},
          F.\ Backs \inst{\ref{inst:KUL}},
          S. A. Brands \inst{\ref{inst:UAmst}},
          F.\ Najarro \inst{\ref{inst:CSIC-INTA}},
          J. Puls \inst{\ref{inst:mun}}, 
          J.S.\ Vink \inst{\ref{inst:armagh}}, 
          P.A.\ Crowther \inst{\ref{inst:sheffield}},
          B.\ Kubátová \inst{\ref{inst:czech}}
          A.A.C.\ Sander \inst{\ref{inst:ari}},
          M.\ Bernini-Peron \inst{\ref{inst:ari}},
          R.\ Kuiper \inst{\ref{inst:UDE}}, 
          R.K.\ Prinja  \inst{\ref{inst:UCL}},
          P.\ Schillemans \inst{\ref{inst:KUL}}, 
          T.\ Shenar \inst{\ref{inst:isr}}, 
          J.Th.\ van Loon \inst{\ref{inst:LJL}}, 
          and
          XShootu collaboration}

   \institute{Institute of Astronomy, KU Leuven, Celestijnenlaan 200D, 3001, Leuven, Belgium \label{inst:KUL}\\
              \email{Olivier.verhamme@kuleuven.be}
            \and
            {Anton Pannekoek Institute for Astronomy, University of Amsterdam, Science Park 904, 1098 XH Amsterdam, The Netherlands \label{inst:UAmst}}
            \and 
             {Departamento de Astrofísica, Centro de Astrobiología, (CSIC-
             INTA), Ctra. Torrejón a Ajalvir, km 4, 28850 Torrejón de Ardoz,
             Madrid, Spain \label{inst:CSIC-INTA}}
            \and 
            {LMU München, Universitätssternwarte, Scheinerstr. 1, 81679 München, Germany \label{inst:mun}}
            \and 
            {Armagh Observatory and Planetarium, College Hill, BT61 9DG Armagh, UK \label{inst:armagh}}
             \and 
            {Department of Physics \& Astronomy, Hounsfield Road, University of Sheffield, Sheffield, S3 7RH United Kingdom \label{inst:sheffield}}
            \and 
            {Astronomický ústav, Akademie v\u{e}d \u{C}eské Republiky, 251 65 Ond\u{r}ejov, Czech Republic \label{inst:czech}}
            \and 
            {Zentrum f{\"u}r Astronomie der Universit{\"a}t Heidelberg, 
            Astronomisches Rechen-Institut, M{\"o}nchhofstr. 12-14, 69120 
            Heidelberg, Germany \label{inst:ari}}
            \and 
             {Faculty of Physics, University of Duisburg-Essen, Lotharstra{\ss}e 1, D-47057 Duisburg, Germany \label{inst:UDE}}
            \and  
             {Dept. of Physics \& Astronomy, University College London, Gower Street, London WC1E 6BT \label{inst:UCL}}
             \and
            {The School of Physics and Astronomy, Tel Aviv University, Tel Aviv 6997801, Israel\label{inst:isr}}
             \and 
             {Lennard-Jones Laboratories, Keele University, ST5 5BG, UK \label{inst:LJL}}
             }

   \date{Received , 05-09-2024; accepted , 17-10-2024}

% \abstract{}{}{}{}{} 
% 5 {} token are mandatory
 
  \abstract
   {Current implementations of mass loss for hot, massive stars in stellar evolution models usually include a sharp increase in mass loss when blue supergiants become cooler than $T_{\rm eff} \sim 20-22$ kK. Such a drastic mass-loss jump has traditionally been motivated by the potential presence of a so-called bistability ionisation effect, which may occur for line-driven winds in this temperature region due to recombination of important line-driving ions.}
   {We perform quantitative spectroscopy using UV (ULLYSES program) and optical (XShootU collaboration) data for 17 OB-supergiant stars in the Large Magellanic Cloud (LMC) (covering the range $T_{\rm eff}$ $\sim 14-32$ kK), deriving absolute constraints on global stellar, wind, and clumping parameters. We examine whether there are any empirical signs of a mass loss jump in the investigated region, and we study the clumped nature of the wind.} 
   %around $T_{\rm eff} \sim 20$ kK.}
   {We use a combination of the model atmosphere code {\sc fastwind} and the genetic algorithm (GA) code Kiwi-GA to fit synthetic spectra of a multitude of diagnostic spectral lines in the optical and ultra-violet (UV).}
   {We find an almost monotonic decrease of mass loss-rate with effective temperature, with no signs of any upward mass loss jump anywhere in the examined region. 
   Standard theoretical comparison models, which include a strong bistability jump thus severely over predict the empirical mass-loss rates on the cool side of the predicted jump.
   %, by up to two orders of magnitude. 
   Another key result is that across our sample we find that on average about 40\% of the total wind mass seems to reside in the more diluted medium in between dense clumps.}  
   {Our derived mass-loss rates suggest that for applications like stellar evolution one should not include a drastic bistability jump in mass loss for stars in the temperature and luminosity region investigated here. The derived high values of interclump density further suggest that the common assumption of an effectively void interclump medium (applied in the vast majority of spectroscopic studies of hot star winds) is not generally valid in this parameter regime.}

    \titlerunning{Constraints on mass-loss rates and clumping parameters
for B supergiants}
    \authorrunning{O.\ Verhamme, J.\ Sundqvist, A.\ de Koter, H.\ Sana }
   \maketitle
%
%-------------------------------------------------------------------

\section{Introduction}
An important process in the life of massive stars are stellar winds driven by radiation intercepted by a multitude of atmospheric spectral lines \citep{castor_radiation-driven_1975} that over time may cause the star to lose a sizeable fraction of their initial mass and angular momentum. This significantly influences the evolution of a massive stars, and, ultimately, the nature and characteristics of its end of life products (e.g., \citealt{heger_how_2003,2012ARA&A..50..107L}).

% The theory behind these radiation driven winds has been first described by \cite{lucy_mass_1970}, \cite{castor_radiation-driven_1975}.
While studying luminous B-stars that experience such line-driven winds, \cite{pauldrach_radiation-driven_1990} and \cite{lamers_terminal_1995} noticed a peculiar behaviour in the ratio of terminal wind speed $\varv_{\infty}$ over surface escape speed $\varv_{\rm esc}$ which seemed to decrease sharply (by a factor of about two) for stars with an effective temperature ($T_{\rm eff}$) cooler than $\sim 21$ kK. 
%JS-note: Not correct, at least from published papers and PhD thesis (only two models computed, but T difference too large to infer jump or decreasing trend).
%The magnitude and approximate temperature location of this jump were confirmed theoretically by \cite{vink_nature_1999, vink_mass-loss_2001}. 
%These authors also predict that the change in terminal wind speed should be associated by a jump in mass-loss rate $\dot{M}$, becoming $3-4$ times stronger at the cool side of this so-called `bi-stability jump'.
Although these early results already hinted at the possibility of a corresponding increase in $\dot{M}$, such an increase was quantified by \cite{vink_nature_1999, vink_mass-loss_2001}. These authors, possibly, identified a recombination of iron (from Fe\,{\sc iv} to Fe\,{\sc iii}) near the sonic point. The increase in the amount of line transitions for Fe\,{\sc iii} resulted in more efficient line-driving, as the underlying physical reason for the mass-loss jump.
%\magenta{Recent independent theoretical developments, however, do not find such a mass-loss jump \citep{bjorklund_new_2023} or do identify one but at a lower temperature \citep{krticka_new_2024} and of a smaller magnitude (see below). It is therefore important to search for empirical evidence for the presence or absence of a bi-stability induced mass-loss jump. As the \citet{vink_mass-loss_2001} theory is the default mass-loss prescription in evolutionary codes including {\sc mesa} \citep{paxton_modules_2010, paxton_modules_2013, paxton_modules_2015} and {\sc genic} \citep[e.g.,][]{2019A&A...627A..24G} effects of bi-stability have been studied in detail. These studies show that if 
%Regarding the latter, the relative contribution of bistability braking and post-main sequence expansion, in combination with internal angular momentum transport efficiencies,   \citep{vink_nature_2010,keszthelyi_modeling_2017,bjorklund_new_2023}.}  
%JS-NOTE: It is important to have this below, since in practise this is what matetras and also since Jorick always points out the second bistability (not analysed here). It is also more linear, and better, for the flow of the intro if the introduction of my and Jo's (note: Robin did the excellent grid-calculations and analysis of results, but he actually didn't develop these models) and Krticka's CMF models are made later, like in previous draft by Olivier.     
Based on these models, \citet{vink_mass-loss_2001} provide \blue{fit-formulae and corresponding} "mass-loss recipes" that are divided in a "hot" and "cool" prescription, where the exact location of the "bistability" divide depends on metallicity $Z$ and the classical Eddington parameter ($\Gamma_{\rm e}$, see equation \ref{eq:gamma_e})\footnote{We note that a second predicted bistability divide also exists, related to recombination from doubly to singly ionised iron, but since this is predicted to lie at lower temperatures than covered by our present observational data set, we do not discuss this further in this paper.}. Depending on these parameters, the cool solution in this recipe may have up to an order of magnitude more mass loss than the hot, resulting in big mass-loss jump, typically located somewhere in the range $T_{\rm eff} \sim 22.5 - 27.5$ kK. This prescription has meanwhile become a standard choice to describe mass-loss from line-driven winds in stellar evolution codes such as Modules for Experiments in Stellar Astrophysics ({\sc MESA}) \citep{paxton_modules_2010, paxton_modules_2013, paxton_modules_2015} 
and the Geneva stellar evolution code {\sc genec} \citep[e.g.,][]{2019A&A...627A..24G}. Studies of the effects of such a bistability mass loss jump in evolution calculations suggest that if this occurs during 
%a phase of evolution that is governed by a nuclear timescale (i.e., during 
the final parts of the main sequence (or during post-main sequence evolution of stars that do not reach the red supergiant stage) substantial mass and angular momentum loss may occur. This then more easily allows for the formation of Wolf-Rayet stars \citep{bjorklund_new_2023} and also causes bistability braking, i.e., a strong reduction of surface rotation speed at $T_{\rm eff}$ cooler than $\sim$ 21\,kK \citep{vink_nature_2010,keszthelyi_modeling_2017,britavskiy_tracing_2024}.  

\begin{figure*}
    \centering
    \subfigure{\includegraphics[width=0.4\textwidth]{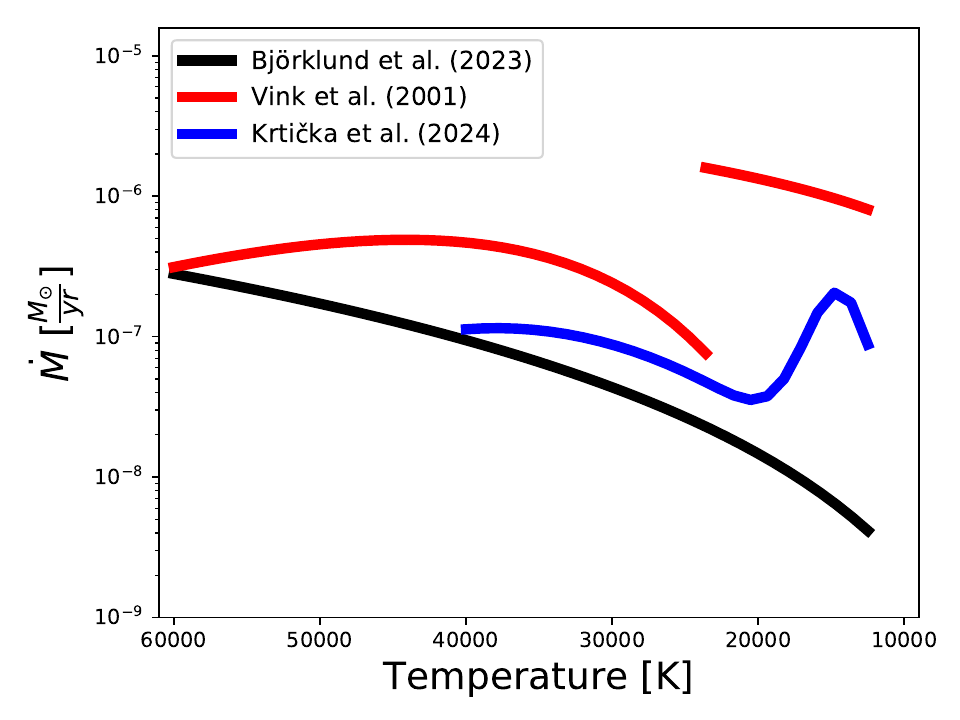}}
    \subfigure{\includegraphics[width=0.4\textwidth]{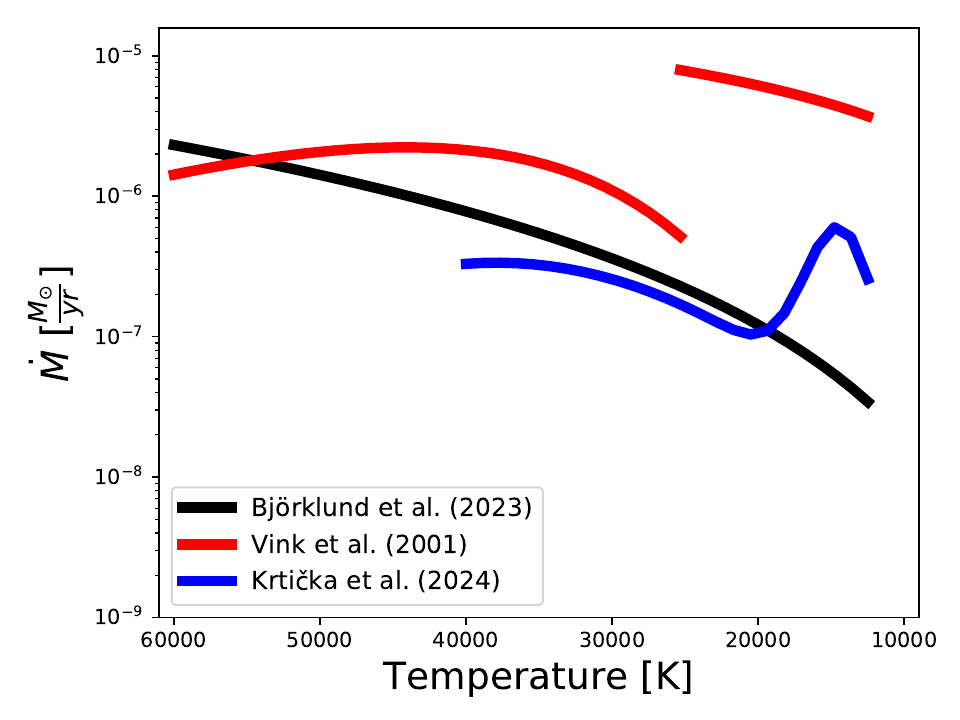}}
    \caption{\blue{Comparison of mass-loss rates from prescriptions by \cite{vink_mass-loss_2001, bjorklund_new_2023, krticka_new_2024} over the temperature range from $60-12.5$\,kK. Key stellar parameters in the left panel are: $\log_{10}(L_\star/L_{\odot}) = 5.3$, $M_\star =\,25 M_{\odot}$ and, $Z = 0.5\,Z_{\odot}$. This results in $\Gamma_{\rm e} = 0.21$. In the right panel adopted properties are $\log_{10}(L_\star/L_{\odot}) = 5.6$, $M_\star = 25\,M_{\odot}$ and, $Z = 0.5\,Z_{\odot}$, yielding $\Gamma_{\rm e} = 0.42$. The full lines signify that the result of the fit-formulate given by these authors are shown rather than the actual underlying model-sets. For the \citeauthor{vink_new_2000} rates, for which a mass-loss jump is identified to cover the regime $\sim 22.5-27.5$\,kK, enforcing a strict discontinuity (i.e. at a single temperature) implies that the size of the $\dot{M}$-jump is somewhat enhanced as compared to the underlying models.}}
    \label{fig:ULLYSES_bi}
\end{figure*}

\blue{Considerable effort has been invested to understand the physical nature and effects of terminal velocity and mass-loss bi-stability.
The \citet{vink_new_2000} models parametrise the wind velocity law $\varv(r)$, apply the  \citet{sobolev_1960} approximation for line transfer, constrain $\dot{M}$ from a global energy balance argument, and employ a modified nebular approximation to compute the state of the gas. In follow-up work, \cite{muller_consistent_2008} and \citet{2012A&A...537A..37M} used their computed Monte-Carlo line force to fit a parametrised radial function. By means of this function then, the equation of motion can be solved in an iterative manner until $\dot{M}$ and $\varv_\infty$ are converged. \cite{vink_fast_2018} applied this technique to luminous stars of different $T_{\rm eff}$ and found sharply decreasing $\varv_\infty$ accompanied by mass-loss rates that increased by more than an order of magnitude when decreasing $T_{\rm eff}$ from 25 to 18 kK (while keeping metallicity, luminosity, and mass fixed, see their Fig. 1). %Recent line-driven wind models by \citep[e.g.,][]{2012A&A...537A..37M,krticka_comoving_2017, bjorklund_new_2021, bjorklund_new_2023,krticka_new_2024} improve on these assumptions. For instance, in \cite{bjorklund_new_2023} the equation of motion is explicitly solved (so without the need for a preset velocity-law and keeping momentum conserved locally) and radiative transfer is solved in the co-moving frame \citep[CMF; see also][]{puls_atmospheric_2020}. Their results show a very different behaviour in the previously predicted by-stability region. Most notably, they do not find any upward increase in mass-loss rate with decreasing $T_{\rm eff}$ (when keeping the stellar luminosity to mass ratio fixed). Key reasons that likely contribute to this difference are that $\dot{M}$ is now set by conditions near the sonic point region employing line opacities that result from CMF radiative transfer.
%From the theoretical side, 
More recently, steady-state line-driven wind models that are based on co-moving frame (CMF) radiative transfer have been calculated \citep{krticka_comoving_2017, sander_coupling_2017, sundqvist_new_2019, bjorklund_new_2021, bjorklund_new_2023, krticka_new_2024}.
These simulations solve the equation of motion without parametrisation, and keep momentum conserved locally throughout the atmosphere and wind.}
%By using co-moving frame (CMF) solutions to the radiative transfer equation these simulations compute the radiative acceleration \citep{puls_atmospheric_2020}, and show a very different behaviour in the previously predicted bistability region. 
Specifically, because of the difficulty of driving wind material through the critical sonic point the models computed by \citet{bjorklund_new_2023} for the B-star regime do not show any upward increase in mass-loss rate with decreasing $T_{\rm eff}$ (again keeping the metallicity and stellar luminosity to mass ratio fixed). The models by \citet{krticka_new_2024} do find a localised and small mass-loss increase (followed by a sharp decrease), but at significantly lower $T_{\rm eff}$ than 21\,kK (see also \citealt{petrov_two_2016}). 
These calculations are also based on CMF radiative transfer and solve the full steady-state equation of motion. However, they scale their radiative accelerations to corresponding calculations based on the Sobolev approximation causing their critical point to shift from the sonic point to the point in the supersonic wind where the radiative-acoustic wave (known as an 'Abbott wave', \citealt{abbott_theory_1980}) speed equals the flow velocity. Additionally, they compute bound-bound rates in the statistical equilibrium equations using this Sobolev approximation, neglecting velocity curvature effects in near-sonic regions \citep{owocki_line-driven_1999} These may be potential causes
%that corresponds to the radiative-acoustic waves known as "Abbott-waves" \citep{abbott_theory_1980, krticka_comoving_2017}; as such, the above-mentioned issues with driving through the sonic point are circumvented. 
%By contrast, in the method by \citet{sundqvist_new_2019} (as well as in the CMF method presented by \citealt{sander_coupling_2017}) the radiative acceleration is computed directly on a discrete adaptive radial grid and thus only depends explicitly on radius. 
%This 
%may be a potential cause 
for the differences in $\dot{M}$-behaviour seen between the (at first glance very similar) models by \citet{krticka_new_2024} and \citet{bjorklund_new_2023}.
Mass-loss rates from the three approaches outlined above are shown in Fig.~\ref{fig:ULLYSES_bi}, illustrating the significant differences in predicted rates for the region under study here. \blue{We point out that what is shown are fit-formulae to the underlying model-sets that not always capture the full $\dot{M}$ behaviour, introducing additional differences. For instance, in the case of the \citet{vink_new_2000} rates these formulae often enhance the size of the jump (which in the underlying model-set typically is about a factor of 5, \citealt{vink_nature_1999}) through enforcing a discontinuous mass-loss jump.}

%On the (evolution-)application side, \citet{bjorklund_new_2023} used MESA to illustrate and discuss the potentially large effect of not including a bistability jump in evolutionary calculations. 
%Another indirect effect in this respect regards spin-down of stellar rotation rates. 
%\cite{vink_nature_2010} note that BSGs below $T_{\rm eff} \sim 22$ kK typically have low ($< 100$ km/s) projected rotational speeds, suggesting this may be due to enhanced braking from increased mass-loss rates associated with the bistability jump. 
%%If the it does not mark the TAMS, but is the result of BSB it needs a core overshooting parameter that is large, to extend the TAMS to lower tempertures. Consequently, B (and even A) supergiants may be MS objects, possibly explaining why there are unexpectely many B supergiants. 
%\cite{keszthelyi_modeling_2017} further point out that if such bistability braking is not happening, it may be hard to explain the generally low spin rates of BSGs. 
%On the other hand, the evolution models presented in \citet{bjorklund_new_2023} do seem to undergo significant braking even in the absence of this bistability mass-loss increase. %(presumably illustrating the sensitivity due %other parameters included in such 1D evolution modeling, e.g. convective core overshooing). 
%%and that a so-far unidentified mechanism may need to be invoked to spin them down sufficiently (all subject to uncertainties regarding dMdt)  

So far, observational evidence of a bistability mass-loss jump has proven difficult to obtain. 
%the order of magnitude jump is more difficult to recover.
%With many not finding a high jump 
\blue{\citet{trundle_understanding_2004} comment that in their sample of small Magellanic cloud (SMC) B-(super)giants, for which they obtained optical spectra, the mass-loss rates of stars at the cool side of the jump are noticeably smaller than the \cite{vink_mass-loss_2001} prescription.}
Additionally, \citet{markova_bright_2008} do not find a jump from an optical H\,${\alpha}$ analysis of B-supergiants (BSGs) and neither do \citet{rubio-diez_upper_2022}, studying continuum infra-red and radio data of bright OB-stars.
\blue{Most recently a large scale study on 116 B-supergiants in the Galaxy, based on optical spectra only, found no evidence for a bistability jump \citep{de_burgos_iacob_2024}.}
\cite{benaglia_testing_2007}, however, do claim tentative signs of a mass-loss increase for stars cooler than $T_{\rm eff} \sim 21$\,kK in their radio analysis. A common denominator of these studies is that they rely solely on diagnostic features for which the associated opacities depend on the square of the density. This makes these mass-loss constraints prone to degeneracies introduced by wind inhomogeneities \citep[or `clumping', see, e.g.,][]{puls_mass_2008}.
By combining spectroscopy in the optical and ultra-violet (UV) (see also \citealt{bernini-peron_clumping_2023}), we here aim to break these degeneracies and derive absolute constraints on mass-loss rates as well as wind clumping properties.
%\red{Indeed, with the completion of the ULLYSES program \citep{roman-duval_ultraviolet_2020} we now have access to the critical UV data, supplemented by optical data from the XShootU collaboration \citep{vink_x-shooting_2023, sana_x-shooting_2024}}.\\

With the completion of the data taking in the Hubble UV Legacy Library of Young Stars as Essential Standards (ULLYSES) \citep{roman-duval_ultraviolet_2020} and XShooting ULLYSES (XShootU) \citep{vink_x-shooting_2023, sana_x-shooting_2024} programs we now a have a unique UV and optical data set of late-O and early- and mid-B type in the Large Magellanic Cloud (LMC). 
This allows us to, for the first time, constrain stellar, mass-loss {\em and} wind-clumping characteristics across the anticipated bistability region for a statistically relevant sample. 
Hence it also permits for the most direct confrontation of empirical results to predictions of mass loss and clumping \citep{driessen_theoretical_2019} in the $14-33$\,kK blue supergiant regime to date.
%In this paper we derive constraints on stellar, wind, and clumping parameters of the BSGs in the LMC present in the ULLYSES and XShootU sample, in order to study the region around $T_{\rm eff} \sim 21$ kK. 
To fit these new data we utilise a combination of {\sc fastwind} (v10.5, \citealt{puls_atmospheric_2005, sundqvist_atmospheric_2018}) and Kiwi-GA \citep{brands_r136_2022}.
%{\sc fastwind} is a unified model atmosphere code which allows us to compute synthetic line profiles of important indicative spectral lines very quickly in comparison to similar codes (e.g., {\sc cmfgen}, \citealt{hillier_01-25_1993}), by not computing all lines over a wide wavelength range but focussing in on a selection. Additionally, in {\sc fastwind} the wind clumping description involves a statistical two-component medium with a non-void "interclump" component (see description in Sect. \ref{sec:fastwind}), so that it is no longer necessary to assume that all the wind mass is contained within the dense clumps. 
%do not need to assume a void interclump medium, but can instead have a variable interclump density.
Previous empirical studies utilising this method for spectroscopic analysis include \cite{hawcroft_empirical_2021, brands_r136_2022,hawcroft_empirical_2024, Backs_inprep} Brands et al. (in prep.). 
From here on, we will refer to \cite{hawcroft_empirical_2021} as H21 and \cite{brands_r136_2022} as B22.

%\JS{ Also, you need to throw a bone to the PoWR group, and insert reference to them somewhere in intro as well...}.   
%This paper utilises a method which combines {sc fastwind} with a GA to aid the fitting process just as has been done in similar studies on different objects \cite{hawcroft_empirical_2021, brands_r136_2022}.
%Here we use the genetic algorithm code (Kiwi-GA \cite{brands_r136_2022}) which allows us to efficiently fit all 18 {\sc fastwind} parameters at the same time. 

%The statistical method which a GA utilises also removes the bias and subjectivity of the 'by-eye' fitting methods while giving a way to compute consistent error margins based on the actual quality of the individual fits. \
%As the previously predicted bistability jump is very large it should be possible to determine if it is present by comparing to the mass-loss rates empirically derived in this paper. 
We first present the dataset utilised to conduct our investigation in Sect.~\ref{sec:data} and introduce {\sc fastwind} (Sect.~\ref{sec:fastwind}) and Kiwi-GA (Sect.~\ref{sec:kiwi-ga}). In Sect.~\ref{sec:results}, we describe the stellar, wind, and clumping parameters obtained from fitting the data, with special attention to mass-loss rates and clumping properties.
In Sect.~\ref{sec:discussion}, we first discuss the influence of some of the choices made during the fitting process, after which we examine the results. 
\blue{Our main findings are that mass-loss does not seem to 'jump' across the bi-stability jump, nor do the clumping properties (which have sizeable uncertainties) show a discontinuous behaviour. 
%JS? Not correct, since we have not detected this at all. Rather it may be a decline, but no jump. 
%We do detect a bi-stability in the terminal velocities, as found previously by \citet{lamers_terminal_1995,2006A&A...446..279C}, though the scatter is large. 
We do find a similar downward trend in terminal wind speeds with effective temperature as has been reported previously, though the scatter in especially the ratio $\varv_\infty/\varv_{\rm esc}$ is large for our sample. We also find that, on average, about 40 \% of the total wind mass seems not to reside in dense clumps, bur rather in the more diluted medium around them. Implications of these results to the wider field are discussed.}

\section{Observations and methodology}
In this section we describe our chosen sample of LMC OB-supergiants (BSGs), the stellar atmosphere and spectral synthesis code {\sc fastwind} used to model the stars in our sample, and the genetic-algorithm (GA) code, Kiwi-GA, used to determine the optimum parameter values and their uncertainties.

\subsection{Dataset}\label{sec:data}

We selected 15 B-supergiant stars and two late O\,I-III stars in the LMC from the {\em Hubble} Space Telescope legacy program ULLYSES \citep{roman-duval_ultraviolet_2020} in which UV spectra for $~$250 sources in the Magellanic Clouds are taken with the Space Telescope Imaging Spectrograph (STIS) and Cosmic Origins Spectrograph (COS) spectrographs. This program is complemented by the Very Large Telescope (VLT)/X-shooter large program XshootU \citep{vink_x-shooting_2023,sana_x-shooting_2024} that collects optical spectra of the same sources. Our dataset covers the spectral range B7 through O8.5, which roughly corresponds to temperatures from 14\,kK to 33\,kK, i.e., the range in which the bistability jump in mass loss is predicted and the terminal wind velocity divided by the escape speed has been seen to change by about a factor two \citep{lamers_terminal_1995}. The reduced STIS spectra used are from ULLYSES data release DR6; the X-shooter spectra are from XShootU DR2. The photometric K-band data used for absolute calibration of the spectra is from \cite{vink_x-shooting_2023}. Star identifications, spectral type and K-band photometry are given in Table~\ref{tab:sample_table}. 
% We were not able to use all B-supergiants in the ULLYSES sample as, although the goal of the ULLYSES program was to avoid binary stars, there were several stars which showed clear signs of radial variability.\\
The Hubble space telescope (HST)/COS has multiple gratings, the G130M/1096 grating has a spectral resolving power of $~6000$ ranging from $940-1240 \AA$. The G130M/1291 and the G160M grating have a resolving power ranging from 11000 to 19000 and cover the wavelength range from $1141-1783 \AA$. 
The HST/STIS spectra obtained using the E140M and the E230M grating cover a range of $1141-1708 \AA$ and $1608-2366 \AA$ respectively. The resolving power of the E140M grating is 48500, and the E230M grating has a resolving power of 30000.
If multiple data sets would cover a modelled feature, the used data was chosen based on signal-to-noise ratio and resolving power.
The optical data has two sections, the VIS and the UVB arm from VLT/X-shooter which have been stitched together by \cite{sana_x-shooting_2024}. The UVB arm covers a spectral range from $3100-5500 \AA$ with a spectral resolving power of 6700 for the chosen slit width of $0.8"$. The VIS arm covers the longer wavelengths from $5500 \AA$ to $8000 \AA$ and has a resolving power of 11400 for the slit width of $0.7"$. 
The visual spectra have a high S/N which is usually higher than 100. The UV data has a lower S/N at around 20.

\begin{table*}[]
    \centering
    \caption{Full sample of stars used in this paper.}

    \begin{tabular}{l|llll}
        Star name & Revised SpT&SpT & $K_s$ (mag) & UV-observation\\
        \hline
        Sk$-67^{\circ} 195$ &B7 Ib& B6 I & 12.68& E230M; G130M; G160M; G185M; G430L\\
        Sk$-68^{\circ} 8$  & &  B5 Ia$^+$ & 11.16& E230M; G130M; G160M; G430L\\
        RMC-109 &&B5 Ia   & 12.09& E230M; G130M; G160M\\
        Sk$-67^{\circ} 78$ & & B3 Ia & 11.35& E230M; FUSE; G130M; G160M\\
        Sk$-70^{\circ} 50$ & & B3 Ia & 11.22& E230M; FUSE; G130M; G160M\\
        Sk$-68^{\circ} 26$ & & B2 Ia &11.27 & E230M; FUSE; G130M; G160M; G230LB; G430L\\
        Sk$-70^{\circ} 16$ &B2 II& B4 I & 13.34& G130M; G160M; G185M; G430L\\
        Sk$-69^{\circ} 52$ & &  B2 Ia & 11.58 & E230M; FUSE; G130M; G160M\\
        Sk$-67^{\circ} 14 $&B1 Ia& B1.5 Ia & 11.90& E140M; E230M; FUSE \\
        Sk$-69^{\circ} 140$ &B1 Ib& B4 I & 13.07& E230M; G130M; G160M\\
        Sk$-66^{\circ} 35$ & BC1 Iab& BC1 Ia & 11.69 & E140M; E230M; FUSE\\
        Sk$-69^{\circ} 43$ &B0.7 Ia&  B0.5 Ia & 12.15& E140M; E230M; FUSE \\
        Sk$-68^{\circ} 41$ &B0.7 Ia& B0.5 Ia & 12.24& E140M; E230M; FUSE\\
        Sk$-68^{\circ} 52 $& &B0 Ia &11.72& E140M; E230M; FUSE\\  
        Sk$-68^{\circ} 155$ &O9 Ia& B0.5 Ia & 12.6& E230M; FUSE G130M; G160M\\
        Sk$-67^{\circ} 107$ &O8.5 II& O9 Ib & 13.06 & E140M; FUSE\\
        Sk$-67^{\circ} 106$ &O8 II& O8 III & 12.39& E140M; FUSE; G230LB; G430L\\
        \hline

    \end{tabular}
    \label{tab:sample_table}
    \begin{tablenotes}
         \item \textbf{Notes}:  The spectral type and photometry is taken from \cite{vink_x-shooting_2023}. We also include revised spectral types based on the XShootu data set from Crowther (in prep.). All objects have been observed using STIS. Here we show which grating was used and whether it was also observed using Far Ultraviolet Spectroscopic Explorer (FUSE).
    \end{tablenotes}
\end{table*}

\subsection{{\sc fastwind}} \label{sec:fastwind}

The unified model atmosphere code {\sc fastwind} solves the NLTE rate equations assuming statistical equilibrium in a spherically symmetric and stationary extended stellar envelope comprising both the photosphere and the outflowing wind. 
We employ here version 10.5 (see \citealt{puls_atmospheric_2005, rivero-gonzalez2012, sundqvist_atmospheric_2018, carneiro_carbon_2018}), which accounts for the accumulative feedback-effects from the multitudes of metallic spectral lines upon the radiation field and atmospheric structure by means of a computationally efficient statistical method.
Some of those chemical elements, that are subsequently used for detailed spectroscopy (here H, He, Si, C, N, O), are treated separately using more precise radiative transfer calculations (for details, \citealt{puls_atmospheric_2005}).  

The atmospheric structure is computed assuming \mbox{(quasi-)hydrostatic} equilibrium in the deeper atmospheric layers, connecting smoothly to an analytic wind outflow with radial velocity field $\varv(r) = \varv_\infty (1 - b R_\ast/r)^\beta$ and average density structure $\langle \rho(r) \rangle = \dot{M}/(4 \pi \varv(r) r^2)$.
\blue{The transition between the (quasi-)hydrostatic atmosphere and the wind is set to occur at a velocity $0.1$ times the gas sound speed at effective temperature.}
Here the input parameters describing the wind are the mass-loss rate $\dot{M}$, wind acceleration parameter $\beta$, and terminal wind speed $\varv_\infty$. $b R_\ast$ is the radius at the photospheric boundary. 
These parameters supplement the effective temperature $T_{\rm eff}$, surface gravity $g$, stellar radius $R_\ast$, and atmospheric chemical abundances in the list of input stellar and wind parameters describing the model atmosphere. 
Regarding elemental abundances we here adopt the solar composition by \cite{asplund_2009}, and scale the overall metallic abundances ($Z$) accordingly to 0.5 times solar composition to match the LMC.
The helium to hydrogen number abundance ratio, $Y_{\rm He}$, as well as some specific metal abundances (carbon, nitrogen, oxygen), are free-parameters in the fitting procedure.
The silicon abundance is set to 7.2 dex (defined as $\log_{10} \epsilon_{\rm Si} = log(N_{\rm Si}/N_H+12$) as in \cite{brott_rotating_2011} (differing slightly from the value given in \citealt{vink_x-shooting_2023}) and is not used as a fit parameter.
This constant Si-abundance has been chosen as we expect the Si-abundance to be nearly constant over the LMC.
Additionally, the GA methodology used here is not optimised to constrain abundances as these will only have a small effect on the overall goodness of fit.
We re-fitted several of the targets spread over the temperature range to determine if including the Si-abundance (and 
micro-turbulence, see below) as a free parameter changes the results of the other parameters. All fits of the test stars had a large uncertainty margin for the Si-abundance which enveloped our baseline of 7.2 dex. 
This suggests that our choice of not including the Si-abundance as a fit-parameter has not influenced our other results. Additionally, the new best fit Si-abundance was within 0.2 dex of the assumed value.

Inhomogeneities in the wind are described with the effective opacity formalism from \cite{sundqvist_atmospheric_2018}, incorporating modifications in opacity associated with clumping in physical and velocity-space. 
This formalism assumes a stochastic two-component wind consisting of over-dense clumps and a rarefied interclump medium. 
Mean densities are related to the root-mean-square (rms) average via a clumping factor $f_{\rm cl} \equiv \langle \rho^2 \rangle / \langle \rho \rangle^2 \ge 1$ and the interclump densities $\rho_{\rm ic}$ set by $f_{\rm ic} \equiv \rho_{\rm ic}/\langle \rho \rangle \ge 0$.
Light-leakage effects in spectral lines ('velocity-porosity') are accounted for by a velocity filling factor $0 \le f_{\rm vel} \le 1$, defined as the fraction of the velocity field covered by the dense clumps. 
Wind clumping is assumed to start at some wind velocity $\varv_{\rm cl,start} > a$, where $a$ is the gas sound speed, and then increase linearly with velocity until the input parameter-values for $f_{\rm cl}$, $f_{\rm ic}$, and $f_{\rm vel}$ are reached at $\varv_{\rm cl,max}$. 
Clump optical depths for spectral lines are calculated from the input parameters using the \cite{sobolev_1960} approximation; that is, we do not assume that clumps are optically thin or thick but instead compute them for all lines
\footnote{All continuum clump optical depths are also computed, following \cite{sundqvist_atmospheric_2018} using a porosity length $h(r)/R_\star = \varv(r)/\varv_\infty$. As we are focused here on line diagnostics, and the porosity length has a negligible influence on the atmospheric structure and ion balance in the atmospheres considered here, we will not discuss this parameter further.} 
according to the structure parameters, to evaluate corresponding effects upon the ionisation balance and spectrum formation. 
The ionisation balance is then calculated for an average effective medium taking into account the clumped and interclumped components.
Moreover, the micro-turbulent velocity is assumed to increase linearly with wind velocity from a fixed photospheric value $\varv_{\rm mic} = 10$ km/s to a maximum value set by an input parameter scaled to the terminal wind speed. 
To quantify the effects of fixing the micro-turbulence to 10 km/s, we refitted a select number of stars over the investigated temperature range while instead fixing $\varv_{\rm mic}$ to 5 km/s, 20km/s, and finally allowing it to be a free parameter. 
Results of these runs are summarised in Fig. \ref{fig:micro_test}. 
In short, we found that the best fit effective temperature and mass-loss rate changed only within the error-margins.    
The CNO abundance overall also did not show any strong changes; typically the oxygen and nitrogen abundance stayed the same within their sizeable uncertainties.
The carbon abundance also generally showed only variations within the error-margin, though for one source the abundance increased by 0.5 dex when lowering $\varv_{\rm mic}$ to 5 km/s and for another source when allowing the micro-turbulence free, the carbon abundance decreases by 1 dex. 
When letting the micro-turbulence free, on average we obtained slightly higher values than the assumed 10 km/s, however as mentioned above this does not significantly impact the overall results in focus here.
%1-$\sigma$ best fitting region covered the chosen 10 km/s value.
% As mentioned above, we tested a few models to instead have the photospheric microturbulence as a free input-parameter. 
% Results from these tests varied, but for most of the test stars the uncertainty region around the best-fit values included 10 km/s. 
%These tests thus suggest that fixing the microturbulence to 10 km/s does not strongly impact our overall results. \JS{It'd be nice if you could illustrate some of this in a figure, at least in Appendix.(Perhaps showing the run with vmic as free parameter, which is quite interesting?)}}

The full method we employ to account for effects of wind clumping is presented and discussed in detail in \cite{sundqvist_atmospheric_2018}, and recent quantitative spectroscopic applications involve H21 and B22. 
Finally, we note that in the limit of a void interclump medium ($f_{\rm ic} \rightarrow 0$), neglecting velocity-porosity, and assuming all clumps are optically thin and follow the mean velocity field, our method recovers the alternative clumping descriptions included as defaults in alternative atmospheric codes such as Potsdam Wolf-Rayet Stellar Atmospheres ({\sc Po\textbf{}WR}) \citep{grafener_line-blanketed_2002,hamann_temperature_2003},  and {\sc cmfgen} \citep{hillier_treatment_1998} (see also discussion in Sect. \ref{cl_inter}). We note, however, that even in this limit, the onset and radial stratification of the clumping parameters we assume in this paper may be different from what is typically assumed in these alternative codes.
%\textbf{mention different fcl stratification}.

We also include X-rays produced by wind embedded shocks, as this turns out to be necessary to reproduce some features in the UV resonance lines, particularly in the surprisingly strong C\,{\sc iv} profiles of cooler B-stars. 
The details of the X-ray implementation can be found in \citet{carneiro_atmospheric_2016}. 
In this description the energy emitted by the hot gas is given by:
\begin{align}
    \epsilon_{\nu} = f_{\rm X}(r) n_{\rm p}(r) \Lambda_{\nu}(n_{\rm e}(r), T_{\rm s}(r))
\end{align}
Here $n_p$ and $n_e$ are the proton and electron density of the stationary pre-shock wind, $T_s$ the shock temperature and $f_X$ defined as $f_X = 16e_s^2$, where $e_s$ is the X-ray volume filling factor, and the factor 16 included to account for the density jump in a strong adiabatic shock (making the allowed range for $f_x$ be 0 to 16).
%An extra advantage of this notation is that is inline with WM-basic, KK09. 
$\Lambda_{\nu}$ is the frequency-dependent volume emission coefficient per proton and per electron. 
The shock temperature is approximated in the strong shock limit by:
\begin{align}\label{eq:jump-T}
    T_s(r) = \frac{3}{16} \frac{\mu m_{\rm H}}{k_{\rm b}}u^2
\end{align}
with u being the shock velocity and $\mu$ is the mean atomic weight.
%Here was assumed that the jump velocity is much larger than the local sound speed ($u^2 >> a_s^2$) resulting a strong shock.
The shock velocity itself is set by:
\begin{align}
    u(r) = u_{\infty} \left[ \frac{\varv(r)}{\varv_{\infty}}  \right]^{\gamma_{\rm s}}
\end{align}
where both the maximum jump speed $u_{\infty}$ and the hardness-parameter $\gamma_s$ are input parameters. 
%and $v_{\infty}$ is the terminal wind speed. %Already defined? 
Finally an onset radius of the X-ray emission is chosen by 
the minimum of $R^{\text{input}}_{\text{min}}$ and the radius at which $\varv_{min} = m_x a$ is reached. 
Here $m_x$ is a parameter which one can tune freely if one would want to change the onset of X-rays.
In this paper we keep $R^{\text{input}}_{\text{min}} = 1.45 R_\star$,  $\gamma_{\rm s} = 0.75$, $m_x = 30.0$ constant. 
$R^{\text{input}}_{\text{min}}$ has been taken from \cite{pauldrach_radiation-driven_1994}, $\gamma_{\rm s}$ is chosen to be in between the values of \cite{krticka_nlte_2009, carneiro_atmospheric_2016} and \cite{pauldrach_radiation-driven_1994}, finally $m_{\rm x}$ is taken from the best-fit value of \cite{pauldrach_radiation-driven_2001}. 
We do fit the filling factor ($f_X)$ and the maximum jump velocity ($u_{\infty}$); effectively this means we are only fitting the overall strength of the X-rays while keeping the onset and hardness constant for each star. %Although they are dependent on other fitting parameters we do fit like the terminal wind speed.  

\subsection{Kiwi-GA}\label{sec:kiwi-ga}

We are trying to constrain many parameters simultaneously in complex systems. Quite generally this means that regular fitting routines may have issues getting around local minima to find robust global solutions at a reasonable computation cost. 
For the problem at hand, making a grid large enough to cover the full parameter space is currently not a practical solution. 
A genetic algorithm (GA) is a way to try and minimise these issues, as it is capable of testing new models that are not decided by gradient descent; the specific genetic algorithm used in this work is Kiwi-GA\footnote{full code available on github \href{https://github.com/sarahbrands/Kiwi-GA}{Kiwi-GA}} \citep{brands_r136_2022}.\\

The base version of Kiwi-GA starts out by computing a first generation of {\sc fastwind} models within the chosen parameter space for all parameters one wants to fit.
For all models in the initial generation a goodness of fit is calculated as

\begin{align}\label{eq:chi2}
    \chi^2 = \sum^{N}_{i = 0} \left(  \frac{\mathcal{F}_{\text{obs, i}}- \mathcal{F}_{\text{mod, i}}}{\mathcal{E}_{\text{obs,i}}} \right)^2.
\end{align}
Here $N$ is the number of points in the spectrum, $\mathcal{F}_{\text{obs, i}}$ is the observed normalised flux, $\mathcal{F}_{\text{mod, i}}$ is the normalised model flux and $\mathcal{E}_{\text{obs,i}}$ is the uncertainty in the observed flux. 
The points in the spectra are modified object to object to minimise the continuum as this might lower $\chi^2$ artificially. Keep in mind, however, that the absence of specific lines also helps to characterise stars.
The best fit will remain the same even when including a lot of continuum, but the error-range might change slightly as it scales with $\chi^2$ (see equation: \ref{eq:RMSEA}, \ref{eq:error_RMSEA}).

The next generation will be formed by combining parameters of the fittest previous generations. 
For example, the effective temperature of one of the best fitting models may be combined with the surface gravity of another well fitting model to create a model in the new generation. 
In an attempt to get a good coverage of the parameter space, two possible mutations (change to the individual parameters) can occur on the parameters.
There is a 70\% chance that a small mutation occurs, and the magnitude of this mutation is determined by a Gaussian centred around the current value with a small width. 
The goal of these small mutations is to give an in-depth exploration of the region around the best fit value.
There is a 30\% chance that a large mutation occurs; these mutations are also normally distributed but the width of the Gaussian for this mutation is 30\% of the full parameter space.
These two mutation changes are not additive, 21\% of the time the parameter is kept constant.
%These larger mutations try to ensure that the GA can 'escape' local minima and gives a better change to find the lowest $\chi^2$ value.
\blue{This method has been compared to the traditional by-eye fitting and a grid-search \citep{sander_x-shooting_2024} using different spectral synthesis codes ({\sc Po\textbf{}WR}, \citealt{grafener_line-blanketed_2002,hamann_temperature_2003, oskinova_neglecting_2007}; CMFGEN, \citealt{hillier_treatment_1998}). They studied the differences in resulting fit parameters when using the different fitting routines on 3 O-stars. Overall the different methods gave comparable results, but differences in $T_{\rm eff}$ up to 3000K were noted and mass-loss also saw variation of around 0.3 dex. We note that since, in addition to the different fitting techniques, also clumping descriptions vary between these codes, consequently that part of these differences may arise because of that (see also discussion in \citealt{sander_x-shooting_2024}).}  
%\JS{For context of THIS study; smooth winds or what clumping descriptions were assumed, sicne they differ?}}

\subsection{2-step process}
Described above is the base version of Kiwi-GA.
However, in an attempt to make the fitting of the 18 parameters more consistent, we split the above routine into 2 steps. 
The motivation being that, even though a GA attempts to not get stuck in local minima, in the normal approach we could clearly see that $T_{\rm eff}$ was not consistently correctly estimated. 
This was apparent from He\,{\sc ii} lines not being present in the data but being quite strong in the best fit.
To solve this issue, we aimed to start the GA in a good estimate of the $T_{\rm eff}$.
To obtain this estimate, we decided to first perform a simple run on a selection of optical lines only (all optical H, He, and Si lines).
This first optical-only run only includes 6 fit parameters ($T_{\rm eff}$, $\log g$, $Y_{\rm He}$, max $\varv_{\rm rot} \sin i$,  $\beta$, $\dot{M}$) and a smooth wind outflow is assumed, which makes this part of the fitting routine very efficient while giving consistent results after only 20 generations with 107 models for each generation. 
% The UV lines added in in the full run are not the prime temperature and surface gravity in highly sen by lines found only in the optical regime (He and Si lines for $T_{\rm eff}$ and Balmer lines for $\log g$) it makes sense to use this first fit as an indication of these parameters.
% No helium lines are added by including the UV lines or the other optical lines so $Y_{\rm He}$ should vary only a very minor amount. 
% The mass-loss rates are only a maximum mass-loss rate as they are only sensitive to $H_{\alpha}$, as we describe in section \ref{not sure} this only gives an upper estimate of the mass-loss rate.
% The $\varv sin i$ includes all broadening and should not be influenced by the UV-lines as they should be consistent with the broadening of the weak Si-lines and the He-lines.

The results of the first run are then used as a simulated first generation for the full run using all 18 parameters ($T_{\rm eff}$, $g_{\rm eff}$, $Y_{\rm He}$, CNO-abundance, upper limit on $\varv_{\rm rot} \sin i$,  $\dot{M}$, $\varv_{\infty}$, $\beta$, $f_{\rm cl}$, $f_{\rm ic}$, $f_{\rm vel}$, $\varv_{\rm cl,start}$, $\varv_{\rm cl,max}$, wind turbulence, $f_X$, and $u_{\infty}$) and all lines (table \ref{tab:linelist}).
Here $g_{\rm eff}$ is the measured surface gravity which is reduced by the rotation ($g_{\rm eff} = g_\star- (\varv_{\rm rot} \sin i)^2/R_\star $ ).
This means the first generation of the full run all have the same 6 parameters of the optical-only run, the others are distributed randomly across the parameter space. 
We limit the variation of $Y_{\rm He}$ and the maximum of $\varv_{\rm rot} \sin i$ to the uncertainty region of the first fit to keep the influence of the UV lines on these parameters limited. 
The full run with 18 parameters is ran for an additional 30-50 generations resulting in a total of 50-70 generations with 107 models for each generation.
The values and uncertainty margins of $Y_{\rm He}$ and max $\varv_{\rm rot} \sin i$ are also derived from the optical-only run. 
This approach ensures that the full run starts with predisposed values at approximately the correct temperature, surface gravity, etc. 
% It also strengthens the effect the chosen optical lines have on these parameters; this is indeed generally a good thing as the chosen optical lines are known primary diagnostics for these parameters. 
Additionally, the first run also gives a good upper limit for $\dot{M}$ before adding all complexities of wind clumping and means we know something might be wrong if we find a higher $\dot{M}$ in the full fits.\\

The consequence of using this method is that, if for some reason the UV lines are formed with very different parameters than the optical lines we could be biasing the fits more heavily. 
However, this should in general not be the case and the opposite is a problem of the 1 step routine where the UV lines, as they have a lot more data points in the spectral window, carry a bigger weight in the fit while they have a bigger uncertainty. 
Another concern could be that we are biasing ourselves to a local minimum which is harder to escape, which is predisposed to a lower clumping factor as our best fit from the optical starts at a smooth wind.
However, after some tests we found consistently lower or equal $\chi^2$ values for the 2 step approach compared to the normal approach after the same total generations.
We note that, although both were run for approximately the same number of generation (50-70), the computation time of the 2-step approach is considerably lower as the simple optical-only run takes less time as we do not include X-rays, clumping , etc. 
% Over the 5 stars we tested both approaches, there was no tendency to lower clumping factors for the 2 step approach with sometimes resulting in slightly lower and sometimes slightly higher clumping factors for the 2 step approach in comparison to the 1 step approach.
% The other fit parameters also changed only within the error margins when changing the fit approach.  
% Figure \ref{fig:mdot_fclump_correlation} shows a good example of how the local minimum is overcome, as you can see the starting $\dot{M}$ by the line of low $\chi^2$ models at a mass-loss rate of -5.8 but highly varying clumping factors which than converges to a higher $f_{\rm cl}$ at lower mass-loss rate.
% Keep in mind that we bias the $f_{\rm cl}$ at 1 because we assumed a smooth wind in the optical-only fit.
% The best fit however, ends up at $f_{\rm cl}$ in the range 6-17. 
% Overall the found $f_{\rm cl}$ is higher than one would expect from theory and sometimes by a lot \citep{driessen_theoretical_2019}, showing again that the induced bias is limited but could still be present.
% Even if the bias is present, it is consistent over all results meaning that any trends in mass-loss rates for instance are still real, but harder to compare to other studies.

\subsection{Uncertainty estimate}
% In this section we will describe the method of determining the confidence intervals on the fit parameters.
% First, we inspect the best fit model and check for significant or systematic deviations with the observed spectrum, focusing on the H\,${\alpha}$ line and the Si lines to check for clear problems with the mass-loss rate and temperature.
% The best fit of the GA is never perfect, e.g., due to missing blended lines as well as the approximations that are made in the atmospheric modeling.  
% For this paper 
We use the new method developed by Brands et al. (in prep) to estimate the error margins of the fits which uses the root mean square error of approximation (RMSEA) \citep{steiger_notes_2016}. 
Starting from the $\chi^2$ value (Eq. \ref{eq:chi2}) we compute the RMSEA as:
\begin{align}\label{eq:RMSEA}
    \text{RMSEA} = \sqrt{\max{\left(\frac{\chi^2 -n_{dof}}{n_{dof}(N-1)},0\right)}}. 
\end{align}
Here $n_{\rm dof}$ are the degrees of freedom and $N$ the total number of data points. 
As the lines are cut at slightly different wavelengths depending on the object the $n_{\rm dof}$ varies around a value of 5000.
When the RMSEA is close to 0, the fits are good. There is no formal cutoff point to decide when a model is no longer acceptable;
in this work the 1-$\sigma$ error cut-off for models is set to:
\begin{align}\label{eq:error_RMSEA}
    \alpha_{\text{RMSEA}} = 1.04 \cdot \min \left({\text{RMSEA}} \right)
\end{align}
RMSEA values of the best fitting model are typically on the order of 0.1-0.01.
The 2-$\sigma$ error is set to 1.09.
These values are calibrated to give similar results to the error-margins in \citet{brands_r136_2022}.
As the error cut-off scales with the best fit, bad fitting stars will have a larger error margin than good fits.

When evaluating the results of this paper we do not place too much weight on the individual best fit value, but instead focus more on the uncertainty region. 
There are many parameters in our fits which can be disentangled only up to a certain point, which the uncertainty takes into account naturally. 
Additionally, {\sc fastwind} (like all current atmospheric models), is most likely not able to account for all multidimensional phenomena (see, e.g., \citealt{schultz_synthesizing_2023}; \citealt{debnath_2d_2024}) in its parameterised 1D descriptions of the various structure parameters.  
For instance, wind clumping is handled as a statistical two-component medium which is unlikely to be a perfect description of reality (see also discussion in Sect. 4.5).  
The genetic algorithm approach is by its nature a statistical approach and although it does try to converge to a 'best-fitting' model the real value of this approach is that we can show the range of parameters for which the deviations in the spectra are small enough that the goodness of fit does not change significantly. This gives us global errors which automatically take into account all degeneracies. 

\subsection{Diagnostic line selection}
The initial line selection in the optical and UV was based on the lines used in \citet{hawcroft_empirical_2021} and \citet{brands_r136_2022}. 
These lines include all prominent H, He, C, N, O, and Si lines in the observed spectra and can be synthesised using {\sc fastwind} V10.5. 
As these papers focused primarily on hotter O-stars, however, a somewhat modified line list has been used here to focus more on the ionisation stages of C, N, O, and Si that are prominent in the cooler B-supergiants. 
For example, in this temperature range there are fewer clear lines from these atoms in the UV domain. 
For the optical line list we also considered the line list used in \cite{trundle_understanding_2004}.
In the appendix we list all lines used in the Kiwi-GA fitting routine (table \ref{tab:linelist}). 
% In the left column the ionisation stage of the line is given, in the middle column the wavelength of all the multiplets of this line are shown, and on the right the name of the diagnostic name of the {\sc fastwind} line is given.
% When lines are close together in wavelength space, or even overlapping, they are fitted in the same spectral window. 
%(but as temperatures change the 'strongest line' is not constant while the name never changes.

\begin{figure*}[h!]
    \centering
    \includegraphics[width=0.9\textwidth]{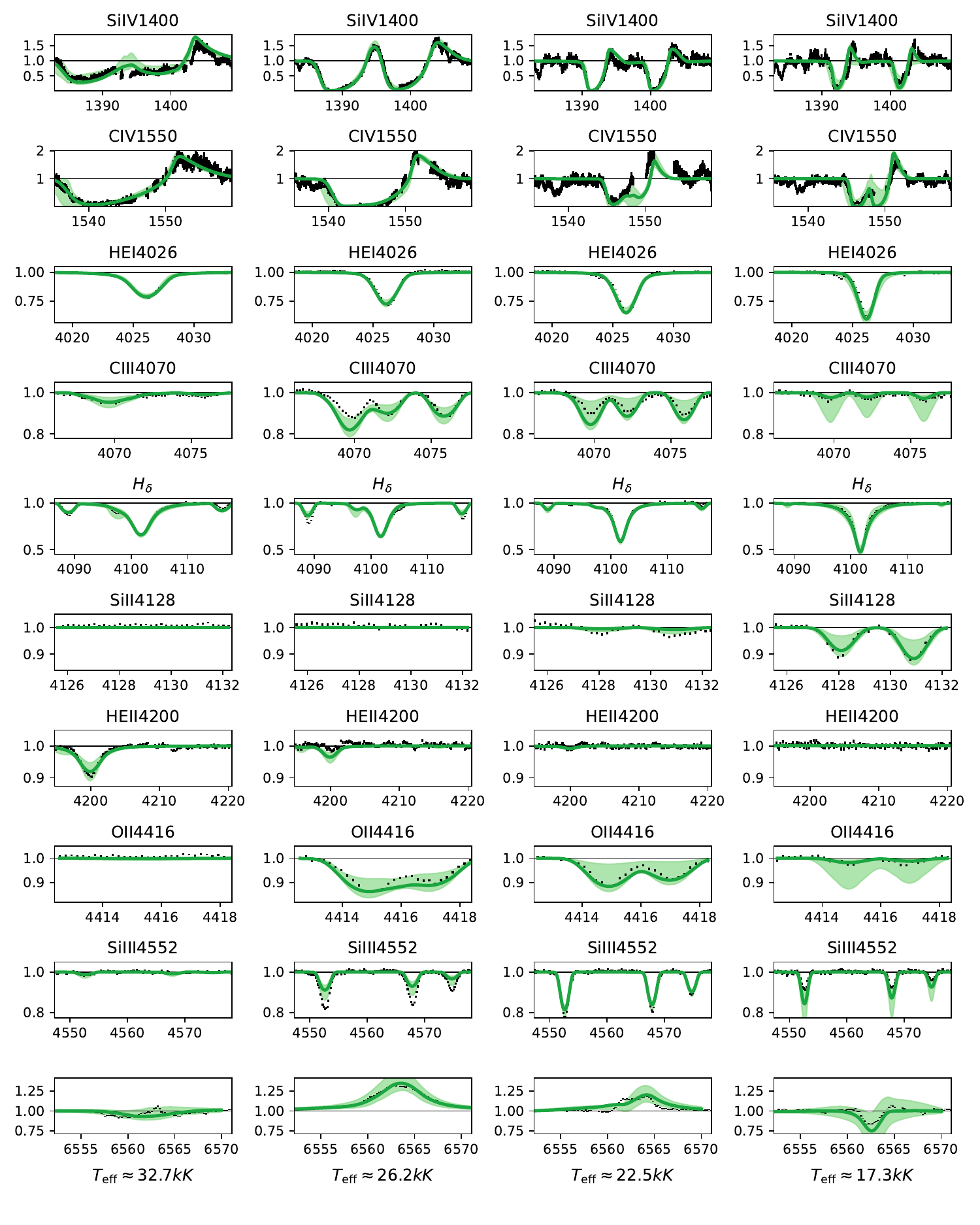}
    \caption{Comparison of diagnostic lines for 4 stars. The stars (Sk$-67^{\circ} 107$, Sk$-68^{\circ}41$,Sk$-69^{\circ}52$,RMC-109) are organised from high (left) to low (right) effective temperature (in spectral type: O8.5 II, B0.7 Ia, B2 Ia, B5 Ia ). The lines are selected due to their roles as good diagnostics for stellar or wind parameters. From top to bottom the lines are sorted by wavelength. Note that the line names here correspond to the names in table \ref{tab:linelist}. The green line shows the best fit and the green shaded region shows the 1-$\sigma$ uncertainty interval.}
    \label{fig:line_compare}
\end{figure*}

\subsection{Stellar parameter determinations}\label{sec:derived}
Not all parameters we are interested in are direct fit-parameters. 
There are a couple of extra parameters we can derive which are of particular interest, including: the mass, the radius, and $\Gamma_{\rm e}$.
The mass can be obtained using
\begin{align}
    M_\star = \frac{g_\star R_\star^2}{G},
\end{align}
where $G$ is Newton's gravitational constant and $g_{\star}$ is the surface gravity corrected for centrifugal effects ($g_\star = g_{\rm eff} +(\varv_{\rm rot} \sin i)^2 /R_\star $).
The stellar radius $R_\star$ is derived by using the fitted temperature and photometrically obtained stellar luminosity in the K-band \citep{vink_x-shooting_2023}, which we de-reddened.
Shifting this apparent magnitude to absolute magnitude is then readily done using the distance modulus to the LMC set to 18.48 \citep{pietrzynski_distance_2019}.
The de-reddening will influence this luminosity anchor very little as the K-band extinction is low and thermal emission by dust only becomes relevant at longer wavelengths. 
%while still high enough in energy to avoid problems with radiation of dust.
When investigating how the K-band reddening influences the mass-loss rate we have computed that a change of 0.1 magnitude in the K-band luminosity results in only an 8\% change in the mass-loss rate. This is thus lower than the expected error due to the depth of the LMC, which is on the order of 16\% \citep{subramanian_depth_2009}, and the typical errors of 
our fits are on the order of a factor 2.
With these values we only need to find the radius for which the luminosity in the chosen band matches the observations.

Obtaining good estimates of stellar masses 
%(or really the ratio $L_\star/M_\star$) 
is also important for the wind analysis, since wind strength scales strongly with the classical Eddington parameter:
\begin{align}\label{eq:gamma_e}
    \Gamma_{\rm e} = \frac{\kappa_e L_\star}{4 \pi c G M_\star },
\end{align}
with $\kappa_{\rm e}$ the electron scattering opacity and c the speed of light; $\Gamma_{\rm e}$ gives the ratio of radiative to gravitational acceleration in the case that the only opacity source is $\kappa_e$. When including these parameters on the various plots illustrating our results, we adopt 
\begin{align}
    \kappa_e = \frac{\sigma_T}{m_{\rm H}} \left(\frac{1+I_{\rm He} \cdot Y_{\rm He}}{1+4 \cdot {Y_{\rm He}}} \right).
\end{align}
Here $\sigma_T$ is the Thomson cross section, $m_{\rm H}$ is the mass of hydrogen, $I_{\rm He}$ is the ionisation of He giving either 1 or 2 free electrons for each He-atom (set to 1 for all B-stars in our sample and 2 for the O-stars); hydrogen is assumed to be fully ionised, and $Y_{He}$ is the helium to hydrogen number abundance ratio. 
Note that $\Gamma_{\rm e}$ does not take into account the line-driving effect of the radiation acceleration, but instead gives a generic radius independent value as a measure of the closeness of the star to the classical Eddington limit.

\section{Results}\label{sec:results}

In this section we present the stellar and wind parameters determined by the fitting method described above, applied to the sample of LMC stars. 
All the  derived parameters and corresponding error estimates are given in tables \ref{tab:stellar_param}, \ref{tab:derivedparam+xray}, and \ref{tab:wind_param}. 
%Among the parameters shown in Table \ref{tab:stellar+windparam} is the helium abundance. 
% In the illustrating plots showing results from the full sample (Figure \ref{fig:velocity+accelration} to \ref{fig:clumping}) we do not highlight which star corresponds to which data point, as we are interested in overall trends in primarily the wind parameters.

Figure \ref{fig:line_compare} illustrates fits for a selection of important lines for four typical stars in our sample. 
The stars are organised from hot to cold, as clearly visible in the He\,{\sc ii} lines which are present for the hottest star, but disappear when moving to the cooler objects. For the cooler stars the Si\,{\sc ii} lines instead become stronger, while the Si\,{\sc iv} line which is present in the red wing of the $H{\delta}$ line vanishes. The H\,${\alpha}$ line for all stars shows clear signatures of wind emission filling in the photospheric absorption component, indicating this line is a good mass-loss indicator throughout our sample (though see discussion below on some complicating factors). 
All four stars show P-Cygni like features in the UV Si\,{\sc iv} and C\,{\sc iv} lines, again indicating the presence of significant wind outflows. 
In this respect, we note that C\,{\sc iv} is clearly present in the wind of objects at around 17kK (although the doublet line there is narrower and weaker than in the other two stars); we discuss this further in Sect. \ref{sec:result_velocities}, as it has consequences for our modelling of the wind ionisation balance and derivation of terminal wind speeds in this regime.

%For all the derived values of each separate star the appendix has a table as well as the fits to the line-profiles.
\subsection{Stellar parameters}

\begin{figure*}
   	\centering
    \subfigure{\includegraphics[width=0.4\textwidth]{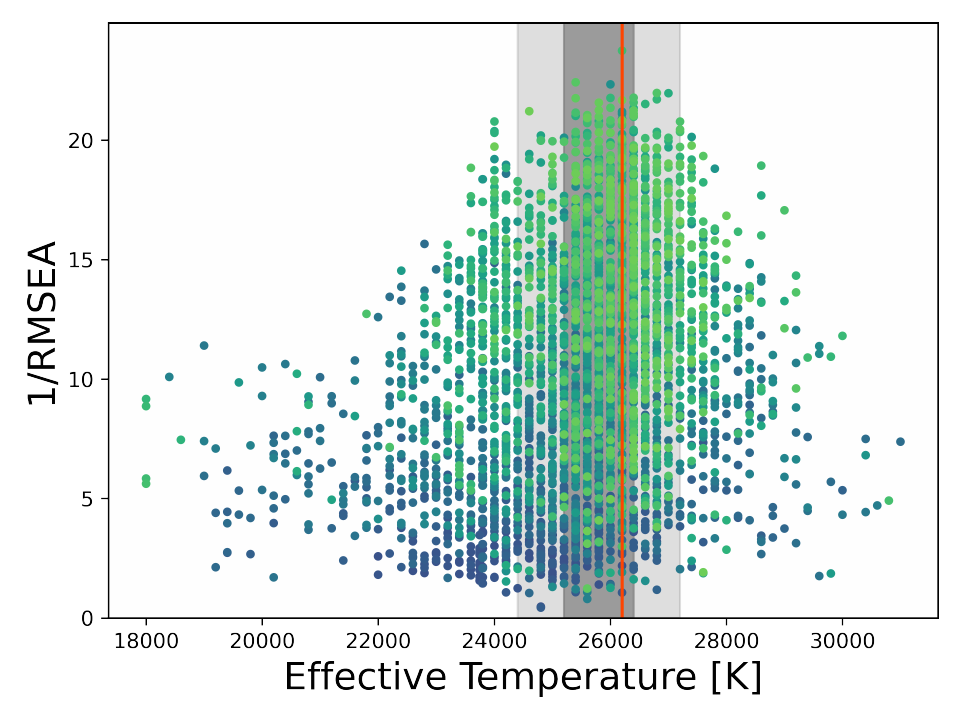}}
   	\subfigure{\includegraphics[width=0.4\textwidth]{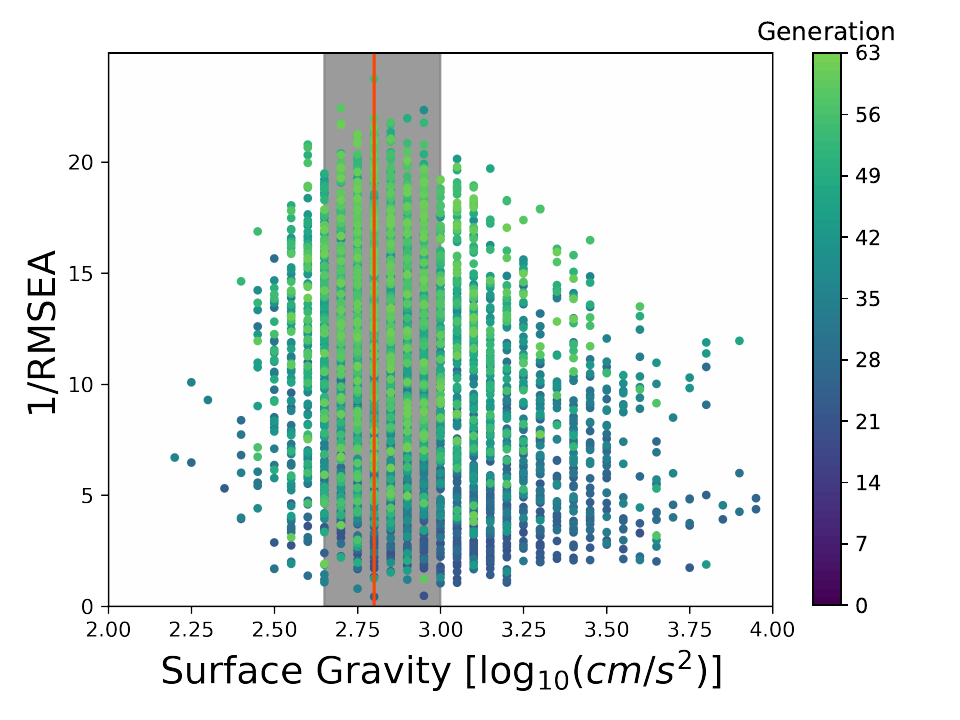}}
   	\subfigure{\includegraphics[width=0.4\textwidth]{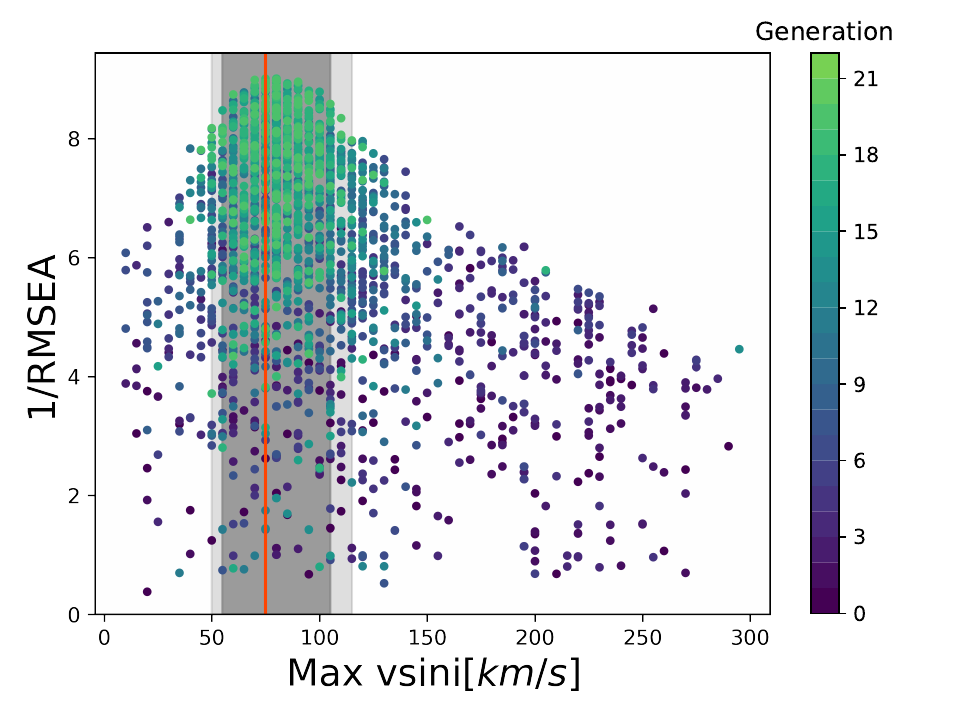}}
    \subfigure{\includegraphics[width=0.4\textwidth]{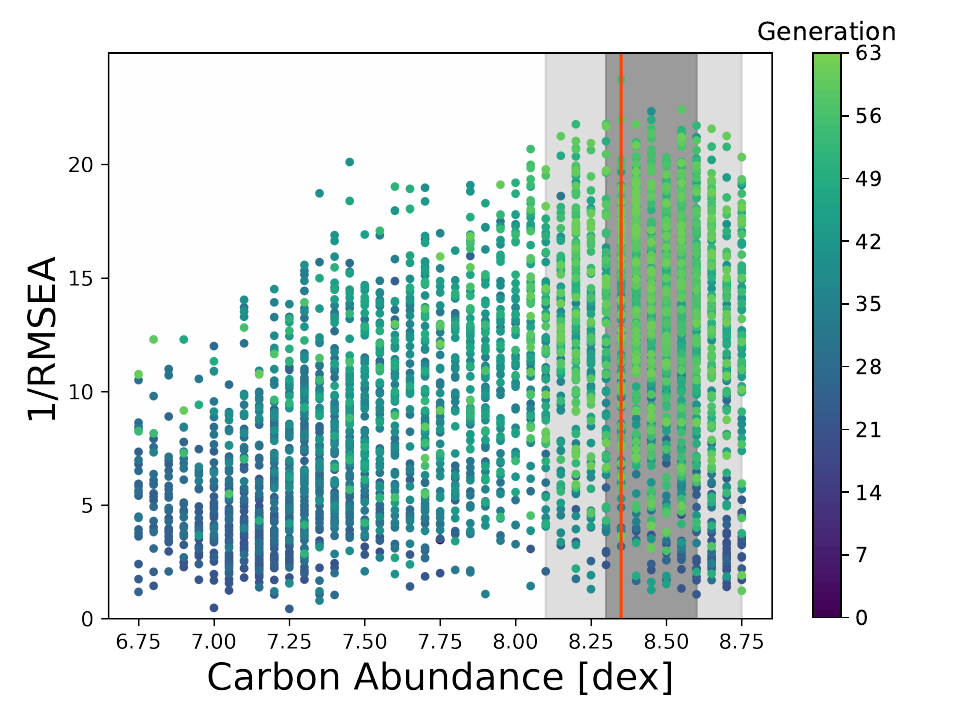}}
    \caption{Distribution of the quality of the fits for stellar parameters of star Sk$-68^{\circ}41$. In dark grey the 1-$\sigma$ interval is shown while the light grey shows the 2-$\sigma$ interval. The red line shows the best fit value. We only display the C-abundance as the other elements have similar fitting curves. As mentioned in section \ref{sec:kiwi-ga} and the following sections we derive the maximum projected rotational velocity from the optical-only fit which is why there are only 21 generations.}
    \label{fig:stellarfit}
\end{figure*}

Although our analysis and discussion will be focused mostly on wind parameters, we start by presenting the stellar parameters in table \ref{tab:sample_table}.
We fit seven stellar parameters in our models, namely the effective temperature  ($T_{\rm eff}$), the effective gravity at the stellar surface ($\log g$), a maximum (see further below) projected equatorial rotational velocity ($\varv_{\rm rot} \sin i$), the surface helium abundance ($Y_{\rm He}$) and the surface carbon, oxygen, and nitrogen (CNO) abundances. 
%Overall, the fit quality is good for most of the stellar parameters. 
Figure \ref{fig:stellarfit} shows stellar parameter distributions of the inverse of RMSEA for models of a characteristic star in our sample. The best fit values are marked with a red line and the 1- and 2-$\sigma$ confidence intervals are the dark and light grey shaded regions. The colour scheme marks which generation in the GA the model stems from. We see that the distributions for effective temperature, surface gravity, and maximum $\varv_{\rm rot} \sin i$ all have well-defined and fairly sharp peaks in their distributions, suggesting good statistical constraints on the parameters. 
The abundances are generally more difficult to constrain, but also these sometimes display relatively peaked distributions, as exemplified for carbon in the figure.

\begin{figure}
    \centering
    \includegraphics[width=0.5\textwidth]{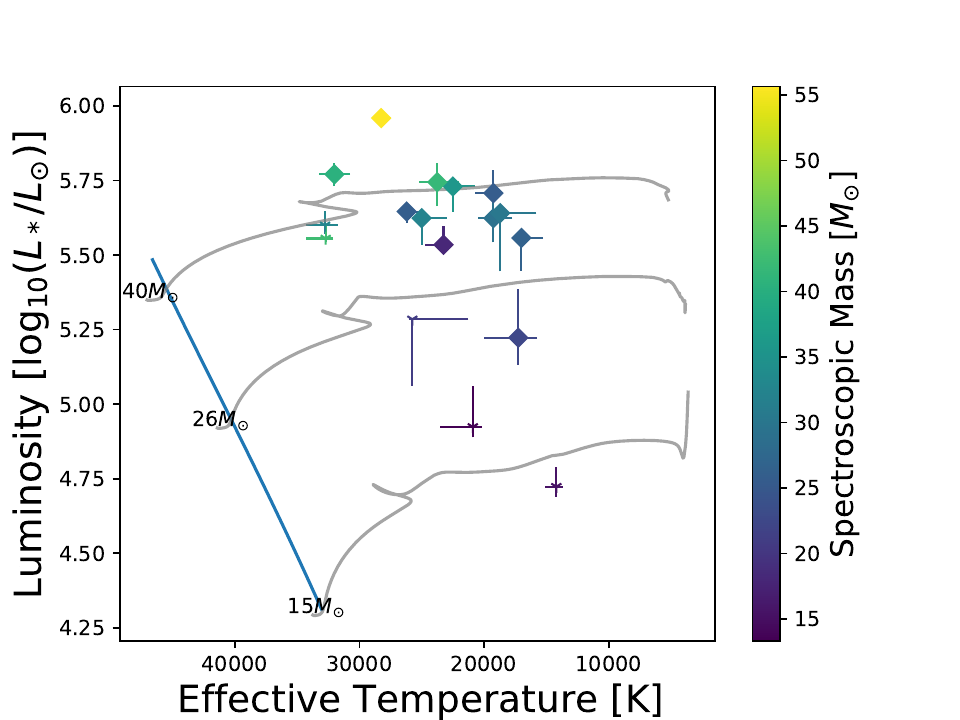}
    \caption{HR-diagram showing all stars in the sample with the colour showing the derived spectral masses. The diamonds mark all the stars with spectral types Ia while the 'v' mark the lower luminosity classes. The stellar evolution tracks are from the MIST data base without any rotation \citep{dotter_mesa_2016, choi_mesa_2016, paxton_modules_2010, paxton_modules_2013, paxton_modules_2015}}
    \label{fig:HR-diagram}
\end{figure}

The sensitivity to effective temperature of the GA is best constrained from the silicon lines, as well as helium lines for stars on the hotter end of our sample.
We note that other lines are sensitive to $T_{\rm eff}$ and that due to the nature of a GA all of these are taken into account when deriving the best fit and the uncertainty margins.
For all stars in the sample we see a strongly peaked distribution for the effective temperature meaning that the combination of stellar lines is adequate for determining $T_{\rm eff}$.
The derived effective temperatures for the sample range from about 15\,kK to 30\,kK with luminosities ranging from $4.7-5.8 \log_{10}(L/L_{\odot})$.
Figure \ref{fig:HR-diagram} shows the stars in the Hertzsprung-Russell Diagram (HRD) with their spectroscopic mass indicated by their colour. We have plotted the evolutionary tracks from the MIST data base \citep{dotter_mesa_2016, choi_mesa_2016, paxton_modules_2010, paxton_modules_2013, paxton_modules_2015} for a 15, 26, and 40 $M_{\odot}$ star in the same window. 
\blue{Comparing these models to our spectroscopic results, the majority of our sample appears to be in the post-main sequence phase of evolution.} 
We note, however, that the evolutionary tracks shown in the figure assume a convective core overshoot parameter $f_{\rm ov, core} = 0.2$ for a step overshoot model \citep{choi_mesa_2016}; if stronger overshooting would be assumed this 
might extend the main sequence so that some of the stars in the sample could still 
be on it \citep{castro_spectroscopic_2014}.

The derived surface gravities are presented in table \ref{tab:stellar_param}. 
For which, in particular $ H{\gamma}, H{\delta}$ are sensitive.
% To give the GA the best chance for determining the surface gravity we include the high level hydrogen Balmer lines ($ H{\gamma}, H{\delta}$), as these lines are affected by Stark broadening (which is sensitive to the electron density, thus gravity, at the surface of the star). 
% The stellar temperature also has an effect on the electron density, but potential degeneracy issues in our fitting are alleviated by other lines which are temperature sensitive while being less surface gravity dependent (e.g., the silicon lines in appendix: table \ref{tab:linelist}). 
% Although the Balmer lines should add a lot of sensitivity for the surface gravity all lines are fitted as one in the GA and therefore all lines with some $\log g$ dependence will contribute to the resulting values.
The fit values for $\log_{10}$ (g \, $\text{cm}/\text{s}^2)$ range from 2.2 to 3.6. 
Only four stars in our sample have a surface gravity $3.0<\log_{10} (g \, \text{cm}/\text{s}^2)$. 

Surface gravity is an important stellar parameter because of its direct influence on the derived spectroscopic mass of the star as described in section \ref{sec:derived}.
The resulting derived spectral masses from these surface gravities lie between between 15 and 75 $M_\odot$, with corresponding stellar radii between 19 and 70 $R_{\odot}$. The latter are shown by the colour scheme in the HR-diagram of figure \ref{fig:HR-diagram}. 
$\Gamma_{\rm e}$ in this sample ranges from 0.10 to 0.46. 
This parameter is important for the analysis below as it represents another key axis over which wind behaviour changes, in addition to the temperature axis. \\

\begin{figure}
    \centering
    \includegraphics[width=0.45\textwidth]{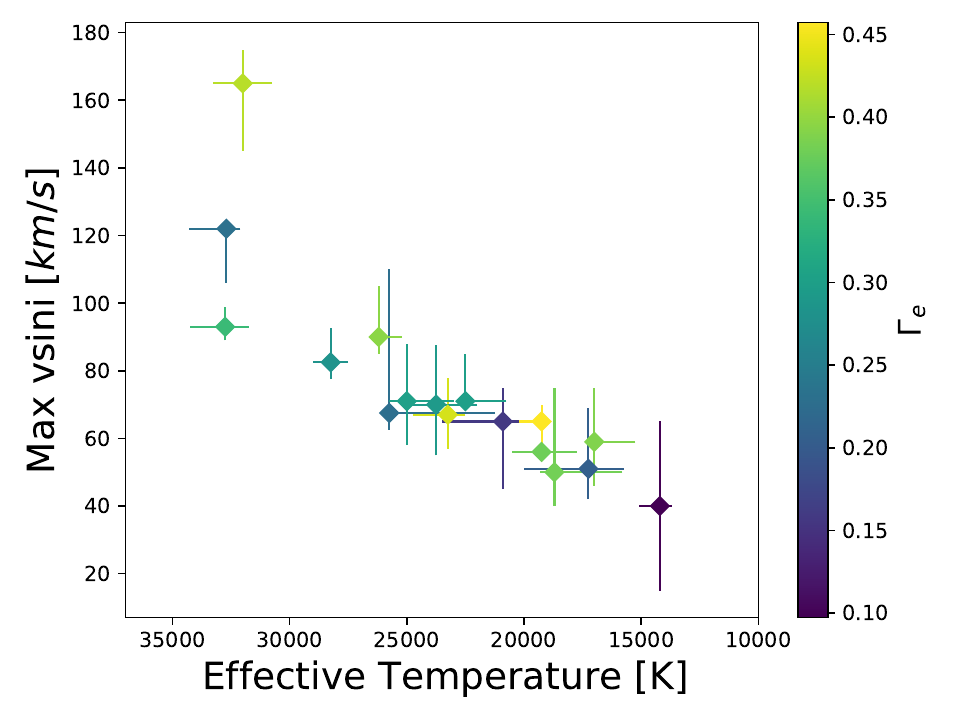}
    \caption{Derived maximum projected rotational velocity in function of effective temperature. Note that this maximum value in practice reflects the combined effect of all sources of macroscopic line broadening (see text).}
    \label{fig:vrot_radius}
\end{figure}

The maximum projected stellar rotation (max $\varv_{\rm rot} \sin{i}$) is the only macroscopic broadening mechanism included in our line fits, meaning we do not attempt here to differentiate between different macroscopic broadening mechanisms (e.g. rotation and 'macro-turbulence').  
%neither micro broadening nor macro broadening is fitted separately. 
As such, the quoted $\varv_{\rm rot} \sin{i}$ values are upper limits, and not the actual projected equatorial rotation velocity (as its measurement is influenced by the broadening mechanisms we do not handle explicitly). 
%\JS{Below not really correct. Look at e.g. Jorick's spin down paper, and also some paper by Zsolt and Jo on evolution effects; i also changed below to Teff since it fits narrative of paper better (than radius, even if also radius plot was nice)...} Rotation is expected to be low as the supergiants have expanded heavily losing rotational velocity during the expansion.
Figure \ref{fig:vrot_radius} shows the derived upper limits of $\varv_{\rm rot} \sin i$ as function of the effective temperature. 
We note that, with the exception of three of the hottest objects, our sample has max $( \varv_{\rm rot} \sin i) < 100 \, \rm km/s$. 
Indeed, it is for these relatively low values that we expect it to be difficult to disentangle rotation broadening from other macroscopic broadening mechanisms. 
Specifically, \cite{sundqvist_rotation_2013} showed that when applying standard techniques to disentangle effects from macro-turbulence $\varv_{\rm mac}$ and $\varv_{\rm rot} \sin{i}$ for massive stars with known very long rotation periods, corresponding to $\varv_{\rm rot} \sin{i} \la 1$ km/s, one still ends up deriving $\varv_{\rm rot} \sin{i} \sim \varv_{\rm mac}$ whenever significant macro-turbulence is present in the observational data. 
This demonstrates the difficulty of present methods to disentangle these two effects, motivating our methodological choice here (indeed, visual inspection of our observed strategic line-profiles often show more Gaussian-like shapes than the rounded shape predicted by broadening due to rotation). That is, our quoted values for max $\varv_{\rm rot} \sin{i}$ may as well reflect the presence of large Gaussian-like turbulent motions and not stellar rotation.      

The average helium abundance is $Y_{\rm He} = 0.1$ (or alternatively $\epsilon_{\rm He}= 11.0$, see below) with the highest and lowest values of the helium abundance 1 dex apart ($\epsilon_{\rm He}\approx 11.4-10.4$) with relative small error margins on the order of 0.2 dex. As one may have noted table \ref{tab:stellar_param} shows values which reach helium abundances below $Y_{\rm He} = 0.08$. 
This might seem problematic as this is below the expected baseline for the LMC at $y=0.091_{-0.012}^{+0.014}$ \citep{russell_abundances_1990}. 
However, the fitting range in all cases reaches values above this 0.08 baseline and as mentioned above, this range is the important property to assess. 
%To make sure these strange abundances do not influence the other parameters, 
We have refitted all problematic stars, as a test, while not allowing the helium abundance to be lower than $Y_{\rm He} = 0.08$ and no values change appreciatively from table \ref{tab:stellar_param}. 
% As a result, although the shown helium abundances might not all reflect the true helium abundance, they do not influence the other parameters and we feel confident that the general trend of low helium abundances is still representative.

Finally, to make the GA sensitive to CNO-abundances, we added atmospheric CNO lines of various ionisation stages. 
\blue{However, because this paper makes an effort to focus on mass-loss rates and not on abundances there are comparatively few data points which are sensitive to the abundance. As such, small changes to a specific abundance only modify the quality of the fit of these specific lines slightly (for instance we include only one NII quintuplet) and typically do not affect significantly the overall fit quality used to define our 'best-fit' model. Consequently, we see high uncertainty regions in CNO abundance as compared to the other stellar parameters.}
The abundances of CNO in this paper are given on the standard scale $\epsilon_{\rm x} = 12+\log_{10}(n_{\rm x}/n_{\rm H})$ where $n_{\rm H}$ is the number abundance of hydrogen abundance and $n_x$ is the number abundance of the element. The average CNO abundances we find for our sample are $\epsilon_{\rm C} = 8.0, \epsilon_{\rm N} = 7.2, \epsilon_{\rm O} = 8.3$. Although these are fitted using different lines they do display similar behaviour for their error margins, which are typically on the order of 0.5-1 dex for all elements. 
Abundances and 1-$\sigma$ error margins for each star in our sample are given in table \ref{tab:stellar_param}. 
\cite{vink_x-shooting_2023} gathered abundances of common elements in the LMC and averaged the abundances found in different stellar populations to find $\epsilon_{\rm C} = 8.01, \epsilon_{\rm N} = 7.03, \epsilon_{\rm O} = 8.4$; in comparison to these results, we appear to have relative agreement.
However, the uncertainty of our derived abundances is usually on the order of 0.5-1 dex.
\blue{5 of the stars in our sample have been studied in the optical regime by \cite{menon_evidence_2024} focusing on the CNO abundances to determine if they are possible binary merger products. All of the stellar parameters (e.g., $T_{\rm eff}$ and $\log {g}$) have overlap in the error-margins, however for some of the CNO abundances we find drastically different best-fit solutions. Due to the very big uncertainty intervals on our CNO-abundances we do still overlap in most cases, but as mentioned above getting close constraints on CNO-abundances has not been our focus.} 
For a small expansion on our derived CNO abundance see figure \ref{fig:cummulative CNO}.

\begin{figure}
    \centering
    \subfigure{\includegraphics[width=0.4\textwidth]{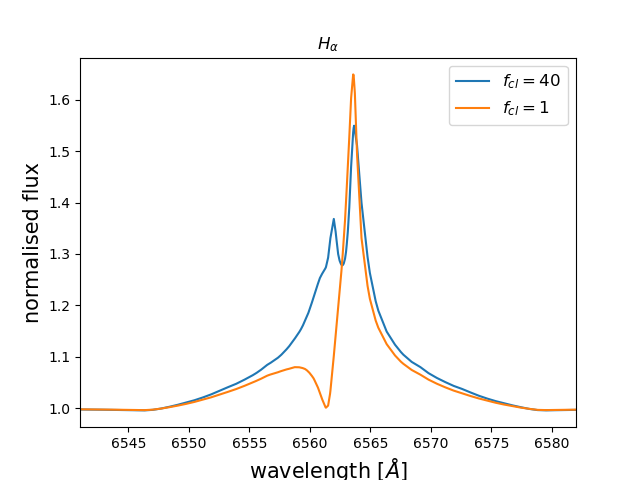}}
    \subfigure{\includegraphics[width=0.4\textwidth]{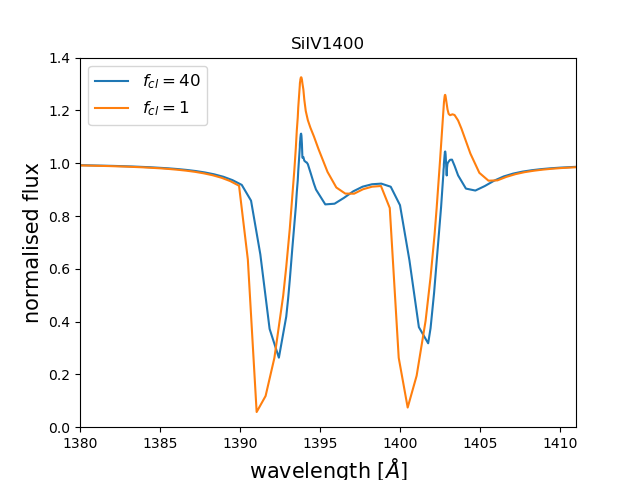}}
    \caption{Clumping influence on the spectral lines. The figures show the influence a change in $f_{\rm cl}$ has on the line profile of on one hand the H\,${\alpha}$ recombination line and on the other hand the Si\,{\sc iv} 1400 resonance doublet. Both model spectra are using almost the same input, they only differ in mass-loss rate and clumping. The blue line both has $\dot{M} = 8 \cdot 10^{-8} M_{\odot}$/yr, and is highly clumped ($f_{\rm cl} =40$). The orange line is  a smooth outflow but has a significantly higher mass-loss rate ($\dot{M} = 5 \cdot 10^{-7} M_{\odot}$/yr); the product $\sqrt{f_{\rm cl}} \dot{M}$ is thus the same for both lines. If we assume an optically thin clumping over the complete line forming region, the 2 H\,${\alpha}$ lines would show perfect agreement. However, this is not the case as we use models that relax the optically thin assumption 
    %making clumping onset effects important.
    and are only clumped above a fitted onset-velocity.}
    \label{fig:Halpha-comp}
\end{figure}

\subsection{Mass-loss rates}\label{sec:results_massloss}
H\,${\alpha}$ is the a sensitive mass-loss diagnostic. 
%The depth of this recombination line is very sensitive to the mass-loss rate.
However, it is important to realise here that since H\,${\alpha}$ is a recombination line in this regime, the line is also sensitive to the wind-clumping factor 
$f_{\rm cl}$. \blue{Additionally, it has been suggested by \cite{petrov_h_2014} that the H\,${\alpha}$ line behaviour and morphology may change with temperature around the supposed bistability jump.} If $f_{\rm cl}$ would be kept constant throughout the complete line-forming region, and clumps would be optically thin, the scaling invariant would simply be $\sqrt{f_{\rm cl}} \dot{M}$. 
However, since H\,${\alpha}$ often is formed in the region where wind clumping starts to become effective in our models (i.e. in regions around $\varv_{\rm cl,start}$ and between $\varv_{\rm cl,start}$ and $\varv_{\rm cl,max}$), deviations from this basic scaling may occur. 
Even so, the degeneracy between $f_{\rm cl}$ and $\dot{M}$ causes problems when trying to determine absolute mass-loss rates using only the strength of H\,${\alpha}$. 
Figure \ref{fig:Halpha-comp} illustrates H\,${\alpha}$'s strong dependency on mass loss and the corresponding degeneracy with $f_{\rm cl}$. 
Both stars in figure \ref{fig:Halpha-comp} have the same input parameters (specifically $T_{\rm eff} = 20$kK, $\log_{10} g = 2.7$, $\varv_{\infty} = 750$ km/s, $f_{\rm ic} = 0.1$, $f_{\rm vel} = 0.5$), except for the mass-loss rate and clumping factor. 
The orange line is for a smooth outflow and $\dot{M} = 5 \cdot 10^{-7} M_{\odot}$/yr, and  the blue line has $f_{\rm cl} = 40$ and $\dot{M} = 8 \cdot 10^{-8} M_{\odot}/$yr. 
The mass-loss rate is thus different for both stars, but the product $\sqrt{f_{\rm cl}} \dot{M}$ is the same. 
One can see that the height of the peak of the $H{\alpha}$ line is similar in the two models.
%reflecting their (almost) identical $\sqrt{f_{cl}} \dot{M}$ products. 
The match is not perfect, however, and the overall shape is different, because the H\,${\alpha}$ line is partly formed in regions where $f_{\rm cl}$ has not reached its maximum value (here we have assumed $\varv_{\rm cl,start} = 0.05$ and $\varv_{\rm cl,max} = 0.1$ for both model stars).
In the optically thin limit and assuming the clumping is constant over the entire velocity range the 2 H\,${\alpha}$ profiles would match perfectly, but the models we use here relax these assumptions. 
%clumping factor and its degeneracy with the mass-loss rate.
%It is thus possible to create an H\,${\alpha}$ line of a star with a lower mass-loss rate that has similar line strength as a line of a high mass-loss rate star by increasing the clumping factor appropriately.
%If these models used optically thin clumping than you could produce the exact same synthetic line spectra by keeping $\sqrt{f_{cl}}\dot{M}$ constant. 
%
%However by changing the clumping the velocity porosity will change the line in other ways.
%The reason for this strong dependency is due to the recombination lines proportionality to $\rho^2$ and as a result the higher density clumps have a big effect on the $H\alpha$ line profile.  
%strength has a degeneracy with mass-loss rate and clumping factor. 
In an attempt to break this 
%$\sim \dot{M} \sqrt{f_{\rm cl}}$ 
mass loss-clumping degeneracy, we fit not only the sensitive H\,${\alpha}$ line but also wind resonance lines in the UV.
These often depend differently on wind clumping as their extinction coefficients 
typically scale as $\sim \langle \rho \rangle$, rather than $\sim \langle \rho^2 \rangle$ as in the case of recombination lines. 
We note that additional dependencies from the ionisation balance and velocity-porosity effects (as well as saturation and interclump density effects) can modify these natural scalings of recombination and resonance lines, making it complicated to sort out all inter-dependencies. 
We show this in the lower panel of figure \ref{fig:Halpha-comp}, displaying the Si IV resonance doublet for the same stars and parameter combinations as the H\,${\alpha}$ line discussed above; it is clear that the dependence on $f_{\rm cl}$ is very different for these lines than for H\,${\alpha}$. 
Specifically for this case, we now observe how the absorption troughs in the doublet lines become weaker for an increasing $f_{\rm cl}$, which here is because the ionisation balance of silicon shifts when strong clumping is introduced. 
In our multi-diagnostic automated fits we utilise these kind of differences in how lines react to derive simultaneous constraints on $f_{\rm cl}$ and $\dot{M}$ (and the other stellar and wind parameters). 

\begin{figure}
    \centering
    \includegraphics[width=0.4\textwidth]{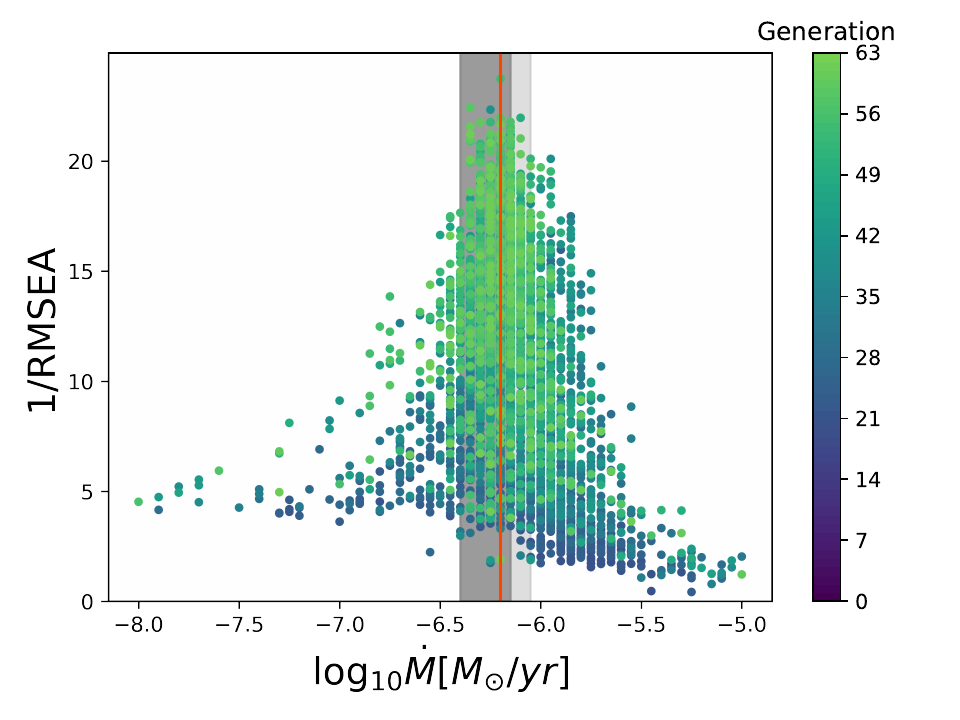}
    \caption{ Quality of all models in a GA run ranked based on the inverse of the RMSEA value as a function of the mass-loss rate. The models for these plot are based on a run trying to fit the star Sk$-68^{\circ}41$ (SpT B0.7 Ia). Each point here is a separate model where the colour defines the generation in which this model was run. The dark grey region highlights the 1-$\sigma$ uncertainty interval and the light grey region is the 2-$\sigma$ uncertainty.}
    \label{fig:mdot_fitdist}
\end{figure}

\begin{figure}
    \centering
    \includegraphics[width=0.4\textwidth]{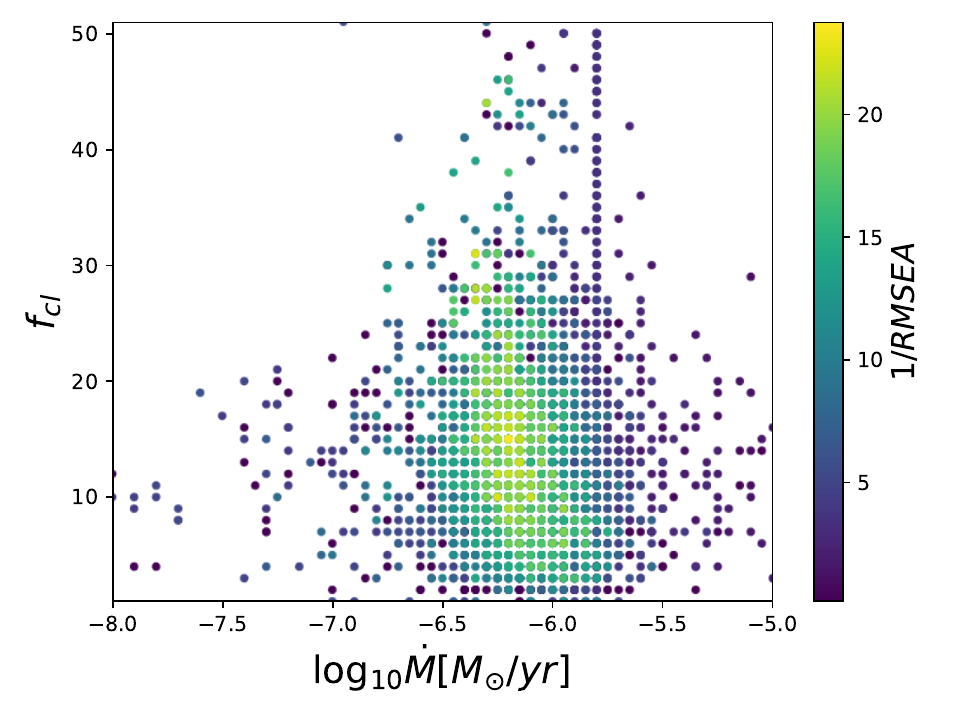}
    \caption{Correlation between the clumping factor and the mass-loss rate for star sk$-68^{\circ} 41$. The colour shows the 1/RMSEA value. }
    \label{fig:mdot_fclump_correlation}
\end{figure}

Overall, our multi-diagnostic GA approach is able to isolate the mass-loss dependence rather well and so derive absolute values of $\dot{M}$. This is demonstrated in Figure \ref{fig:mdot_fitdist}, showing the distribution of the inverse of RMSEA for models of the same star but with different mass-loss rates.  
The more peaked this distribution is, the better the GA can distinguish the different parameter values from each other.%and as a result gives an indication on the statistical error of the GA methodology.
The GA method takes into account the interactions of all degeneracies automatically when deriving uncertainty intervals.
In the cases where the clumping factor stays somewhat degenerate with the mass-loss rate we see a broadness in both the mass loss and derived clumping factor. 
This is more common in the low temperature stars.
Best-fitting mass-loss rates for all stars in our sample are shown in figure \ref{fig:Mass-loss rate emp} and vary from $\sim 10^{-6} M_{\odot} / \text{yr}$ for the hotter stars to $\sim 10^{-8} M_{\odot}/ \text{yr}$ for the cooler stars. 
Figure \ref{fig:mdot_fclump_correlation} further displays the correlation between $\dot{M}$ and $f_{\rm cl}$ for the fit of star Sk$-68^{\circ}41$, showing 1/RMSEA for each combination of clumping factor and mass-loss rate (note that one such combination may have several models due to variation in other parameters). 
Due to the sparse sampling there is some stochasticity in the RMSEA values.
The figure illustrates that although the uncertainty in $f_{\rm cl}$ is rather high, no severe degeneracy issues between $f_{\rm cl}$ and $\dot{M}$ seem to be present in our fits as there is a clear optimal region.
%Keep in mind that on one combination of clumping factor and mass loss multiple models are present with another combination of other parameters, here we plot the best fitting model for this combination of clumping factor and mass-loss rate. 

%It seems like this line is mostly unchanging with mass loss and only has a clumping effect.
%believe that divergent effect of mass loss and clumping on these different lines  allows us to break the degeneracy we see in the H\,${\alpha}$ line.   

\subsection{Wind speeds, accelerations, and turbulence}\label{sec:result_velocities}

\begin{figure}
   	\centering
    \subfigure{\includegraphics[width=0.4\textwidth]{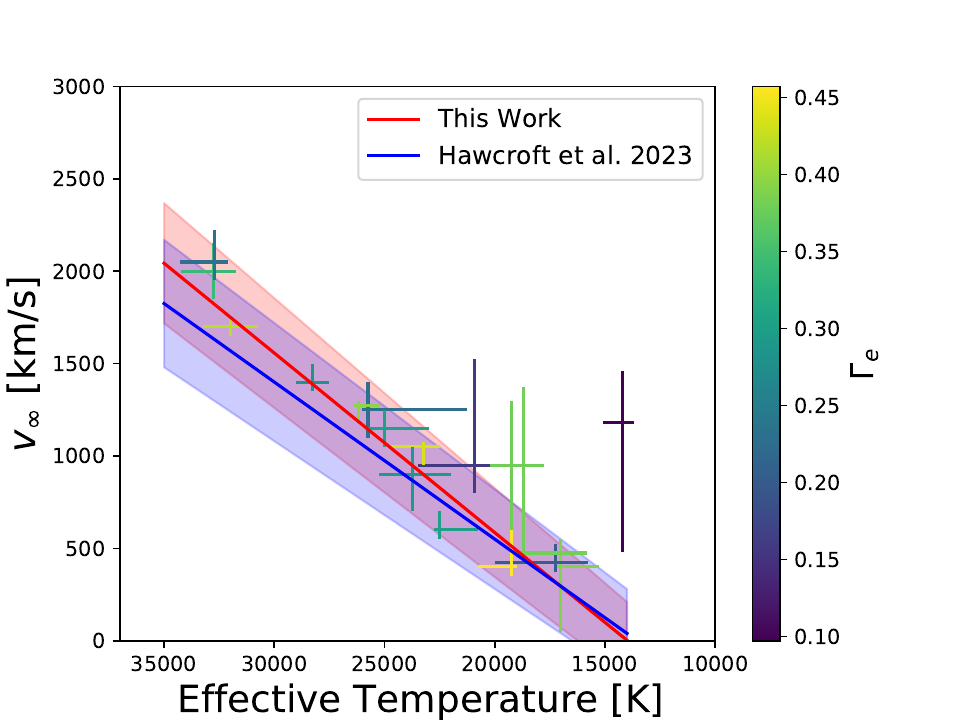}}
   	\subfigure{\includegraphics[width=0.4\textwidth]{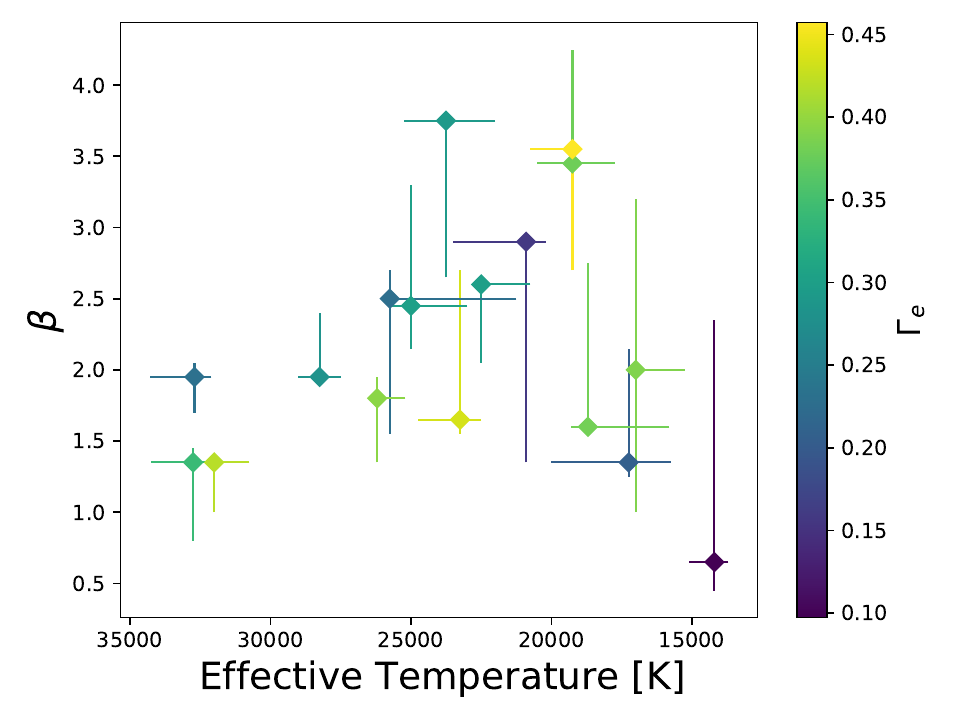}}
   	% \subfigure{\includegraphics[width=0.45\textwidth]{images/teff_vs_windturbkms.png}}
    \subfigure{\includegraphics[width=0.4\textwidth]{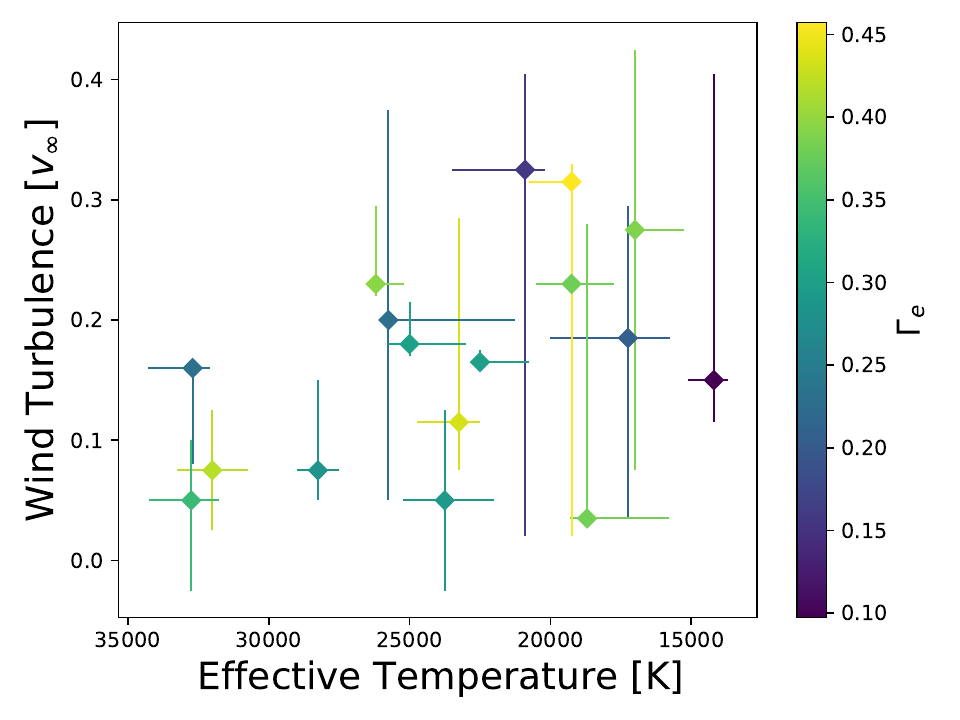}}
    \caption{Derived terminal velocities, wind acceleration parameter $\beta$, and maximum wind micro-turbulence, as function of effective temperature. colour for all panels shows the $\Gamma_{\rm e}$ value. The first panel also shows a linear fit to the terminal velocities found in this work in red compared to the fit in blue from \cite{hawcroft_x-shooting_2024}.}
    \label{fig:velocity+accelration}
\end{figure}

The lines that provide a good diagnostic of terminal wind speed to the GA fits are Si\,{\sc iv} 1400 and C\,{\sc iv}1550. 
These lines often appear as distinct 'P-Cygni' profiles, allowing for fairly straightforward visual verification of the fitted $\varv_\infty$ from the broadness of their blue-shifted absorption. 
The dependence of the strength of these lines on temperature can be seen in figure \ref{fig:line_compare}, where a few characteristic lines are shown for four stars ranging from high (left)  to low (right) effective temperature. 
The observed C\,{\sc iv} and Si\,{\sc iv} lines for higher $T_{\rm eff}$ are well-developed, strong or even saturated towards the blue absorption edge. 
The features become less broad when reducing the temperature, which generally points to lower terminal velocity. 
In the standard modelling of the cooler stars in our sample, C\,{\sc iv} is actually no longer present in the wind. 
Nonetheless, as seen in figure \ref{fig:line_compare}, some of these cool stars still display significant observed P-Cygni profiles for C\,{\sc iv}1550 (albeit weaker than for the hotter objects). 
To reproduce this behaviour in the modelling, we have included X-rays in the {\sc fastwind} simulations also in this temperature region (previous versions did not consider X-ray ionisation below $T_{\rm eff} \sim 25$ kK), which then can raise the wind ionisation balance and produce enough C\,{\sc iv} to explain the observed lines. 
The influence of these X-rays upon the determination of $\varv_\infty$ in this parameter range is further discussed below. 
The terminal wind speeds resulting from our fits of the full sample are displayed in panel 1 of figure \ref{fig:velocity+accelration} as a function of effective temperature. 
For the coolest star in our sample (Sk$-67^{\circ}195$) all UV-lines which are sensitive to the terminal wind speed are no longer present. As a result, we do not get good constraints on the terminal wind speed this is also shown in the fit-distribution in figure \ref{fig:app_sk67-195}, which is flat.

The wind acceleration is parametrised by the $\beta$ value in the analytic velocity law (see previous section). 
% Essentially, the lower the value of $\beta$ the faster the wind accelerates in particularly the near photospheric regions.
As can be seen in panel 2 of figure \ref{fig:velocity+accelration}, we find relatively high values $\ga 1.5$, 
with a sample average $\langle \beta \rangle = 2.1$. This is significantly higher than values derived for hotter O-stars by B22, who found an average 
$\beta \la 1.0$ in that regime, \blue{but comparable to the optical only BSG study in the Galaxy by \cite{haucke_wind_2018} and the UV+optical SMC BSG study by \cite{bernini-peron_x-shooting_2024}}. 
Overall, the slower wind acceleration we find for BSGs compared to the faster wind acceleration found for early O-stars agree well with a range of previous studies (overview in \citealt{puls_mass_2008}).
The error margins on these values are appreciable but still keep the 1-$\sigma$ error interval clear above 1.
A larger scatter is seen for the cooler  stars in the sample.  

%While we do not fit the turbulent velocity in the star, we do fit a micro turbulent velocity when synthesising the spectral lines. 
%This wind turbulence velocity introduces a depth dependent extra velocity broadening dependent on the mass terminal wind speed of the star. 
%The broadening effect increases from the micro turbulence at the base of the star (10 km/s) to the wind turbulence velocity at the the point where the wind reaches $\varv_{\infty}$. 
The inferred values of maximum wind micro-turbulence range from 0.05 to 0.4 in units of $\varv_{\infty}$, with a sample average 0.2. The uncertainty interval of this fit parameter is generally high with typical errors around 0.1, but with outliers reaching error margins as high as 0.4 on this scale. We note that this value then also has a significant effect of the derived terminal wind speed, as turbulence adds to the blueward extension of the modelled line-profile.
%\textbf{influence of vturb on vinf (jo puls)} 
% When values are converted to physical units (km/s) a decreasing trend with decreasing $T_{\rm eff}$ is present, but this can to a large extent be explained by the corresponding trend in terminal wind speed presented above. 

%\textbf{TODO:Figure difference X-ray no xray+rerun first star with x-ray and newer ones without}

\subsection{Clumping}
\begin{figure*}
    \centering
    \subfigure{\includegraphics[width=0.45\textwidth]{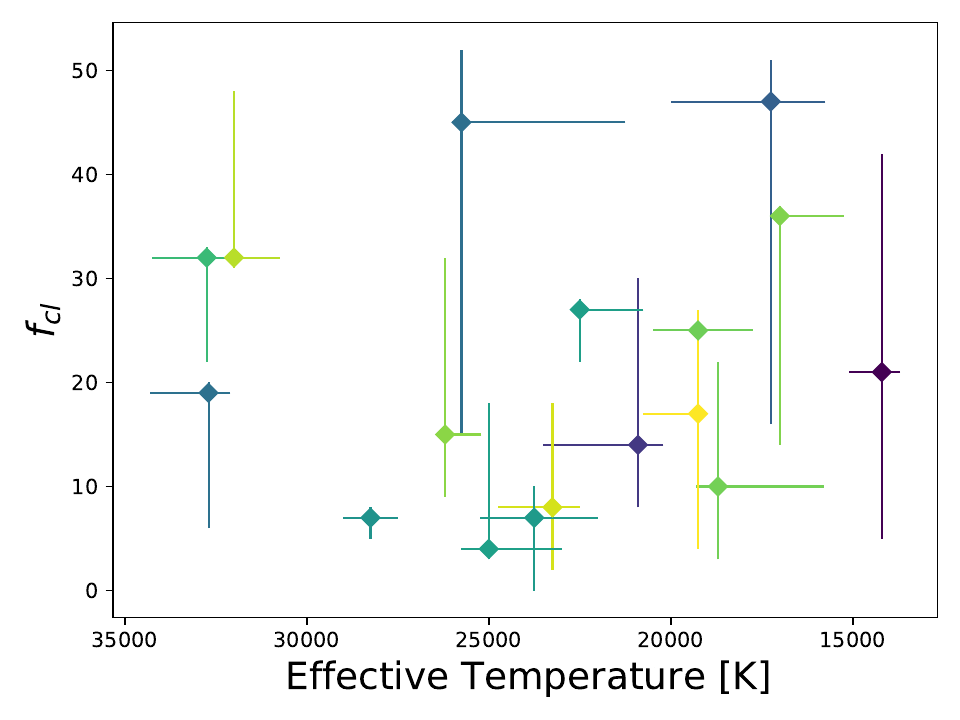}}
    \subfigure{\includegraphics[width=0.45\textwidth]{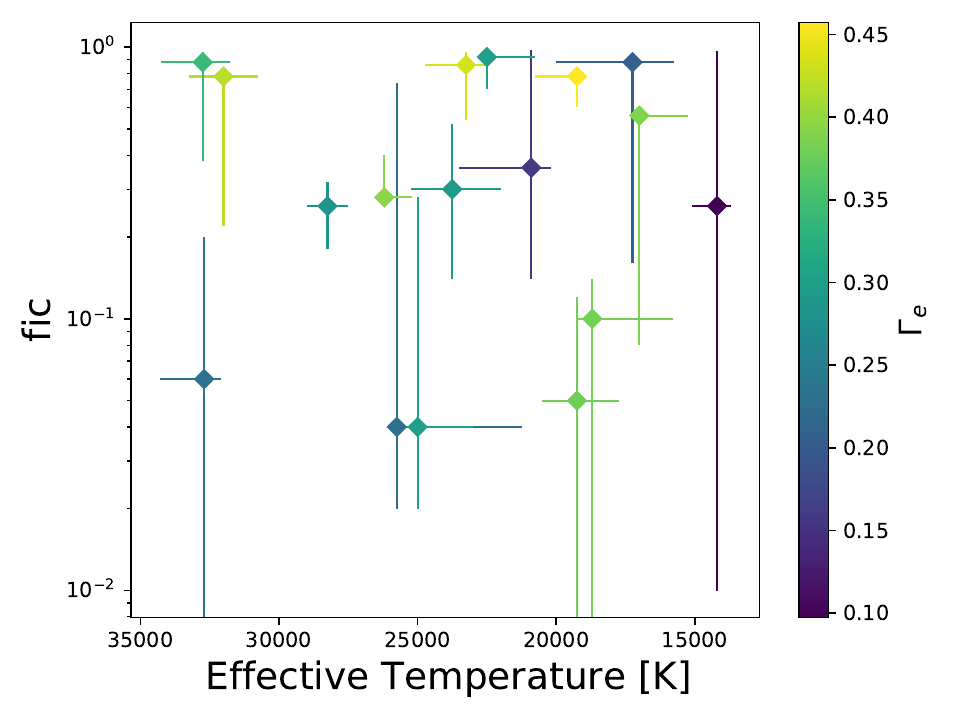}}
    \subfigure{\includegraphics[width=0.45\textwidth]{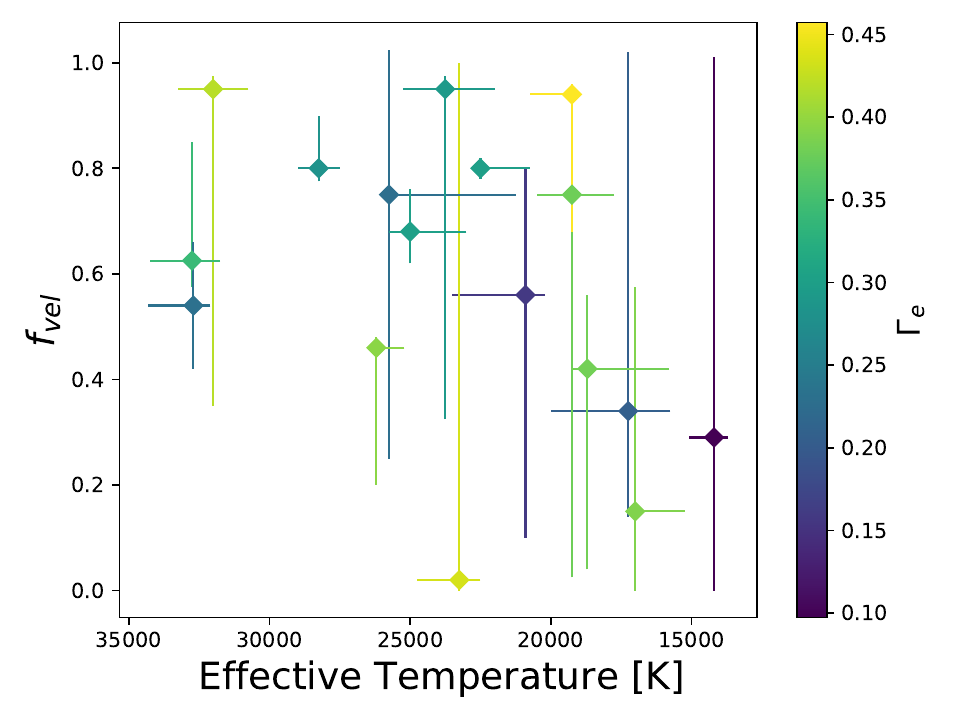}}
    \subfigure{\includegraphics[width=0.45\textwidth]{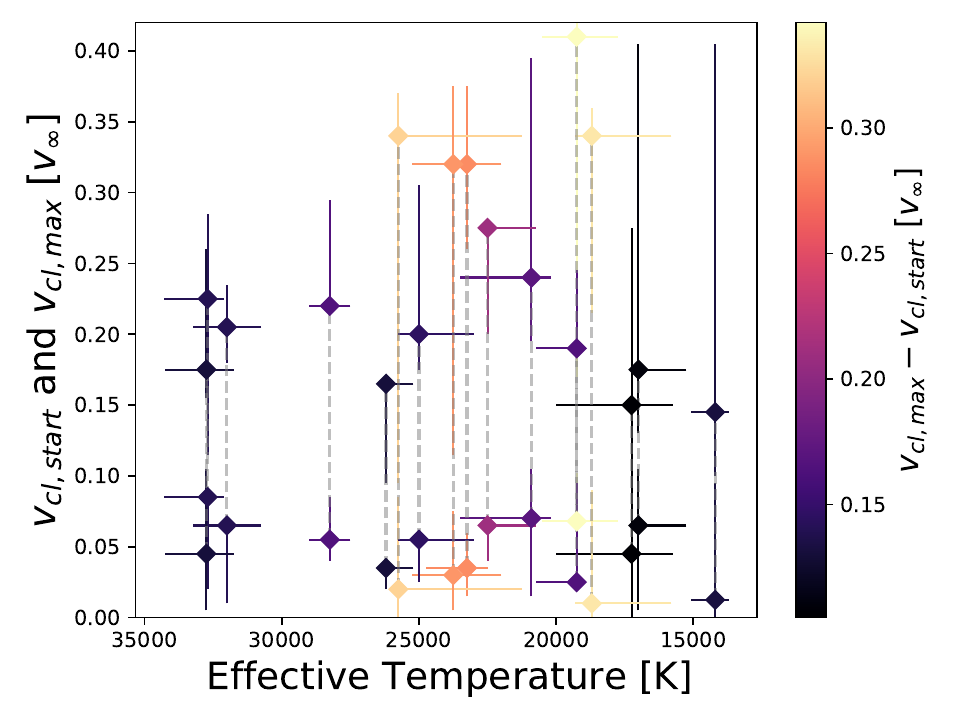}}
    \caption{Clumping parameters with respect to the effective temperature. The panels from top left to bottom right show the clumping factor, interclump density (in units of mean density), velocity filling factor and the onset clumping velocity ($\varv_{\rm cl,start}$) and the velocity of maximum clumping ($\varv_{\rm cl, max}$). The 3 first panels colour according to $\Gamma_{\rm e}$ for the stars, while in the last plot the colour indicates the difference between $\varv_{\rm cl, start}$ and $\varv_{\rm cl, max}$. For each star in this plot the 2 points are connected by a dashed line (not visible in some objects as the error margins are covering the dashed line). }
    \label{fig:clumping}
\end{figure*}

\begin{figure*}
   	\centering
    \subfigure{\includegraphics[width=0.4\textwidth]{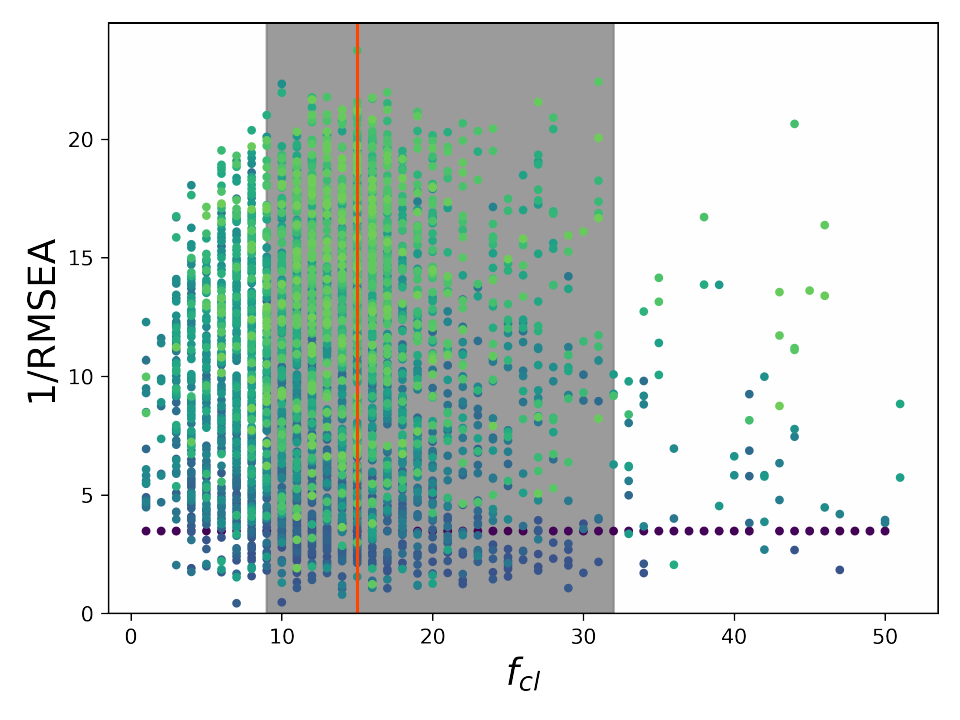}}
   	\subfigure{\includegraphics[width=0.4\textwidth]{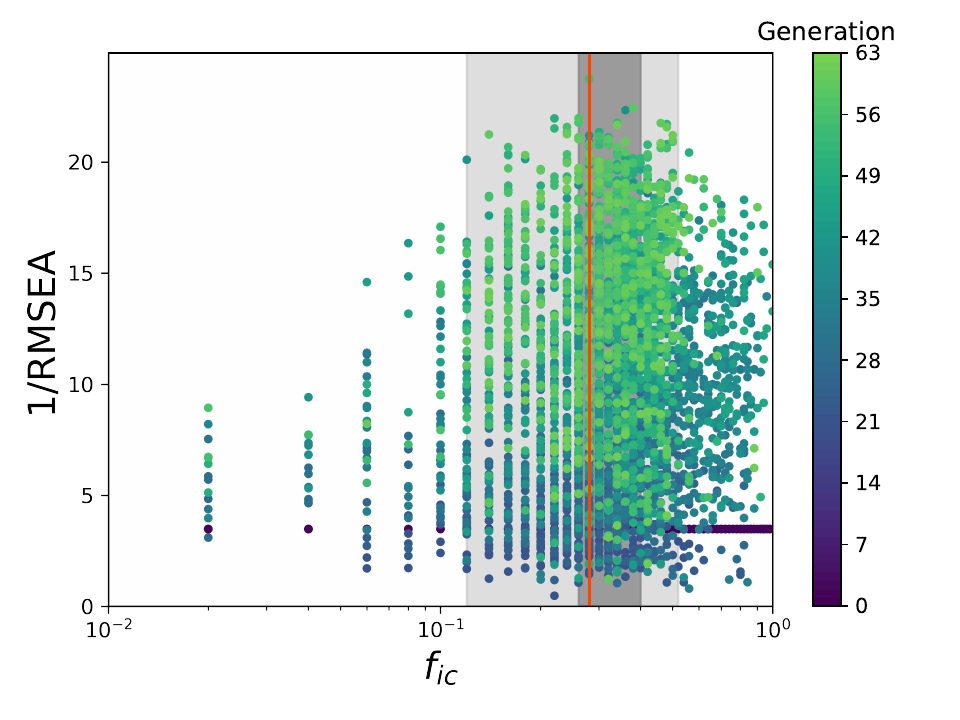}}
   	\subfigure{\includegraphics[width=0.4\textwidth]{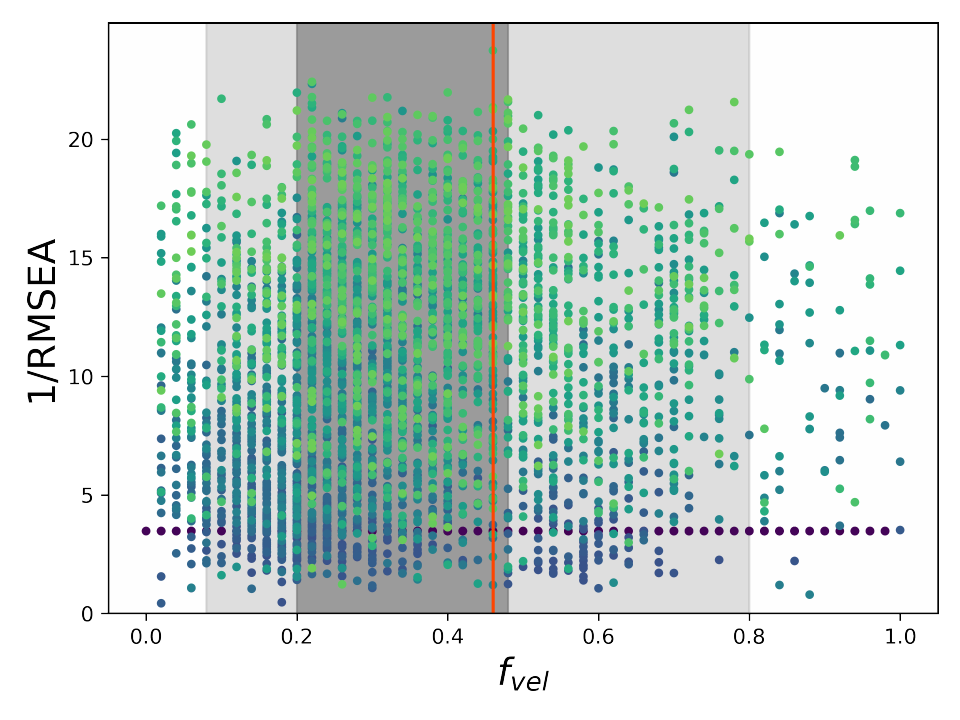}}
    \subfigure{\includegraphics[width=0.4\textwidth]{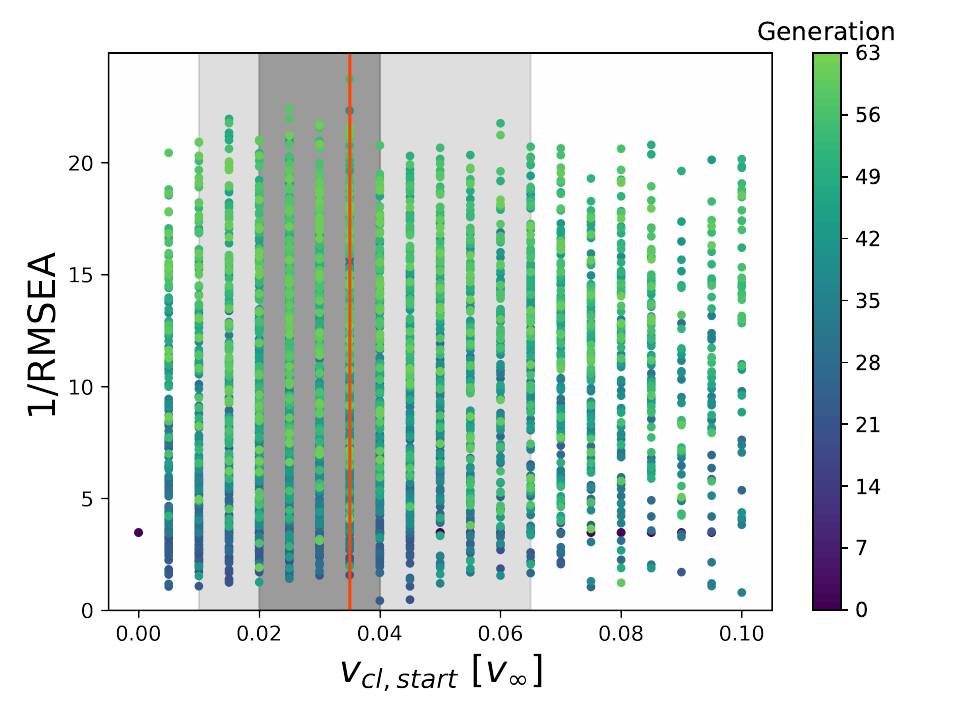}}
    \caption{Distribution of the quality of the fits for clumping parameters. We show a typical distribution of the models at end of a GA fitting run. Here the colour indicates the generation of the model and the grey areas indicate the 1- and 2-$\sigma$ uncertainty intervals.}
    \label{fig:clumpfit}
\end{figure*}

To recapitulate, in addition to the clumping factor ($f_{\rm cl}$), we fit the interclump-density parameter $f_{\rm ic}$, the velocity filling factor $f_{\rm vel}$, the onset of clumping $\varv_{\rm cl,start}$, and the wind speed at which maximum values for the clumping parameters are reached $\varv_{\rm cl,max}$. 
All parameter-values displayed in figures and quoted in text are the maximum values, applied in our models from $\varv_{\rm cl,max}$ to the outer boundary. 
As also found in the previous studies by H21 and B22 it is generally challenging to obtain simultaneous constraints on these parameters.  

Figure \ref{fig:clumping} shows the determined clumping parameters as function of effective temperature. 
The plots are generally characterised by significant scatter and large error-bars, illustrating the general challenge of obtaining these parameter values \blue{or finding a trend with temperature.}
%JS: I marked this off in results, since better in discussion, as discussed late week 
%This lack of trend in structure with temperature has also been noted by \cite{parsons_optically_2024}.}
Within 1-$\sigma$ errors, clumping factors range between the lower limit 1 up to $\sim$ 40 for our sample and velocity filling factors between very low values $\la 0.1$ all the way up to the upper limit unity. 
While the scatter in interclump density is high, this parameter does seem to show some preference for values $\ga 0.1$. 
$\varv_{\rm cl,start}$ is generally preferred to have wind clumping start rather close to our lower allowed bound at $\varv_{\rm cl,start} \approx a$, although the scatter in this parameter is high with typical values within 1-$\sigma$ ranging from $\varv_{\rm cl,start}/\varv_\infty \la 0.01$ up to $\sim 0.1$. $\varv_{\rm cl,max}$ typically is about $\sim 0.2 \varv_\infty$, with similarly broad ranges including a few outliers at the high end reaching values of 0.35; the 1-$\sigma$ interval for this parameter ranges from 0.05 to 0.15 $\varv_\infty$.
Additionally, panel four in figure \ref{fig:clumping} uses colour to show the difference between $\varv_{\rm cl,max}$ and $\varv_{\rm cl,start}$, where we find that most stars have values $\varv_{\rm cl,max} -\varv_{\rm cl,start} \approx 0.2$ in units of $\varv_\infty$.

\begin{table*}
\centering
    \caption{Best fit photospheric parameters of the sample stars.}\label{tab:stellar_param}
    \begin{tabular}{l|cccccccc}
         object& SpT&$T_{\rm eff}$[K] &  $g_{\rm eff}$ [$\log_{10}(cm/s^2)$]& $y_{\rm He}[n_{He}/n_H]$ & $\varv_\text{max, vsini} [km/s]$ &C [dex] &N [dex] &O [dex] \\
         \hline
Sk$-67^{\circ}195$&B6 I&$14200_{-500}^{+900}$&$2.45_{-0.2}^{+0.2}$&$0.04_{-0.02}^{+0.05}$&$40_{-25}^{+25}$&$7.1_{-0.4}^{+1.0}$&$6.2_{-0.35}^{+1.75}$&$8.05_{-0.75}^{+1.05}$\\
Sk$-68^{\circ}8$&B5 Ia$^+$&$17000_{-1750}^{+250}$&$2.15_{-0.25}^{+0.05}$&$0.05_{-0.03}^{+0.08}$&$59_{-13}^{+16}$&$7.35_{-0.4}^{+1.1}$&$7.85_{-1.0}^{+0.1}$&$8.45_{-0.6}^{+0.95}$\\
RMC-109&B5 Ia&$17250_{-1500}^{+2750}$&$2.45_{-0.35}^{+0.4}$&$0.05_{-0.0}^{+0.08}$&$51_{-9}^{+18}$&$8.5_{-0.2}^{+0.25}$&$6.35_{-0.5}^{+0.6}$&$7.6_{-0.25}^{+0.4}$\\
Sk$-67^{\circ}78$&B3 Ia&$19250_{-1500}^{+1250}$&$2.35_{-0.1}^{+0.2}$&$0.08_{-0.02}^{+0.03}$&$56_{-2}^{+14}$&$7.2_{-0.5}^{+0.6}$&$7.85_{-1.05}^{+0.15}$&$8.9_{-1.4}^{+0.2}$\\
Sk$-70^{\circ}50$&B3 Ia&$18700_{-2900}^{+600}$&$2.3_{-0.35}^{+0.2}$&$0.08_{-0.02}^{+0.01}$&$50_{-10}^{+25}$&$7.85_{-0.05}^{+0.5}$&$7.3_{-1.1}^{+0.8}$&$7.9_{-0.2}^{+0.7}$\\
Sk$-68^{\circ}26$&B2 Ia&$19250_{-250}^{+1500}$&$2.2_{-0.05}^{+0.2}$&$0.16_{-0.05}^{+0.01}$&$65_{-11}^{+5}$&$8.45_{-0.55}^{+0.25}$&$7.4_{-0.35}^{+0.05}$&$8.3_{-0.05}^{+0.55}$\\
Sk$-70^{\circ}16$&B2 II&$20900_{-700}^{+2600}$&$2.85_{-0.2}^{+0.25}$&$0.1_{-0.05}^{+0.04}$&$65_{-20}^{+10}$&$8.2_{-0.25}^{+0.15}$&$8.35_{-0.55}^{+0.5}$&$8.82_{-1.02}^{+0.03}$\\
Sk$-69^{\circ}52$&B2 Ia&$22500_{-1750}^{+250}$&$2.6_{-0.1}^{+0.05}$&$0.26_{-0.09}^{+0.01}$&$71_{-1}^{+14}$&$8.35_{-0.1}^{+0.05}$&$7.55_{-0.05}^{+0.35}$&$8.45_{-0.2}^{+0.05}$\\
Sk$-67^{\circ}14$&B1 Ia&$23250_{-750}^{+1500}$&$2.55_{-0.05}^{+0.35}$&$0.16_{-0.12}^{+0.04}$&$67_{-10}^{+11}$&$8.4_{-0.8}^{+0.05}$&$7.15_{-0.05}^{+1.5}$&$8.7_{-0.45}^{+0.05}$\\
Sk$-69^{\circ}140$&B1 Ib&$25750_{-4500}^{+250}$&$3.05_{-0.4}^{+0.15}$&$0.11_{-0.02}^{+0.06}$&$68_{-5}^{+42}$&$8.45_{-0.3}^{+0.35}$&$6.1_{-0.2}^{+1.45}$&$7.8_{-0.25}^{+1.3}$\\
Sk$-66^{\circ}35$&BC1 Iab&$23750_{-1750}^{+1500}$&$2.75_{-0.1}^{+0.55}$&$0.18_{-0.08}^{+0.01}$&$70_{-15}^{+18}$&$8.4_{-0.4}^{+0.35}$&$7.45_{-1.15}^{+0.4}$&$8.5_{-0.4}^{+0.5}$\\
Sk$-69^{\circ}43$&B0.7 Ia&$25000_{-2000}^{+750}$&$2.85_{-0.25}^{+0.05}$&$0.14_{-0.02}^{+0.09}$&$71_{-13}^{+17}$&$7.95_{-0.25}^{+0.15}$&$7.35_{-0.2}^{+0.6}$&$9.25_{-0.9}^{+0.15}$\\
Sk$-68^{\circ}41$&B0.7 Ia&$26200_{-1000}^{+200}$&$2.8_{-0.15}^{+0.2}$&$0.16_{-0.04}^{+0.01}$&$90_{-5}^{+15}$&$8.35_{-0.05}^{+0.25}$&$7.6_{-0.05}^{+0.65}$&$9.25_{-0.5}^{+0.05}$\\
Sk$-68^{\circ}52$&B0 Ia&$28250_{-750}^{+750}$&$3.1_{-0.1}^{+0.25}$&$0.13_{-0.03}^{+0.02}$&$82_{-5}^{+10}$&$8.55_{-0.1}^{+0.15}$&$7.5_{-1.05}^{+0.2}$&$7.75_{-0.4}^{+0.05}$\\
Sk$-68^{\circ}155$&O9 Ia&$32000_{-1250}^{+1250}$&$3.2_{-0.25}^{+0.2}$&$0.12_{-0.02}^{+0.03}$&$165_{-20}^{+10}$&$7.1_{-0.05}^{+1.05}$&$7.15_{-0.05}^{+0.7}$&$8.25_{-0.05}^{+0.65}$\\
Sk$-67^{\circ}107$&O8.5 II&$32750_{-1000}^{+1500}$&$3.35_{-0.25}^{+0.15}$&$0.08_{-0.01}^{+0.06}$&$93_{-4}^{+6}$&$8.05_{-0.1}^{+0.25}$&$6.55_{-0.4}^{+0.1}$&$7.6_{-0.25}^{+1.05}$\\
Sk$-67^{\circ}106$&O8 II&$32700_{-600}^{+1600}$&$3.5_{-0.15}^{+0.5}$&$0.11_{-0.03}^{+0.01}$&$122_{-16}^{+1}$&$7.7_{-0.45}^{+0.2}$&$6.25_{-0.1}^{+0.7}$&$7.75_{-0.05}^{+0.45}$\\
\hline
    \end{tabular}
    \begin{tablenotes}
         \item \textbf{Notes}: For $y_{\rm He}$ and $\varv_\text{max, vsini}$ we show the errors of the optical only fits, as their fit range has been constrained for the full UV and optical fit.
    \end{tablenotes}
\end{table*}

\begin{table*}
\centering
\caption{Derived and X-ray parameters of all objects in our sample.}\label{tab:derivedparam+xray}
    \begin{tabular}{l|cccccccc}
         object &  $M_\text{spec} [M_\odot]$ & $M_\text{evo} [M_\odot]$ & $\log_{10}(L/L_\odot)$ & Radius$ [R_\odot]$ & $\Gamma_{\rm e}$ & $u_\infty [km/s]$ & $\log_{10}$($f_x$) & $\log_{10}(L_x/L_\text{bol})$\\
         \hline
Sk$-67^{\circ}195$&$15.43_{-4.49}^{+5.4}$&N/A&$4.72_{-0.03}^{+0.06}$&$38.33_{-1.32}^{+0.71}$&$0.1_{-0.02}^{+0.05}$&N/A&N/A&N/A\\
Sk$-68^{\circ}8$&$26.45_{-6.89}^{+1.93}$&$32.0^{+2.15}_{-3.28}$
&$5.56_{-0.11}^{+0.0}$&$69.89_{-0.0}^{+4.13}$&$0.39_{-0.09}^{+0.03}$&$175.0_{-175.0}^{+325.0}$&$-1.95_{-0.1}^{+0.95}$&$-10.34_{-23.41}^{+3.31}$\\
RMC-109&$22.56_{-9.48}^{+21.33}$&$22.0^{+4.01}_{-2.34}$
&$5.22_{-0.09}^{+0.16}$&$46.18_{-3.64}^{+2.21}$&$0.21_{-0.08}^{+0.07}$&$505.0_{-375.0}^{+87.5}$&$0.4_{-0.65}^{+0.65}$&$-6.1_{-3.03}^{+0.0}$\\
Sk$-67^{\circ}78$&$29.22_{-2.1}^{+9.28}$&$30.0^{+2.49}_{-1.17}$
&$5.62_{-0.08}^{+0.06}$&$58.82_{-1.76}^{+2.44}$&$0.38_{-0.1}^{+0.0}$&$625.0_{-125.0}^{+325.0}$&$0.25_{-0.95}^{+0.05}$&$-5.69_{-1.38}^{+0.02}$\\
Sk$-70^{\circ}50$&$30.23_{-10.44}^{+11.38}$&$31.2^{+2.74}_{-2.77}$
&$5.64_{-0.2}^{+0.03}$&$63.55_{-1.0}^{+6.67}$&$0.38_{-0.09}^{+0.02}$&$175.0_{-25.0}^{+775.0}$&$-0.2_{-1.3}^{+0.05}$&$-8.35_{-0.41}^{+1.39}$\\
Sk$-68^{\circ}26$&$25.71_{-2.2}^{+10.19}$&$38.4^{+3.3}_{-2.31}$
&$5.71_{-0.0}^{+0.08}$&$64.79_{-2.39}^{+0.0}$&$0.46_{-0.12}^{+0.17}$&$600.0_{-25.0}^{+275.0}$&$-1.4_{-0.65}^{+0.55}$&$-7.53_{-0.21}^{+0.66}$\\
Sk$-70^{\circ}16$&$13.33_{-3.61}^{+5.16}$&$17.0^{+1.71}_{-1.03}$
&$4.92_{-0.04}^{+0.14}$&$22.29_{-1.43}^{+0.39}$&$0.16_{-0.03}^{+0.05}$&$517.5_{-125.0}^{+50.0}$&$0.88_{-0.03}^{+0.35}$&$-5.72_{-2.44}^{+0.32}$\\
Sk$-69^{\circ}52$&$35.69_{-0.55}^{+0.0}$&$36.0^{+2.18}_{-3.24}$
&$5.73_{-0.08}^{+0.0}$&$48.67_{-0.0}^{+2.02}$&$0.3_{-0.01}^{+0.0}$&$675.0_{-425.0}^{+25.0}$&$-1.9_{-0.05}^{+0.1}$&$-7.92_{-0.98}^{+0.0}$\\
Sk$-67^{\circ}14$&$18.0_{-0.0}^{+18.22}$&$31.4^{+2.01}_{-1.95}$
&$5.53_{-0.03}^{+0.06}$&$36.39_{-1.1}^{+0.46}$&$0.43_{-0.19}^{+0.0}$&$425.0_{-350.0}^{+25.0}$&$-1.0_{-0.45}^{+1.7}$&$-6.82_{-2.68}^{+0.0}$\\
Sk$-69^{\circ}140$&$20.80_{-10.08}^{+7.72}$&$37.8^{+2.76}_{-2.94}$
&$5.28_{-0.22}^{+0.0}$&$22.25_{-0.0}^{+2.46}$&$0.23_{-0.1}^{+0.11}$&$350.0_{-150.0}^{+550.0}$&$-0.1_{-0.45}^{+0.35}$&$-7.51_{-1.18}^{+0.88}$\\
Sk$-66^{\circ}35$&$41.62_{-3.96}^{+85.6}$&$38.80^{+3.25}_{-2.86}$&$5.74_{-0.08}^{+0.06}$&$44.42_{-1.3}^{+1.72}$&$0.29_{-0.19}^{+0.05}$&$150.0_{-75.0}^{+50.0}$&$0.05_{-0.25}^{+0.35}$&$-8.42_{-1.73}^{+0.28}$\\
Sk$-69^{\circ}43$&$32.34_{-10.1}^{+0.0}$&$32.8^{+2.27}_{-3.07}$
&$5.62_{-0.09}^{+0.02}$&$34.88_{-0.39}^{+1.51}$&$0.3_{-0.03}^{+0.08}$&$650.0_{-200.0}^{+25.0}$&$-1.7_{-0.25}^{+0.5}$&$-7.07_{-0.78}^{+0.0}$\\
Sk$-68^{\circ}41$&$25.82_{-3.97}^{+10.38}$&$35.0^{+1.69}_{-1.89}$
&$5.65_{-0.04}^{+0.0}$&$32.58_{-0.0}^{+0.58}$&$0.39_{-0.11}^{+0.05}$&$950.0_{-725.0}^{+25.0}$&$-1.3_{-0.05}^{+0.35}$&$-6.42_{-1.55}^{+0.0}$\\
Sk$-68^{\circ}52$&$75.65_{-6.61}^{+41.34}$&$56.00^{+2.38}_{-3.85}$&$5.96_{-0.02}^{+0.02}$&$40.2_{-0.4}^{+0.41}$&$0.28_{-0.09}^{+0.02}$&$225.0_{-175.0}^{+25.0}$&$-0.7_{-0.2}^{+1.05}$&$-7.83_{-3.99}^{+0.0}$\\
Sk$-68^{\circ}155$&$40.37_{-13.31}^{+13.57}$&$44.20^{+2.71}_{-3.74}$&$5.77_{-0.04}^{+0.04}$&$25.22_{-0.43}^{+0.45}$&$0.42_{-0.09}^{+0.18}$&$375.0_{-350.0}^{+325.0}$&$-0.55_{-0.05}^{+0.45}$&$-7.15_{-13.26}^{+0.98}$\\
Sk$-67^{\circ}107$&$32.82_{-11.14}^{+6.54}$&$37.4^{+0.0}_{-0.0}$
&$5.6_{-0.03}^{+0.05}$&$19.77_{-0.41}^{+0.26}$&$0.34_{-0.06}^{+0.13}$&$475.0_{-375.0}^{+25.0}$&$0.35_{-0.25}^{+0.45}$&$-5.86_{-2.62}^{+0.0}$\\
Sk$-67^{\circ}106$&$42.46_{-9.15}^{+67.04}$&$36.0^{+1.96}_{-1.54}$
&$5.56_{-0.01}^{+0.05}$&$18.85_{-0.59}^{+0.18}$&$0.23_{-0.13}^{+0.07}$&$375.0_{-25.0}^{+625.0}$&$-1.4_{-0.4}^{+0.2}$&$-8.11_{-0.01}^{+1.03}$\\
\hline

    \end{tabular}
    \begin{tablenotes}
        \item \textbf{Note}: $\Gamma_{\rm e}$ is the classical Eddington parameter, $u_{\infty}$ is the maximum jump speed used to determine the X-ray characteristics, $\log_{10} (f_x)$ is defined as $16e_s^2$ with $e_s$ the X-ray volume filling factor. $L_x/L_{\rm bol}$ is the X-ray luminosity scaled to the bolometric luminosity.   
    \end{tablenotes}
\end{table*}

\begin{table*}
    \centering
    \caption{Best fit wind parameters including optically thick clumping.}\label{tab:wind_param}
    \begin{tabular}{l|cccccccccc}
         object& $\log_{10} \dot{M} [M_{\odot}/yr]$ & $\varv_{\infty} [km/s]$ & $\beta$ & $f_{\rm cl}$ & $f_{\rm ic}$& $f_{\rm vel}$ & $\varv_\text{cl, start} [\varv_\infty]$& $\varv_\text{cl, max} [\varv_\infty]$ & $\varv_\text{turb}$ [$\varv_\infty$]  \\
         \hline
Sk$-67^{\circ}195$&$-7.45_{-1.5}^{+0.6}$&\grey{$1180_{-700}^{+280}$}&$0.65_{-0.2}^{+1.7}$&$21.0_{-16.0}^{+21.0}$&$0.26_{-0.25}^{+0.71}$&$0.29_{-0.29}^{+0.72}$&$0.01_{-0.01}^{+0.08}$&$0.14_{-0.04}^{+0.26}$&$0.15_{-0.04}^{+0.26}$\\
Sk$-68^{\circ}8$&$-7.1_{-0.35}^{+0.3}$&$400_{-350}^{+150}$&$2.0_{-1.0}^{+1.2}$&$36.0_{-22.0}^{+1.0}$&$0.56_{-0.48}^{+0.04}$&$0.15_{-0.15}^{+0.43}$&$0.06_{-0.06}^{+0.04}$&$0.18_{-0.04}^{+0.23}$&$0.28_{-0.2}^{+0.15}$\\
RMC-109&$-7.75_{-0.55}^{+0.5}$&$425_{-50}^{+100}$&$1.35_{-0.1}^{+0.8}$&$47.0_{-31.0}^{+4.0}$&$0.88_{-0.72}^{+0.04}$&$0.34_{-0.2}^{+0.68}$&$0.04_{-0.04}^{+0.06}$&$0.15_{-0.05}^{+0.13}$&$0.18_{-0.15}^{+0.11}$\\
Sk$-67^{\circ}78$&$-6.82_{-0.12}^{+0.28}$&$950_{-400}^{+350}$&$3.45_{-0.55}^{+0.8}$&$25.0_{-21.0}^{+1.0}$&$0.05_{-0.06}^{+0.07}$&$0.75_{-0.72}^{+0.03}$&$0.07_{-0.02}^{+0.03}$&$0.41_{-0.24}^{+0.06}$&$0.23_{-0.14}^{+0.1}$\\
Sk$-70^{\circ}50$&$-6.95_{-0.3}^{+0.4}$&$475_{-25}^{+900}$&$1.6_{-0.05}^{+1.15}$&$10.0_{-7.0}^{+12.0}$&$0.1_{-0.12}^{+0.04}$&$0.42_{-0.38}^{+0.14}$&$0.01_{-0.01}^{+0.08}$&$0.34_{-0.13}^{+0.02}$&$0.04_{-0.01}^{+0.25}$\\
Sk$-68^{\circ}26$&$-6.8_{-0.15}^{+0.35}$&$400_{-50}^{+200}$&$3.55_{-0.85}^{+0.05}$&$17.0_{-13.0}^{+10.0}$&$0.78_{-0.18}^{+0.02}$&$0.94_{-0.26}^{+0.02}$&$0.02_{-0.01}^{+0.07}$&$0.19_{-0.01}^{+0.05}$&$0.32_{-0.3}^{+0.02}$\\
Sk$-70^{\circ}16$&$-7.7_{-1.15}^{+0.3}$&$950_{-150}^{+575}$&$2.9_{-1.55}^{+0.05}$&$14.0_{-6.0}^{+16.0}$&$0.36_{-0.22}^{+0.62}$&$0.56_{-0.46}^{+0.24}$&$0.07_{-0.06}^{+0.04}$&$0.24_{-0.04}^{+0.16}$&$0.32_{-0.3}^{+0.08}$\\
Sk$-69^{\circ}52$&$-6.7_{-0.05}^{+0.1}$&$600_{-50}^{+100}$&$2.6_{-0.55}^{+0.05}$&$27.0_{-5.0}^{+1.0}$&$0.92_{-0.22}^{+0.02}$&$0.8_{-0.02}^{+0.02}$&$0.06_{-0.02}^{+0.01}$&$0.28_{-0.08}^{+0.01}$&$0.16_{-0.01}^{+0.01}$\\
Sk$-67^{\circ}14$&$-6.25_{-0.3}^{+0.05}$&$1050_{-100}^{+25}$&$1.65_{-0.1}^{+1.05}$&$8.0_{-6.0}^{+10.0}$&$0.86_{-0.32}^{+0.1}$&$0.02_{-0.02}^{+0.98}$&$0.04_{-0.02}^{+0.02}$&$0.32_{-0.06}^{+0.05}$&$0.12_{-0.04}^{+0.17}$\\
Sk$-69^{\circ}140$&$-7.3_{-0.35}^{+0.3}$&$1250_{-150}^{+150}$&$2.5_{-0.95}^{+0.2}$&$45.0_{-30.0}^{+7.0}$&$0.04_{-0.02}^{+0.7}$&$0.75_{-0.5}^{+0.27}$&$0.02_{-0.02}^{+0.06}$&$0.34_{-0.25}^{+0.03}$&$0.2_{-0.15}^{+0.18}$\\
Sk$-66^{\circ}35$&$-6.4_{-0.15}^{+0.25}$&$900_{-200}^{+150}$&$3.75_{-1.1}^{+0.05}$&$7.0_{-7.0}^{+3.0}$&$0.3_{-0.16}^{+0.22}$&$0.95_{-0.62}^{+0.03}$&$0.03_{-0.02}^{+0.05}$&$0.32_{-0.2}^{+0.05}$&$0.05_{-0.08}^{+0.08}$\\
Sk$-69^{\circ}43$&$-6.3_{-0.25}^{+0.05}$&$1150_{-100}^{+100}$&$2.45_{-0.3}^{+0.85}$&$4.0_{-1.0}^{+14.0}$&$0.04_{-0.02}^{+0.24}$&$0.68_{-0.06}^{+0.08}$&$0.06_{-0.03}^{+0.0}$&$0.2_{-0.03}^{+0.1}$&$0.18_{-0.01}^{+0.04}$\\
Sk$-68^{\circ}41$&$-6.2_{-0.2}^{+0.05}$&$1275_{-100}^{+25}$&$1.8_{-0.45}^{+0.15}$&$15.0_{-6.0}^{+17.0}$&$0.28_{-0.02}^{+0.12}$&$0.46_{-0.26}^{+0.02}$&$0.04_{-0.02}^{+0.0}$&$0.16_{-0.07}^{+0.01}$&$0.23_{-0.01}^{+0.06}$\\
Sk$-68^{\circ}52$&$-5.85_{-0.05}^{+0.15}$&$1400_{-50}^{+100}$&$1.95_{-0.05}^{+0.45}$&$7.0_{-2.0}^{+1.0}$&$0.26_{-0.08}^{+0.06}$&$0.8_{-0.03}^{+0.1}$&$0.06_{-0.02}^{+0.03}$&$0.22_{-0.01}^{+0.07}$&$0.08_{-0.02}^{+0.08}$\\
Sk$-68^{\circ}155$&$-6.2_{-0.05}^{+0.15}$&$1700_{-50}^{+50}$&$1.35_{-0.35}^{+0.05}$&$32.0_{-1.0}^{+16.0}$&$0.78_{-0.56}^{+0.02}$&$0.95_{-0.6}^{+0.03}$&$0.06_{-0.06}^{+0.01}$&$0.2_{-0.02}^{+0.03}$&$0.08_{-0.05}^{+0.05}$\\
Sk$-67^{\circ}107$&$-6.4_{-0.1}^{+0.25}$&$2000_{-150}^{+150}$&$1.35_{-0.55}^{+0.1}$&$32.0_{-10.0}^{+1.0}$&$0.88_{-0.5}^{+0.08}$&$0.62_{-0.05}^{+0.22}$&$0.04_{-0.04}^{+0.06}$&$0.18_{-0.02}^{+0.09}$&$0.05_{-0.08}^{+0.05}$\\
Sk$-67^{\circ}106$&$-6.35_{-0.2}^{+0.15}$&$2050_{-100}^{+175}$&$1.95_{-0.25}^{+0.1}$&$19.0_{-13.0}^{+1.0}$&$0.06_{-0.06}^{+0.14}$&$0.54_{-0.12}^{+0.12}$&$0.08_{-0.06}^{+0.01}$&$0.22_{-0.11}^{+0.06}$&$0.16_{-0.08}^{+0.01}$\\

 \hline
    \end{tabular}
\end{table*}

\section{Discussion}\label{sec:discussion}

%\subsection{Stellar parameters}
%Some general issues with fitting specific lines?
%I think we can hold on this for now. 

\subsection{Is there a generic mass loss jump in the range $T_{\rm eff} \sim 15 - 35$ kK?}

A key focus of the present analysis is to investigate whether there are empirical signs of a general increase in $\dot{M}$ when moving from hotter to cooler stars in the region $T_{\rm eff} \sim 15.0 - 27.5$ kK. Previous empirical studies attempting to tackle this question have often been hampered by the mass loss-clumping degeneracy of optical recombination lines like H\,${\alpha}$ (e.g., \citealt{markova_bright_2008}).
Although \citeauthor{markova_bright_2008} do not find a jump comparable to the predicted jump \citep{vink_mass-loss_2001} as there is some indication that the clumping factor might reversely decrease with temperature as well \citep{driessen_theoretical_2019} this was considered to not be fully conclusive.   
As outlined in the previous section, here we break this degeneracy by a detailed account of wind clumping and by considering a multitude of diagnostic lines in the optical and UV. 
This allows us to derive absolute empirical constraints on $\dot{M}$. 

\begin{figure}
    \centering
    \subfigure{\includegraphics[width=0.45\textwidth]{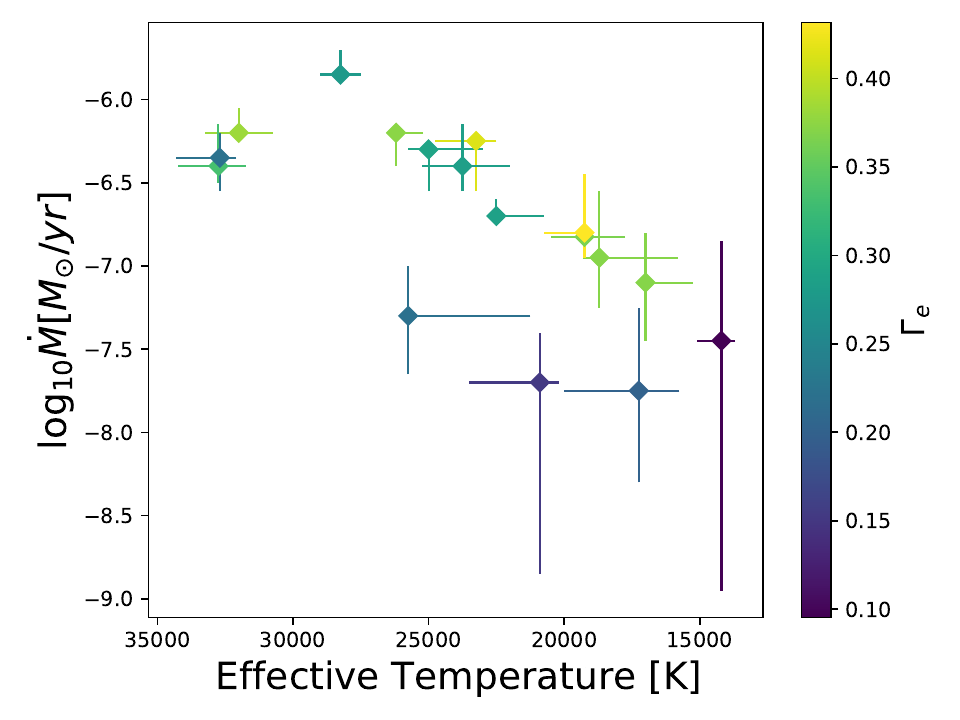}}
    \subfigure{\includegraphics[width=0.45\textwidth]{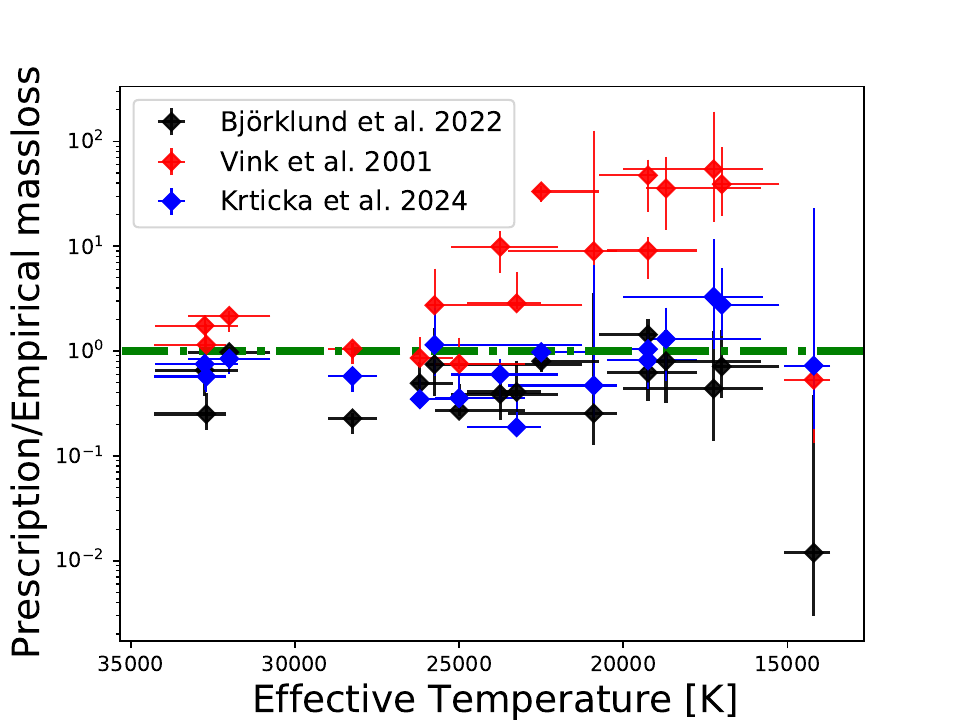}}
    \caption{Derived mass-loss rates for our sample as function of effective temperature. The top plot shows the derived mass-loss rates and he colour shows the derived classical Eddington parameter for each star. The bottom plot shows the ratio between the theoretical prescriptions by \cite{vink_mass-loss_2001, bjorklund_new_2023, krticka_new_2024} and the empirically found values. The green line indicates where these are equal.}
    \label{fig:Mass-loss rate emp}
\end{figure}

Figure \ref{fig:Mass-loss rate emp} shows the empirically derived mass-loss rates of our sample as function of effective temperature. 
%As mentioned above, mass loss is typically also a strong function of the Eddington parameter $\Gamma_{\rm e} \sim L_\star/M_\star$. 
\blue{We stress that this figure is showing stars which not only differ in $T_{\rm eff}$ but also in their $L_\star/M_\star$ ratios, where the latter have a big influence on the mass-loss rate. This ratio is shown efficiently by colour coding the stars according to their $\Gamma_{\rm e}$ values.}
Two stars with the same $T_{\rm eff}$ may have different $\dot{M}$ depending on their individual values of $\Gamma_{\rm e}$, where typically the star with the higher value also has a higher mass-loss rate. 
%This must thus be taken into account when interpreting a plot of $\dot{M}$ vs. $T_{\rm eff}$.
Therefore, we indicate $\Gamma_{\rm e}$ with a colourbar in Figure \ref{fig:Mass-loss rate emp}. 
%This value is added as this shows an additional determining factor of the star as this is a value shows the amount of force exerted by radiation on the potentially escaping compared to the force pulling on this same mass.
%This is the main value determining the behaviour of stars.
The figure demonstrates that there is a clear general downward trend in $\dot{M}$ when moving from the hotter to cooler objects in our sample. 
If we inspect only stars with similar $\Gamma_{\rm e}$ values such a trend persists. 
For objects with higher Eddington ratios (green and yellow in figure), there is a fairly small but noticeable downward trend in the region $T_{\rm eff} \sim 25 - 15$ kK. 
For objects with somewhat lower Eddington ratios (purple in figure) we also observe that the collection of stars above $T_{\rm eff} \sim 30$ kK have higher mass-loss rates than corresponding objects around and below $T_{\rm eff} \sim$ 20 kK. That is, in our current sample there are no empirical signs  of an upward jump (or upward trend) in $\dot{M}$ with decreasing $T_{\rm eff}$ in this region. \\

%In thinner crosses the estimated value of 2 different mass loss prescriptions. 
%In red the prescription from \cite{vink_mass-loss_2001} is shown and in black the prescription from \cite{bjorklund_new_2023} is given.
%There is a small increase of the mass-loss rates of the B-stars compared to the O-stars, but this is as a result the higher $\Gamma_{\rm e}$ of these stars which is besides the effective temperature an important value in determining the mass-loss rate.
%When comparing these low $\Gamma_{\rm e}$ stars with the other cooler low $\Gamma_{\rm e}$ stars the trend from O-star to B-star is clearly downward.
%The general trend of the stars with higher $\Gamma_{\rm e}$ from 25000K to 15000K is clearly downwards and there is no big jump towards higher mass-loss rates.  

\blue{We next compare our empirically derived rates (figure \ref{fig:Mass-loss rate emp}) to the predictions by \citet{vink_mass-loss_2001}, \cite{krticka_new_2024} and \citet{bjorklund_new_2023}. As discussed in the introduction, the $\dot{M}(T_{\rm eff})$ behaviour for a fixed mass and luminosity differs between these prescriptions.  
%From high to low temperature, all show an overall downward trend in mass loss, however, the \cite{vink_mass-loss_2001} rates show an upward jump in mass loss of a factor $5$ at about $T_{\rm eff} \sim 22-25$\,kK. At this temperature, no such jump is present in the other two sets of predictions. Both the \cite{vink_mass-loss_2001} and \cite{krticka_new_2024} rates show (indications of) an upward jump once the temperature gets as low as $T_{\rm eff} \sim 13-15$ kK. 
In the context of the present study, the main difference in the predictions of these schemes is that \cite{vink_mass-loss_2001} predict significantly higher mass-loss rates for stars below $T_{\rm eff} \sim 22-25$ kK than for hotter stars, whereas \cite{bjorklund_new_2023} do not predict such a jump but instead a steady decrease of mass-loss rates with decreasing temperature. The \cite{krticka_new_2024} rates do not exhibit such a monotonic decrease of mass loss with decreasing temperature, but rather has a characteristic bump at $T_{\rm eff} \sim 13-15$ kK. This bump is very localised and predicts a rather modest mass loss increase (followed by a steep decrease), which also lies at the edge of our observational coverage range. As such, it is not possible to scrutinise this particular low-temperature feature in the mass-loss behaviour for the present set of stars (spectral types O8.5-B7).}

The comparisons between \cite{vink_mass-loss_2001}, \cite{bjorklund_new_2023}, and \cite{krticka_new_2024} are shown in the bottom panel of Figure \ref{fig:Mass-loss rate emp}, where the ordinate displays predicted $\dot{M}$ divided by the empirical $\dot{M}$ derived in this paper, again plotted as function of effective temperature. 
All predicted rates have been computed by means of the 'mass-loss recipes' provided in the corresponding papers, using the stellar parameters derived for each star ($L_\star$, spectroscopically derived $M_\star$, $T_{\rm eff}$, $\varv_{\infty}$), and a metallicity reflecting that of the LMC. 
Here we corrected for the higher solar metallicity used in \cite{vink_mass-loss_2001} compared to the \cite{bjorklund_new_2023} rates as described in \cite{sundqvist_new_2019}.
As such, additional dependencies on for instance $\Gamma_{\rm e}$ are naturally accounted for in these comparisons.    
%also divided the model predictions from \cite{vink_mass-loss_2001} and \cite{bjorklund_new_2023} by the empirically derived value in figure \ref{fig:Mass-loss rate div}.
%This removes the other dependencies of the mass-loss rate such as luminosity and mass. 
\blue{At temperatures above $\sim 22.5$\,kK 
%both the rates of \citeauthor{vink_mass-loss_2001} and \citeauthor{bjorklund_new_2023} 
all three prescriptions are, on average, fairly aligned with the empirical rates. However, at lower temperature the predictions by \citeauthor{vink_mass-loss_2001} jump up as a result of the bistability jump causing their mass-loss recipe to over-predict}
%The effect from a predicted mass-loss jump in the region becomes very prominent in this figure. 
%Namely, while the mass-loss rates from \citet{vink_mass-loss_2001} are on average fairly aligned (only somewhat higher, see below) with our empirical rates for stars above $T_{\rm eff} \sim 22.5$ kK, below this threshold (where the mass-loss jump is predicted) this mass-loss recipe overpredicts 
the empirical rates by more than an order of magnitude. 
The recipe by \citet{bjorklund_new_2023} shows no such jump, and instead displays a rather constant trend with $T_{\rm eff}$, aligned with the trend inferred for the empirical rates. 
For the coolest star in our sample the \cite{bjorklund_new_2023} rates are much lower than our best-fit values.
This might be an outlier or due to the extrapolation of the \cite{bjorklund_new_2023} model grid. For stars hotter than 26 kK, two stars agree very closely with the observation while two underestimate by about a factor 8.
When comparing the observations to the \cite{krticka_new_2024} prescription we see that they also agree fairly well. 
However, in contrast to the \cite{bjorklund_new_2023} prescription the \cite{krticka_new_2024} prescriptions oscillate between over- and under-predictions. 
%The localised increase around $13-14$ kK which \citeauthor{krticka_new_2024} predicts is not possible to investigate with our current sample as we do not reach cool enough temperatures, and the error-margins on mass-loss increase substantially in this region as the rates are generally lower.} 

% \magenta{However, the error margins for the empirical rates of these stars are very large (e.g., the lower 1-$\sigma$ limit for the mass-loss estimate of the coolest star is more than an order of magnitude lower than the best-fit value), complicating a meaningful comparison.}

%This could be due to the aforementioned lack of stars below 15kK used to compute the \cite{bjorklund_new_2023} prescription
%(meaning the comparison in practise is an extrapolation). On the other hand, we note also that the error margins are very large for these stars (e.g., the lower 1$\sigma$ limit for the empirical $\dot{M}$ of the coolest star in our sample is more than an order of magnitude lower than the best-fit value).  

%Here it is clear that both prescriptions have some problems reproducing the exact values found by fitting the observations.
%However, one model describes the general trend of the mass-loss rate quite well while the other has a very different trend over temperature compared to the empirical data.
%The sudden mass loss jump predicted by the Vink prescription around the 22500K region is the most obvious departure from the empirical trend, but even in the hotter regions the trend is off as the hot stars have too high a mass-loss rate which declines steeply to right before the jump. 
Although the prescription by \citet{bjorklund_new_2023} reproduces the empirically found trend rather well, we observe from the bottom panel of figure \ref{fig:Mass-loss rate emp} an almost constant underestimation with a relative offset of $\langle |\dot{M}_{\rm P} - \dot{M}_{\rm E}| /\dot{M}_{\rm E} \rangle \sim 0.5$ from our best-fit values, where $\dot{M}_{\rm P}$ denotes the prescription results and $\dot{M}_{\rm E}$ is the empirical mass-loss rate. This average is weighted with the error-margin.
%A value of 0 would indicate a perfect match, while a value 0.5 indicates that on average the prescription either underestimates or overestimates the observed value by $0.5  \dot{M}_{\rm E}$.}
% We define the average absolute offset as the weighted average of the division between $\dot{M}_{\rm P} / \dot{M}$ or the inverse if this value is smaller than unity. This slightly more complicated weighted average is not biased when a prescription has over and under estimations which would counteract each other in a regular average and therefore gives the actual expected offset from observations when taking a random star.}
%computing the average across our sample by dividing the \cite{bjorklund_new_2023} prescription by the empirical gives a value of 0.5. 
%\JS{Think here how to do with standard deviations, does it make sense, or do we need the further uncertainties?}. 
For the prescriptions by \citet{vink_mass-loss_2001} we split such average comparisons into stars with $T_{\rm eff}$ above and below the jump. 
For stars above this threshold we find $\langle |\dot{M}_{\rm P} - \dot{M}_{\rm E}| /\dot{M}_{\rm E} \rangle \sim 0.8$ and for stars below $\langle |\dot{M}_{\rm P} - \dot{M}_{\rm E}| /\dot{M}_{\rm E} \rangle \sim 20$. 
\blue{%This means that, on average, on the cool side of the jump the \cite{vink_mass-loss_2001} rates overestimate the mass-loss rate by $20  \dot{M}_{\rm E}$. Because at the cool side of the supposed jump the \cite{vink_mass-loss_2001} rates always overestimate this equates to an overestimation by a factor of 21. 
The \cite{krticka_new_2024} prescription fluctuates between over and under estimating but there is no clear cut-off temperature to split the prescription. 
If we compute the average relative offset for the full sample, we find $\langle |\dot{M}_{\rm P} - \dot{M}_{\rm E}| /\dot{M}_{\rm E} \rangle \sim 0.5$. Which is exactly the same as the \cite{bjorklund_new_2023} offset and close to the \cite{vink_mass-loss_2001} offset for the hot stars.}

%
%
%\blue{The \citeauthor{krticka_new_2024} prescription on average is underestimating the empirical values by only  $\langle \dot{M} / \dot{M}_E \rangle \sim 0.9$. 
%Although, because this prescription changes from over to under estimating this average is a bit deceiving as a random star will have a bigger offset, either under or over-estimating. To this end we computed the average of the absolute offset from the empirical mass loss, by inverting the value if it is below unity. This new average disregards whether the prescription is over or under estimating. This new average gives an offset of  $\langle \dot{M}/\dot{M}_E \rangle \sim 2$. }

%The averages of the offset have also been plotted in figure \ref{fig:Mass-loss rate div} to highlight how far the different prescriptions are offset to the observations. 
%Here we have made two averages for the Vink prescription one before the jump and one after. 
%The Bjorklund rate gives solution 0.51 times the found mass-loss rate on average, the Vink rate above 22500K overestimates by a factor 2.1 while the Vink rate below 22500K overestimates by a factor 39.3. 
Again these comparisons and simple averages show quite clearly that the sudden increase in mass-loss rate predicted at the supposed bi-stability jump by the \cite{vink_mass-loss_2001} prescription is not present in these observational results.
%This causes a big deviation of up to 2 orders of magnitude in mass-loss rate between the \cite{vink_mass-loss_2001} and the observational data for the cooler star. The \cite{bjorklund_new_2023} rates on the other hand seem to be under-predicting the mass-loss rate of the stars. 
%With a big difference being that this offset by a factor of 2.0 is consistent over the temperature range studied here.

\blue{We may further compare our results to other empirical studies. 
\cite{bernini-peron_clumping_2023} show a study of 4 BSGs (B2-B5) in the Galaxy with access to UV spectra and therefore also try to constrain the clumping factor. The mass-loss rates acquired in that study are also higher for similar effective temperatures as expected due to the higher metallicity. They do find some localised increase when compared to 4 literature stars at hotter temperature, but this is hard to verify due to not covering the supposed bistability jump with the stars from their sample. 
In contrast \cite{markova_bright_2008} assume a smooth wind outflow and derive mass-loss rates for Galactic BSGs using H\,${\alpha}$. 
%If these star in reality have clumped winds, 
Within our current spectroscopic formalism for wind clumping, these mass-loss values should be interpreted as upper limits. Assuming effects of clumping do not change the inferred rates of mass loss significantly across the region, we may still examine the mass-loss trend with effective temperature. 
%the clumping factor of these stars is about constant (regardless of its value) we may still examine the mass-loss trend with effective temperature. 
%as such their $\dot{M}$ are upper limits (see previous sections).  
%Here they study 8 Galactic supergiants with spectral class ranging from B0.5 to B9 using optical spectroscopy. 
%As mentioned before they assume a smooth wind to find the mass-loss rates.
%The found mass-loss rates in this paper ranged from $10^{-5.82} \frac{M_{\odot}}{yr}$ for the hottest stars down to $10^{-7.22} \frac{M_{\odot}}{yr} $ for the cool B-stars with a temperature of 11000K. 
%The luminosity range of the sample is comparable with the highest luminosity stars being at $10^{5.9} L_{\odot}$ down to $10^{4.6} L_{\odot}$.
Over their investigated temperature range the derived mass-loss rates decrease monotonically, showing no signs of an increase around 22.5\,kK. The derived $\dot{M}$ in these Galactic stars are generally higher than those we infer here for LMC stars. As the luminosity ranges are comparable, this is due to the higher metallicity in the Milky Way. 
%This difference in mass-loss rate is expected as the higher metallicity in the galaxy should result a higher mass-loss rate as well as the fact that they can not account for clumping resulting in a naturally higher estimate value of the mass-loss rate.
%The general trend with temperature is consistent with our results, though, showing no indication of a mass-loss jump. 
More recently, \citet{rubio-diez_upper_2022} examined winds of OB-stars, including 13 Galactic BSGs, by means of radio and far-infrared continuum observations. They derive values of $\dot{M}$ for the BSGs in their sample that are significantly higher than here. This is to be expected as, first, their measured mass-loss rate is an upper limit; second, they targeted very luminous objects, and, third, they consider a Galactic sample. 
%the B0 to B4 (27000K-15000K) stars range from $10^{-4.89} \frac{M_{\odot}}{yr}$ to $10^{-6.30} \frac{M_{\odot}}{yr}$. 
%but again that is to be expected due to their Galactic sample (and in this case also because they focused on very luminous objects). 
%this being a Galactic sample, but also because the goal of \cite{rubio-diez_upper_2022} was to find upper limits to the mass-loss rates. 
Compared to this study, their general trend with effective temperature is similar. 
%(save for the offset). A
%Again the general trend with effective temperature remains constant with our findings; a 
A comparison between their rates and the predicted \cite{vink_mass-loss_2001} rates therefore too shows a large discrepancy at temperatures below 22.5\,kK, with the predictions higher than the empirical rates by between 1 and 2 orders of magnitude (for their estimates of maximum mass loss).}

\blue{Very recently, \cite{de_burgos_iacob_2024} used optical spectra of 116 Galactic BSGs in the temperature range 15 to 35 kK to derive upper limits to mass-loss rates (upper limits because they used smooth wind models); they did not find any signs of an upward trend in mass-loss rate with decreasing temperature in the predicted bistability region.}  
%The \citeauthor{krticka_new_2021} rates seem to agree best with their sample, while the \citeauthor{bjorklund_new_2023} under predicts by a factor 6 and the \citeauthor{vink_mass-loss_2001} over predicts by a factor 2 or 3. \JS{Ok, fine, but again you need to focus on difference relevant to THIS study. Now you have done it just as we have discussed we do NOT want to do it; that is, the way it is written it is implied that their results are done and derived in the same way as in this paper... (which is not the case as they have small sample (4 stars), do not account for light leakage effects from porous medium in velocity space, assume a void interclump medium, fit 'by eye' (or?)...).}}

% \begin{figure}
%     \centering 
%     \includegraphics[width=0.45\textwidth]{images/devide_Vink+Bjorklund.pdf}
%     \caption{Difference between the empirically found mass-loss rates and 2 known mass loss prescription by dividing the prescription values by the empirically derived values. In red the values for the \cite{vink_mass-loss_2001} mass-loss rates are shown. In black the \cite{bjorklund_new_2023} prescription is used. We also add the average offset in the doted lines where the \cite{vink_mass-loss_2001} prescription is separated in 2 parts one below and one above the jump derived as in \cite{vink_mass-loss_2001}. The green line marks the line where prescription and observation agree.}
%     \label{fig:Mass-loss rate div}
% \end{figure}

\subsection{Terminal wind speeds}

\begin{figure}
    \centering
    \includegraphics[width=0.45\textwidth]{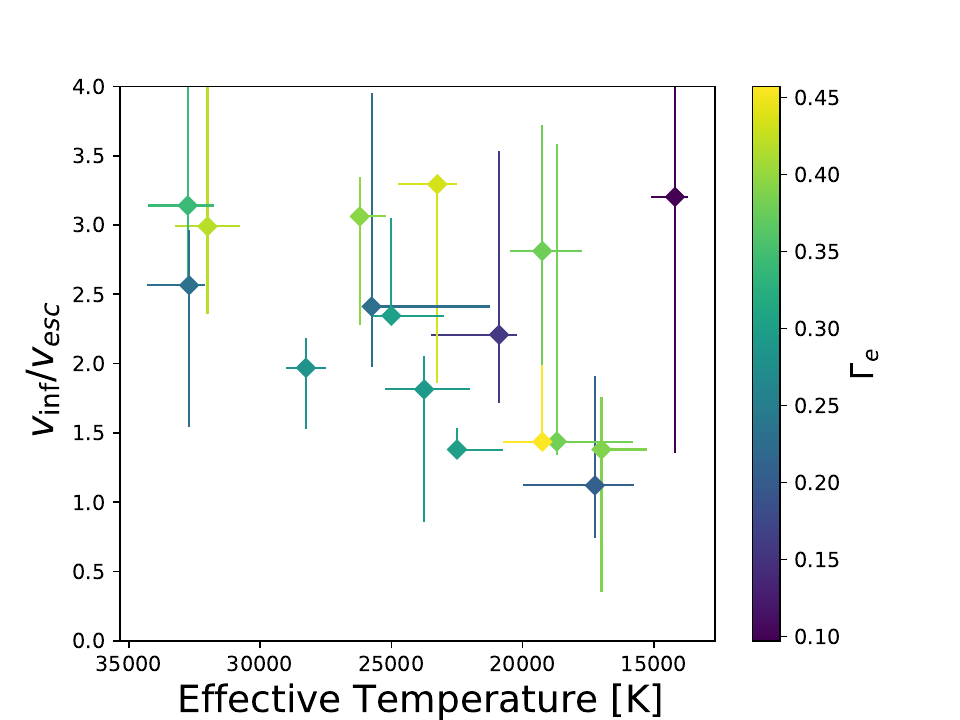}
    \caption{Terminal wind speed divided by escape speed plotted over temperature. The mass used to determine the escape speed is the effective spectral mass.}
    \label{fig:vinf_vesc}
\end{figure}

%The terminal wind speed is measured most sensitively from the width of the blue shifted P-cygni absorption side of the C\,{\sc iv}1550 line.
As can be seen in the top panel of figure \ref{fig:velocity+accelration}, the resulting terminal wind speeds suggest a linear dependence on effective temperature. 
Fitting a linear function $\varv_{\infty} = a T_{\rm eff} -b$, we find best fit values $a = (9.7 \pm 0.6) 10^{-2}$km (sK)$^{-1}$ and $b = 1360 \pm 130$ km s$^{-1}$.
As mentioned in the previous section, the linear behaviour as well as the numerical fit values are in line with the recent findings of \citet{hawcroft_x-shooting_2024}, who fitted $a = (8.5 \pm 0.5) 10^{-2}$ and $b=1150 \pm 170$km s$^{-1}$ for their larger sample. 
Compared to the prescription from the modelling efforts by \citet{vink_metallicity-dependent_2021} we notice that the discontinuity at the supposed bi-stability range does not agree with the observed values. Instead we see a continues increase of $\varv_{\infty}$ with $T_{\rm eff}$.
%(but focusing on finding this sole parameter instead of the 18 parameters we derive here). 
%The similarity in the found behaviour of the terminal wind speed in function of velocity is also highlighted in figure \ref{fig:velocity}. \\
The classical study of UV P-Cygni lines by \citet{lamers_terminal_1995} found a sudden decrease in the ratio $\varv_{\infty}/\varv_{\rm esc}$ around $T_{\rm eff} \sim 21$ kK, a behaviour which has often been associated with the presence of a 
bistability jump, whereas other studies have found a more gradual decrease (e.g., \citealt{markova_bright_2008}). Here $\varv_{\rm esc}^2 = 2 G M_\star (1-\Gamma_{\rm e})/R_\star $ is the effective (i.e., reduced by Thomson scattering) escape speed from the stellar surface.  
%has been observed where the terminal wind speed divided by the escape speed ($v_{\infty}/ v_{esc}$) suddenly drops from about 2.6 to 1.3 \citep{lamers_terminal_1995}.
%As this is around the temperature where the bistability jump is predicted to be this behaviour has been associated with the bistability jump.
%The potential higher mass loss is associated with a lower terminal wind speed due to winds which are optically thick in the lymann continuum.
%This causes the radiation pressure to be produced by the maby metal lines at the Balmer continuum. 
%Resulting in a lower value of the 'force multiplier $\alpha$ giving a lower terminal velocity.
Figure \ref{fig:vinf_vesc} shows our measured $\varv_{\infty}/ \varv_{\rm esc}$ against effective temperature. The figure is dominated by large scatter in the inferred values, reflecting also  uncertainties in the stellar parameters used to compute $\varv_{\rm esc}$. We do indeed observe some indication of a downward trend in the range $T_{\rm eff} \sim 25 - 17$ kK, although also in our relatively small sample at least some outliers to this trend clearly exists on the cold end. Actually, one may detect a similar behaviour also in the above-mentioned larger study by \citet{hawcroft_x-shooting_2024}, if one zooms in on the region around $T_{\rm eff} \sim 20$ kK for LMC stars (see their Figure 5). However, in that study this quite subtle 'by-eye' detected trend is statistically washed away by the large number of stars with hotter effective temperatures and the scarcity of stars with $T_{\rm eff} < 20$ kK.

\blue{Nonetheless, although uncertainties are still large, there may be signs of the general trends observed by \cite{crowther_physical_2006} present in this sample as well. 
This trend is similar to the trend by \cite{lamers_terminal_1995} but instead of a sudden downward jump upwards in $\varv_{\infty} / \varv_{\rm esc}$ a more gradual decrease with decreasing $T_{\rm eff}$ is seen instead. 
%Modelling efforts by \cite{vink_fast_2018} have exhibited the same gradual increase in $\varv_{\infty}/\varv_{\rm esc}$ over the same temperature range.
}

%Figure \ref{fig:vinf_vesc} has relatively large uncertainties associated with it due to the uncertainty on the spectral mass. 
%Although the uncertainties are quite large the general behaviour observed by \cite{lamers_terminal_1995} is still observed here. 
%Most interestingly the behviour remains true at very different $\Gamma_{\rm e}$ values as well.
%The high $\Gamma_{\rm e}$ stars are not consistently getting higher $v_{\infty}/ v_{esc}$ as one might have expected. 
% \subsection{CNO abundances}
% need to make the plots to compare to evolutionary predictions

\subsection{X-rays}

%The inclusion of X-rays has been motivated, as mentioned above, to obtain fits of the P-Cygni profiles of the C\,{\sc iv}1550 doublet in the cooler stars. 
Without X-ray ionisation the later-type stars in our sample show very little triple ionised carbon in their winds and outer atmospheres, resulting in absent or very weak C\,{\sc iv}1550 lines, without any P-Cygni like wind signatures. 
Observations, in contract, show that some cooler stars in the sample have strong C\,{\sc iv}1550 lines with clear wind signatures.% (while also having effective temperatures below 20 kK). 
To reproduce this, we added X-rays produced due to shocks in the wind also in this temperature region, as outlined in Sect. 2.2. 
To ensure all results are comparable even the stars that fit well without X-rays have been fitted with X-rays and those are the results noted here, with the exception of the coolest star in the sample for which the X-rays caused convergence issues.
This also allows us to see if there is a noticeable improvement in fitting by including X-rays. 
Those hotter stars that do not need X-rays to excite carbon or the cool stars with no noticeable C\,{\sc iv} give X-ray values with much higher error margins or very low values and the fit quality stays comparable.

\begin{figure}
    \centering
    \subfigure{\includegraphics[width=0.4\textwidth]{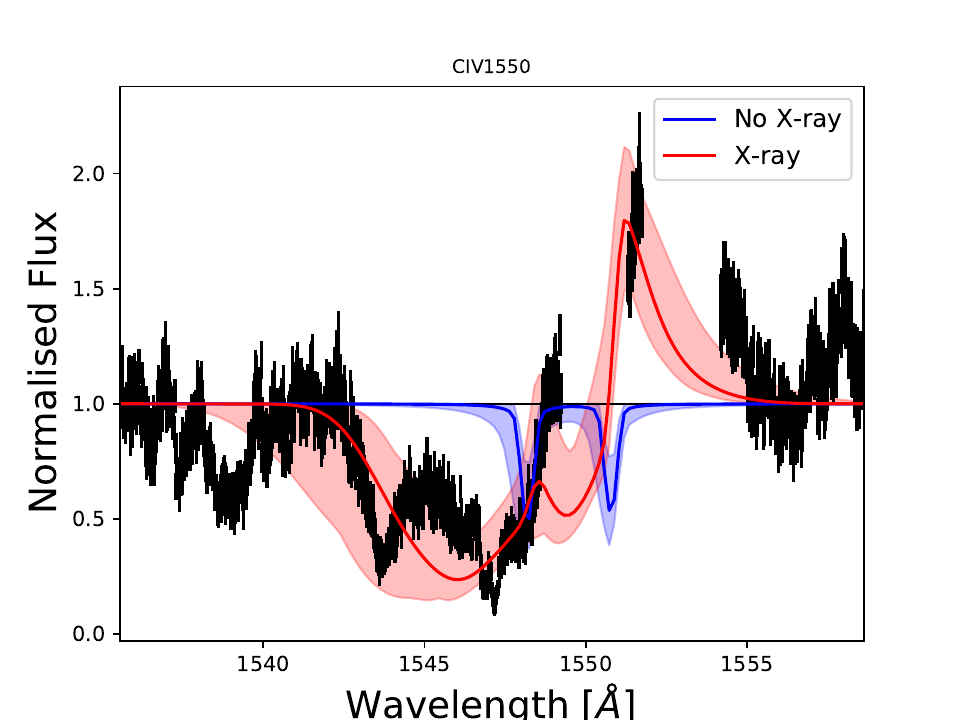}}
    \subfigure{\includegraphics[width=0.4\textwidth]{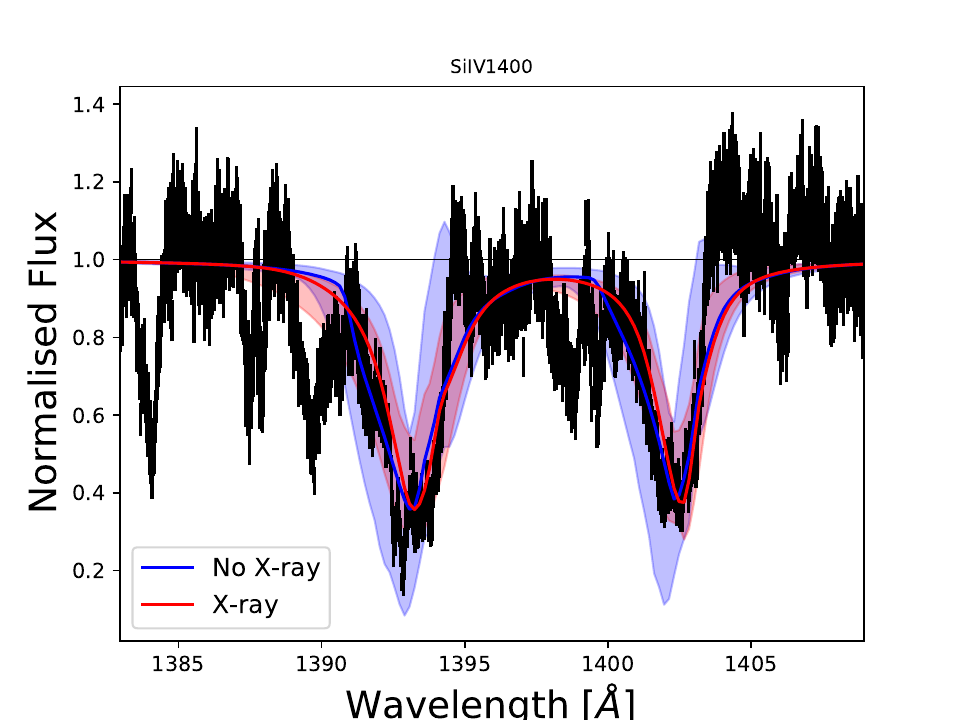}}
    \caption{Three versions of the same doublet line per panel, in black we show the data from Sk$-70^{\circ}16$ a B2 II star in blue the best fit when not allowing for X-rays and in red the best fit when allowing for X-rays. The top panel shows the C\,{\sc iv} 1550 line and the bottom panel shows the Si\,{\sc iv}1400 line. }
    \label{fig:x-ray_linedif}
\end{figure}

Figure \ref{fig:x-ray_linedif} shows the difference the X-rays make to the ability to fit the C\,{\sc iv} and S\,{\sc iv} wind resonance lines of sk$-70^{\circ}16$. The best-fit $\varv_\infty$ of the model including X-ray ionisation is 1000 km/s while the one without gives $\varv_\infty = 500$ km/s.
Notice how this big change in $\varv_{\infty}$ and the addition of X-rays as a whole did not influence the Si\,{\sc iv} line noticeably while transforming the C\,{\sc iv} line substantially, showing that the Si\,{\sc iv} line is a poor indicator of $\varv_{\infty}$.
Although introducing X-rays forces us to consider even more free input-parameters in our modeling, the upside of the addition is that it does not significantly affect any of the other parameters within our sample. 
For example, $T_{\rm eff}$ is typically slightly lowered when X-rays are included, but the reduction is always within the 1-$\sigma$ uncertainty of the originally found effective temperature. 
The issue is often that while weak, unsaturated UV resonance lines may still show clear signatures of line formation in the lower wind, they can be insensitive to $\varv_\infty$ due to very low optical depths in the outermost wind. 
To properly capture this one thus needs to model also the detailed wind ionisation balance, which is typically not done in the simplified large-scale approaches that work well on strong and (almost) saturated UV wind lines (e.g., \citealt{hawcroft_x-shooting_2024, lamers_terminal_1995}).  

%When discussing the strength of the x-ray luminosity we will mostly be using the X-ray luminosity over the bolometric luminosity in $\log_{10}$ scale $\log_{10}(L_x/L_{bol})$. 
We find an average $\log_{10}(L_x/L_{bol}) =-7.3$ for our sample, however with very large error margins of $\pm 1$ to $\pm 2.5$.
For individual stars the large errors cause the best fit range to reach down to $\log_{10}(L_x/L_{bol}) <-8$.
Our average value is on the same order as the often-quoted empirical relation $\log_{10}(L_x/L_{bol}) = -7.2 \pm 0.2$ found from X-ray observations of O-stars \citep{Rauw_2015}.
\blue{ \cite{crowther_x-ray_2022} have studied the X-ray properties of specifically O-stars in the LMC and found a similar relation  $\log_{10}(L_x/L_{bol}) = -6.90 \pm 0.65$.
It also agrees well with the X-ray luminosity range found by \cite{bernini-peron_clumping_2023} ($\log_{10}(L_x/L_{bol}) \approx [-7, -8.3]$).
When computing the range of shock temperatures using equation \ref{eq:jump-T}, we note typical shock temperatures ranging from $10^5$ K, at the X-ray onset radius, to $5 \times 10^6-10^7$ K at the terminal wind speed.}
Note again that the reasoning behind adding X-rays here was solely to reproduce the C\,{\sc iv}1550 doublet in a few cooler stars but implemented in all for standardised comparison (see above). 
For the rest of the stars in our sample including X-rays does not improve the best fit quality. 
As such, the corresponding fit parameters become redundant and the uncertainties of X-ray luminosities extremely high. 
% For example, we have 2 stars with very high best fit $\log_{10}(L_x/L_{bol}) \approx -6$ in our sample, but with 1-$\sigma$ error bars that reach down to $\log_{10}(L_x/L_{bol}) \approx -9$. 
%These 2 stars also are part of the lower luminosity stars in the sample with luminosities 5 times lower than the rest of the sample. 
%As a result the actual X-ray luminosities of these 2 stars is not as remarkable to the rest of the sample.

%\JS{Add some discussion on strength of X-rays here, and physical interpretation.} 

%Nonetheless, this line often times gets used in this temperature domain to determine the terminal wind speed as it is a still present at lower temperatures, which we have now seen might not be the best idea as we can get vastly different terminal wind speeds with nearly identical Si\,{\sc iv}1400 lines. 
%\textbf{TODO: Figure on xray}\\

\subsection{Clumping}
\label{cl_disc}
\begin{figure}
    \centering
    \subfigure{\includegraphics[width=0.35\textwidth]{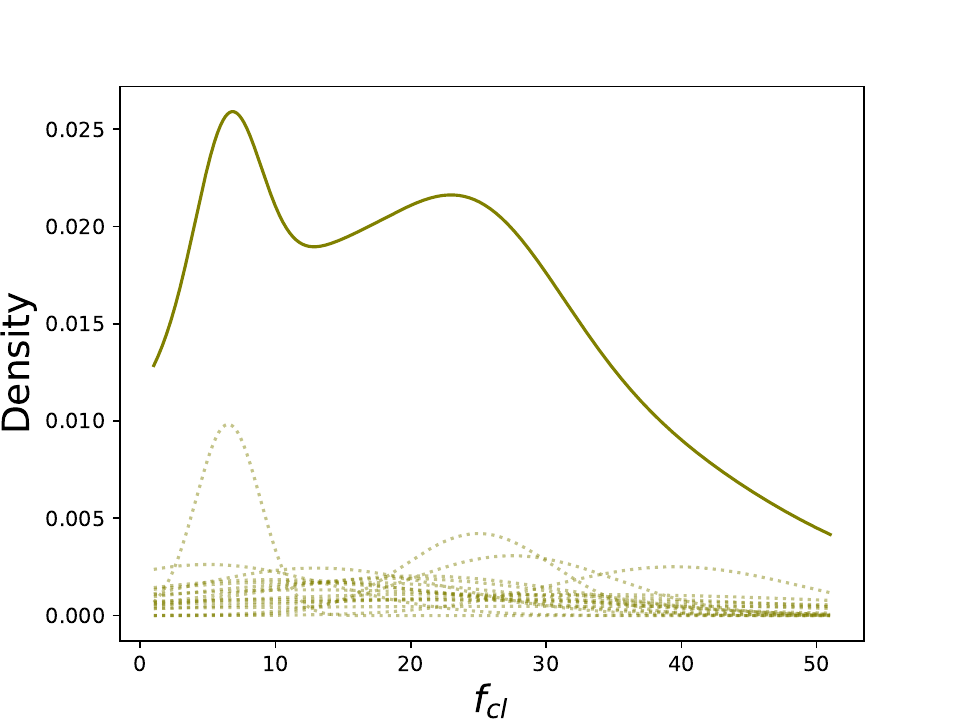}}
    \subfigure{\includegraphics[width=0.35\textwidth]{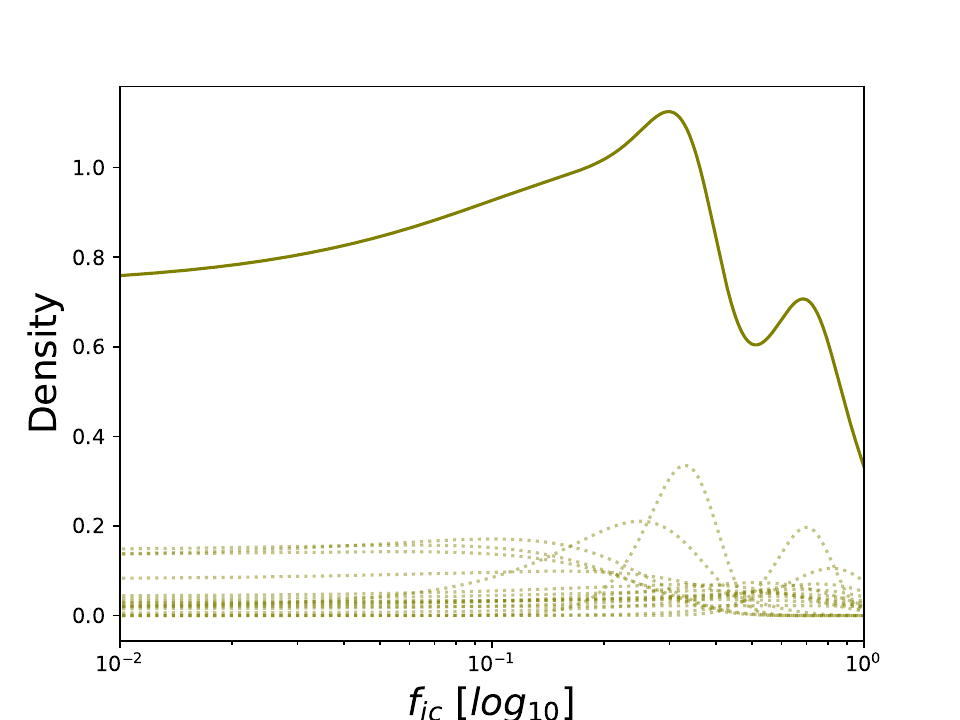}}
    \subfigure{\includegraphics[width=0.35\textwidth]{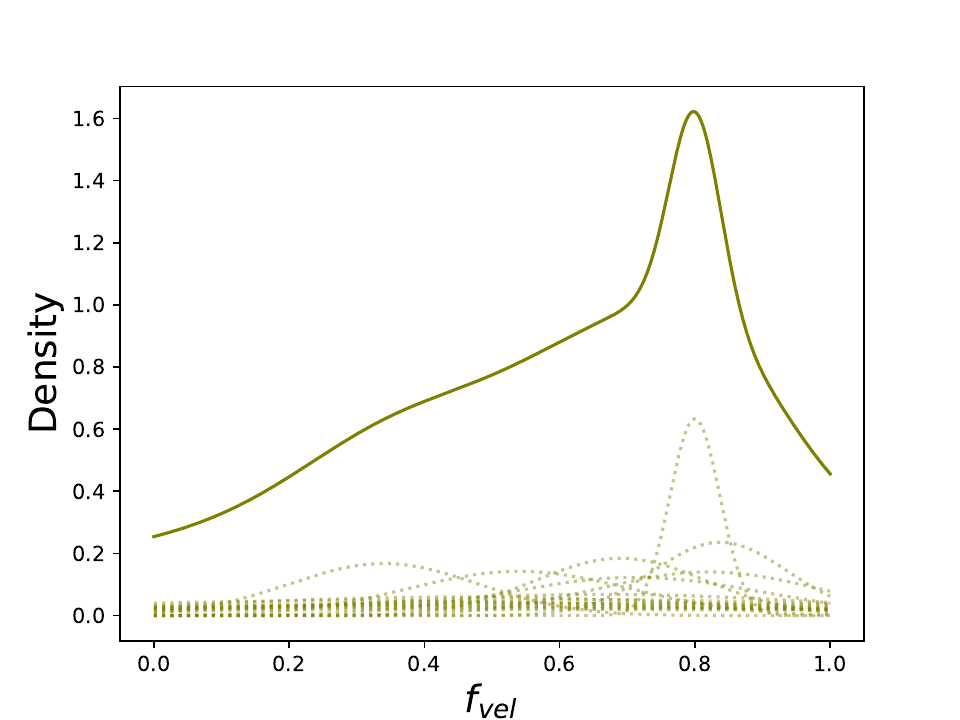}}
    \caption{Approximate kernel distribution of the clumping parameters. In full lines we show the normalised approximate kernel distribution of the clumping parameter ($f_{\rm cl}$), the interclump density ($f_{\rm ic}$), and the velocity filling factor ($f_{\rm vel}$). In dashed lines the contributions of the separate fits contributing to the distribution. }
    \label{fig:gauss_fclump-fic-fvel}
\end{figure}
Figure \ref{fig:clumping} shows that clumping is ubiquitous, however, the scatter and error margins of the clumping parameters are large, and our sample is rather small, making robust quantitative interpretations challenging.
The Pearson correlation coefficient to find correlations between all clumping factors, the $\beta$ parameter, $T_{\rm eff}$, and mass-loss rate, showed no significant correlations for the clumping parameters with any of the other parameters. However, the aforementioned trends, such as the mass-loss rate, $T_{\rm eff}$ correlation and the $\varv_{\infty}$, $T_{\rm eff}$ correlation are reflected by the Pearson coefficient.

As by eye we perceived possible trends for the subsample of stars less than 30\,kK, we did a similar test for those stars. However, this did not reveal significant trends in any of the combinations save for the ($f_{\rm cl}$, $T_{\rm eff}$) and ($f_{\rm cl}$, $\dot{M}$) relations. 
First off we see a tenuous increase in $f_{\rm cl}$ for decreasing $T_{\rm eff}$ \footnote{The linear fit gives $f_{\rm cl} = (-2.1 \pm 0.5)10^{-3} \cdot T_{\rm eff} + (65 \pm 14)$ with a Pearson coefficient of $-0.258 \pm 0.241$.}.
From line de-shadowing instability (LDI) simulations \citep{driessen_theoretical_2019} we expect a downward jump of $f_{\rm cl}$ as we go from O-stars to B-stars, but assuming our observed trend is real it could explain why this jump is difficult to observe. 
Although, we want to highlight again that this observed trend is rather tenuous. 
Because of the clear correlation between $\dot{M}$ and $T_{\rm eff}$, we also find a slightly stronger correlation between $f_{\rm cl}$ and $\dot{M}$\footnote{A linear fit between the mass-loss rate and the clumping factor results in: $f_{\rm cl} = (-17 \pm 3) \log{\dot{M}} - (85 \pm 21)$ with a Pearson coefficient of $-0.369 \pm 0.238$.}. This is inline with the trends observed in O-stars by \cite{brands_r136_2022}, although they also highlight the delicate nature of this result.

Instead of correlations, we can study the general trends of the parameters not in respect to another. 
Figure \ref{fig:gauss_fclump-fic-fvel} shows an approximate kernel distribution where each separate fit result is interpreted as a Gaussian with the uncertainty margin as the width of the Gaussian and each result is scaled by its $1/\chi^2$ value. 
Adding all these Gaussians together yields a distribution which is normalised to unity.
%JS-Very nice dicussion below !! 
\blue{This Gaussian allows us to talk about the behaviour of this parameter over the full sample when no clear trend is present. 
The clumping factor in our sample has a preference to around 20, but while for low clumping factors the likelihood stays strong, due to a single good fit, the tail towards the higher clumping factors decreases rapidly.
The interclump density is less clearly peaked, but still has a 
visible peak around 0.4. The apparent long tail towards lower values also has to do with the displayed logarithmic axis, which visually prioritises small values and compresses the higher end of the axis.
The velocity filling factor shows one of the possible pitfalls of this method, where one data point is dominating the distribution due to fitting well and having a low error-margin. The rest of the distribution is still useful, however, where we see a broad distribution with a relatively steep drop-off towards the lowest values.}
%JS- Very discussion above !! 
%Finally to get errors on these values we fit this distribution with a new Gaussian which gives us an estimate of the error on this value.
%Using this procedure, the values for each of the parameters are $\langle f_{\rm cl} \rangle = 19 \pm 17$, $\langle f_{\rm ic} \rangle = 0.33\pm 0.45$ and, $\langle f_{\rm vel} \rangle = 0.68 \pm 0.40$.
% \OV{Although our method is nice we actually get cleaner results by just taking weighted averages: $\langle f_{\rm cl} \rangle = 21 \pm 10$, $\langle f_{\rm ic} \rangle = 0.39\pm 0.24$ and, $\langle f_{\rm vel} \rangle = 0.65 \pm 0.18$} \JS{Ok, but then we should use this (weighted averages) to do the averages of course? You can just say that the Gaussian kernels visualise the spread in a good way. Also, for weighted averages you find 40 \% on average, but you have not updated key sentences in abstract, conclusions and so?}
% For all of these there is a very wide spread on these values, most notably is the value distribution for $\langle f_{\rm vel} \rangle$ which is dominated by one well determined result. \blue{The gaussian fit finds better correspondence by ignoring this outlier, resulting in reliable results}
% \blue{The large error is partly due to the error on the individual results themselves, but mainly due to the large spread in the individual best-fit values.
% The values quoted above are reached at a certain velocity ($\varv_{\rm cl, max}$) by slowly increasing from the smooth values at a velocity of $\varv_{\rm cl, start}$. 

\begin{figure}
    \centering
    \subfigure{\includegraphics[width = 0.35\textwidth]{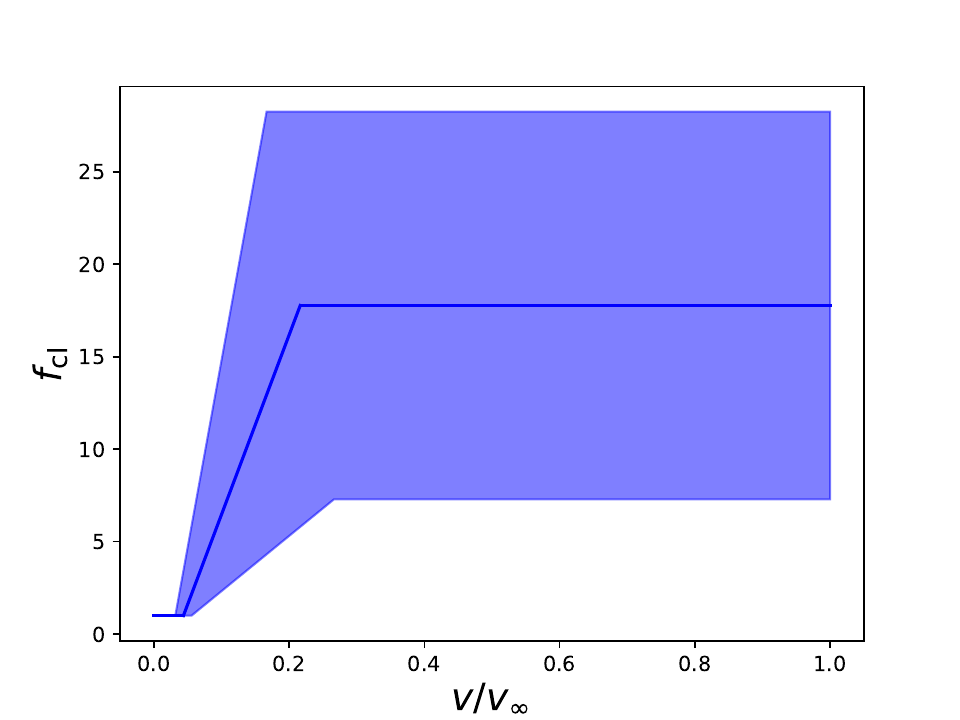}}
    \subfigure{\includegraphics[width = 0.35\textwidth]{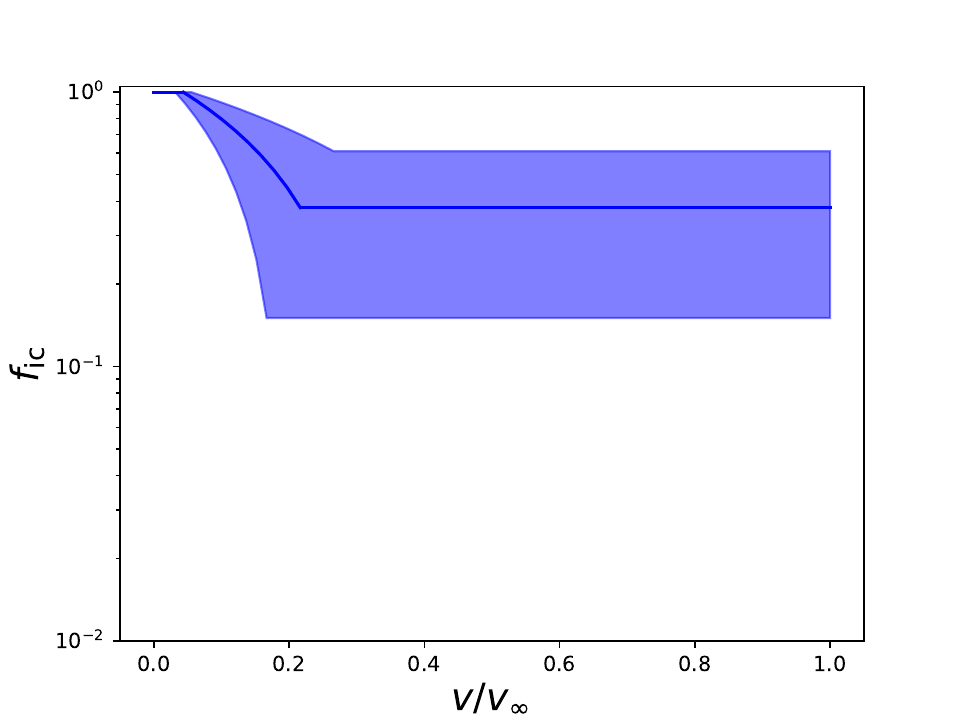}}
    \subfigure{\includegraphics[width = 0.35\textwidth]{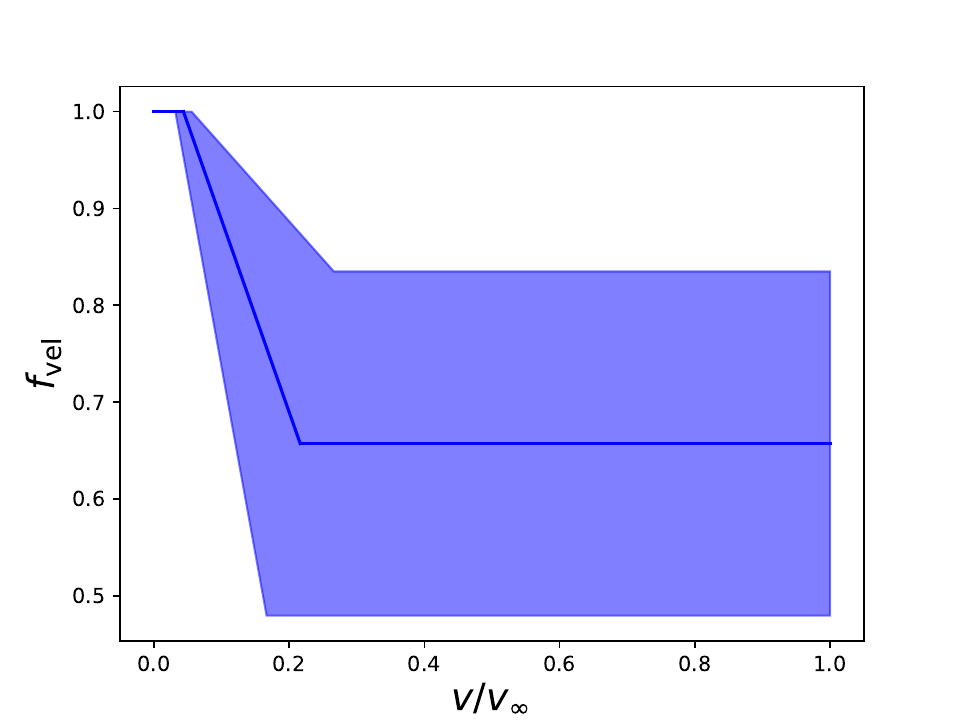}}
    \caption{\blue{Change of the 3 clumping parameters as a function of $\varv/\varv_{\infty}$. The dark blue line shows the averages, with the blue shaded region showing the $1-\sigma$ region of these averages.}}
    \label{fig:clump_over_v}
\end{figure}
%as well as some potential methodological issues further discussed below. 

\blue{When taking a weighted average of the three clumping parameters we find:  $\langle f_{\rm cl} \rangle = 18 \pm 10$, $\langle f_{\rm ic} \rangle = 0.38\pm 0.23$ and, $\langle f_{\rm vel} \rangle = 0.66 \pm 0.18$. 
These values are representing the part of the wind where it is most structured; at low velocities (below the gas sound speed) the atmosphere is always assumed smooth. How the three clumping parameters on average change with velocity is shown in figure \ref{fig:clump_over_v}.}
The high mean value inferred for the interclump density sticks out, suggesting that we can expect that the interclump component comprises close to 40\% of the average wind density.
This is in stark contrast to alternative descriptions of wind clumping that are based on an effectively void interclump component. 
%Moreover, especially in combination with the characteristic values of the other clumping parameters, the  high 
Non-zero $f_{\rm ic}$ have also been found in the previous studies (of O-stars) by H21 and B22. 
These best fits, although quite a bit lower to the value quoted here (H21 $f_{\rm ic} = 0.13 \pm 0.08$, B22 $f_{\rm ic} = 0.13_{-0.13}^{+0.15}$), seem at least not-pointing towards zero. 
As further discussed in the next section, this changes significantly the interpretation of the clumping parameters in terms of the relations between clumping factors, clump over-densities, and volume filling factors, and also invalidates the commonly applied assumption that the complete wind mass is compressed into dense and small clumps. 
%do pose questions regarding the general validity of modeling wind clumping effects in spectroscopic studies through a two-component medium ansatz; we discuss this potential generic issue in Sect. X. 

The inferred mean value of the (maximum) clumping factor is higher than in models by \cite{driessen_theoretical_2019} \blue{and also higher than the Galactic observational study by \cite{bernini-peron_clumping_2023} where they find clumping factors of at most 2. 
We do note that both the fitting method and clumping description in their empirical study are different to ours, but the large difference is still noteworthy.} \blue{Our derived clumping factors are, in contrast,} lower than the sample averages for Galactic O-supergiants by H21 ($f_{\rm cl} \approx 25$) and for LMC O and WNh stars by B22 ($f_{\rm cl} \approx 29 \pm 15$) who use the same clumping model.
%Except for one outlier at the cool end, there seems to be a small tendency in the data that objects with lower $T_{\rm eff}$ display somewhat lower values. 
A downward trend with $T_{\rm eff}$ would be consistent with the linear analysis and non-linear instability simulations by \cite{driessen_theoretical_2019}, but establishing such a potential trend empirically would require a (significantly) larger sample than investigated here. 
In view of these empirical uncertainties and the fact that the theoretical clumping-study by \cite{driessen_theoretical_2019} only used 1D simulations neglecting the influence of a turbulent stellar surface \citep{jiang_2015,debnath_2d_2024} on the clumpy outflow, it is at this point not meaningful to discuss more quantitative comparisons of empirical and theoretical clumping factors in this regime. 

The empirical data suggest that clumping on average starts at low wind velocities within our sample, with several of the $1\sigma$ error ranges extending down towards our lower allowed bound at $\varv \approx a$. While this is qualitatively consistent with previous empirical studies for hotter O-stars (\citealt{puls_2006,cohen_2011, hawcroft_empirical_2021, brands_r136_2022}), corresponding 1D instability simulations \citep{sundqvist_2013}, and \blue{3D radiation transfer studies \citep{surlan_macroclumping_2013}} also here we caution against more quantitative interpretations due to the fact that current wind clumping implementations may not be ideally suited for turbulent layers near the photosphere (see discussion in \citealt{debnath_2d_2024}).

Finally, the fits for the velocity filling factor suggest that velocity-porosity effects indeed play a role in the spectrum formation of these winds, with a mean value that seems reasonable in view of general theoretical expectations (see discussion in \citealt{sundqvist_atmospheric_2018}). The scatter is high, however, and no significant trends can be identified from the present data sample.  
We note in this context that there exists an independent  observational indication for velocity-porosity (parametrised in the present study by a velocity filling factor) in BSG winds, namely via direct comparison of the depth of the blue and red absorption dips of unsaturated UV resonance doublets \citep{prinja_ultraviolet_2013}. \cite{parsons_optically_2024} investigated this for stars in the ULLYSES sample and found similarly that there are clear indications for velocity-porosity across the sample but no significant trend with temperature. They also find very large scatter in their inferred optical depth ratios, as well as a strong temporal variation. Such temporal variations are not taken into account here as we combine two observations (UV+OPT) not taken simultaneously. 

Comparing the obtained optical depths from \cite{parsons_optically_2024} to the velocity filling factors derived here quantitatively is challenging, but equation 23 from \cite{sundqvist_2014} does allow us to approximate the vorosity factor ($f_{\rm vor}$) from the optical depth of the blue and red absorption dips.
This $f_{\rm vor}$ is 
%defined as the average clump velocity span divided by the velocity separation between the clump centers. 
%This verocity can be converted into 
%velocity filling factor ($f_{\rm vel}$) using 
then related to $f_{\rm vel}$ by 
equation 14 in \cite{sundqvist_atmospheric_2018}. 
The empirically derived blue and red optical depths from \citet{parsons_optically_2024} then provides an independent estimate of $f_{\rm vel}$, which can be directly compared to the values derived in this paper. The key qualitative point from this analysis is that velocity-porosity clearly is required in order to explain the blue to red optical depth ratios measured by \citet{parsons_optically_2024}. However, due to high uncertainties in these measured ratios, we still find significant quantitative error margins in $f_{\rm vel}$ (see Fig. \ref{fig:parson_Test}). Nonetheless, within these (admittedly very large) error margins, the two methods seem to provide results that overall are in reasonable agreement. This lends some support to the ability of the fast method utilised by \citet{parsons_optically_2024} for identifying stars where velocity-porosity is important in their UV resonance wind line formation.

\subsection{On the interpretation and validity of current wind clumping methods for spectroscopic studies}
\label{cl_inter}
%\textbf{mention ionisation average (or in method)}
The generally high values found for the interclump density challenge some common assumptions and interpretations regarding wind clumping in massive stars. 
Namely, in most previous studies it has been assumed that the interclump medium is effectively void, so that all wind mass is contained within clumps and the relation between the clumping factor and clump volume filling factor $f_{\rm vol}$ is simply $f_{\rm cl} = f_{\rm vol}^{-1}$, accompanied by a characteristic clump over-density $D = \rho_{\rm cl}/\langle \rho \rangle = f_{\rm cl}$. But for the typical values of $f_{\rm ic} \equiv \rho_{\rm ic}/ \langle \rho \rangle$ found here, these simple relations no longer hold. Instead we must use the full relations for the stochastic two-component medium \citep{sundqvist_atmospheric_2018}: 
\begin{equation}
    f_{\rm vol} = \frac{(f_{\rm ic}-1)^2}
    {f_{\rm cl} + f_{\rm ic}^2 - 2 f_{\rm ic}}
\end{equation} 
\begin{equation} 
D = \frac{\rho_{\rm cl}}{\langle \rho \rangle} = \frac{1-f_{\rm ic}(1-f_{\rm vol})}{f_{\rm vol}}
\end{equation} 
for a mean wind density 
\begin{equation} 
    \langle \rho \rangle = f_{\rm vol} \rho_{\rm cl} + (1-f_{\rm vol}) \rho_{\rm ic} 
\end{equation}
and clumping factor 
\begin{equation} 
    f_{\rm cl} \equiv \frac{ \langle \rho^2 \rangle} {\langle \rho \rangle^2} = 
    \frac{f_{\rm vol} \rho_{\rm cl}^2 + (1-f_{\rm vol}) \rho_{\rm ic}^2}
    {(f_{\rm vol} \rho_{\rm cl} + (1-f_{\rm vol}) \rho_{\rm ic})^2}.
\end{equation} 
Let us first point out that while the above involves the quantities $f_{\rm vol}$, $f_{\rm cl}$, $f_{\rm ic}$, and $D$, only two of these are independent. This is readily seen from the equation for mean density, which we may write in normalised form $1 = f_{\rm vol} D + (1-f_{\rm vol}) f_{\rm ic}$. Thus, if one chooses to set, say, $f_{\rm ic}$ and $D$, $f_{\rm vol}$ follows accordingly, and thereby also $f_{\rm cl}$ from the relations above. That is, while in our modeling we choose the input parameters $f_{\rm cl}$ and $f_{\rm ic}$ we may as well have chosen any pair of the above four quantities and then calculated out the others. Similarly, in the previously standard method of simply assuming $f_{\rm ic} = 0$, only one of the above parameters is necessary to set; e.g. for an input $f_{\rm vol}$ we see directly that indeed $f_{\rm vol}^{-1} = D = f_{\rm cl}$.  

To illustrate the situation when $f_{\rm ic} \ne 0$, we take our derived best-fit values for star Sk$-67^{\circ}14$; $f_{\rm ic} = 0.86$ and $f_{\rm cl} = 8.0$. 
This yields a clump volume filling factor $f_{\rm vol} = 0.003$ that is now much lower than $f_{\rm cl}^{-1} = 0.13$. 
Similarly we find for the characteristic clump over-density $D = 51$, whereas interpreting this the standard way would yield $D = f_{\rm cl} = 8$. 
Moreover, identifying mass-contributions from the clumps and the interclump medium from the equation for $\langle \rho \rangle$ one gets for the former $D f_{\rm vol} = 0.15$ and the latter $(1-f_{\rm vol})f_{\rm ic} = 0.85$; that is, we find here that most of the wind mass is actually not contained in the dense clumps, but rather in the interclump medium. 
Although this star lies on the extreme end of our sample, similar re-interpretations are necessary also when taking the weighted averages of our sample $f_{\rm cl} =18 \pm 10$ and $f_{\rm ic} = 0.38 \pm 0.23$, whereby we obtain $f_{\rm vol} = 0.022 \pm 0.020$ (vs. $f_{\rm cl}^{-1} = 0.06 \pm 0.03$), $D = 28 \pm 27$ (vs. $f_{\rm cl} = 18 \pm 10$) and that the interclump medium contributes $37 \pm 22$\% of the wind mass. These general tendencies arise because we typically find a significant interclump density together with a rather high rms over-density $\sqrt{\langle \rho^2 \rangle}/\langle \rho \rangle$ from the observational data. 
This combination forces clumps to be confined into very small volumes, so that relative contributions from the clump and interclump media to the total wind mass become approximately $1-f_{\rm ic}$ and $f_{\rm ic}$, respectively. Additionally, since typically $f_{\rm vel}$ is significantly higher than $f_{\rm vol}$, it means that these spatially very confined clumps within our formalism are quite spread out in velocity space. 

% Because we determine the error-margins on the sample averages by fitting a symmetrical Gaussian to distributions which (as one can clearly see in figure \ref{fig:gauss_fclump-fic-fvel}) are not symmetrical, this most likely results in an overestimate of the error-margins. Indeed, when propagating these errors it is possible for the uncertainty margins to even extend into the non-physical domain, such as a cumulative clump overdensity that ranges into values below 1 or volume filling factor below 0. 
% Therefore, when assessing the cumulative $f_{\rm cl}$, $f_{\rm ic}$, and $f_{\rm vel}$ we encourage to also simply visually review figure \ref{fig:gauss_fclump-fic-fvel}.

H21 find values $f_{\rm ic} \sim 0.15-0.3$ from their sample of Galactic O-supergiants. B22 split their sample of O and WNh stars into two categories according to stellar luminosity, and find sample averages for their low and high luminosity stars $f_{\rm ic} =0.07 \pm 0.06$ and $f_{\rm ic} = 0.28 \pm 0.21$, respectively. 
Our sample average here is thus noticeably higher, but still in overall qualitative agreement with these previous results for high luminosity O-stars regarding the possibility of a significant non-clump wind mass (we note in this respect that, originally, the discussion in \citealt{sundqvist_atmospheric_2018}, based largely on 1D simulations, argued for low "standard" values on order $\sim$0.01-0.2 for this parameter). The increase of $f_{\rm ic}$ in our study compared to H21 and B22 could be due to the different temperature region or due to slight differences in the fitting routine such as different fitting assumptions. \blue{On the other hand, in an independent study \cite{surlan_macroclumping_2013} used 3D Monte-Carlo radiation transfer, in combination with the 1D code PoWR, for a Galactic O-star analysis and showed that the inclusion of interclump density was needed to reproduce the P\,{\sc v} 1118 doublet. They got best correspondence for relatively high interclump density ($0.1-0.4<\rho>$), however including the interclump density reduced the effect of the clumping factor on the line profiles. These counteracting effects probably also increase the error-margin of our clumping parameters, but due to the global fits of the GA method such inter-dependencies are all taken into account.}

Overall, the general picture that emerges is different from that which has been assumed in spectroscopic studies of massive-star wind clumping previously; a significant fraction of the wind mass seems actually not to be contained in small-scale clumps, but rather in the medium in between these. 
While this is in stark contrast with the assumptions underlying the vast majority of previous spectroscopic studies, it actually correlates rather well with recent theoretical results from multi-dimensional radiation-hydrodynamic simulations. The 3D models of Wolf-Rayet stars by \cite{moens_first_2022} indeed find that approximately half of the wind mass is contained in parcels having densities lower than the mean density of the wind, illustrating again the general issue with assuming a wind completely dominated by dense clumps. 

However, in these recent simulations it is also found that the density distributions do not resemble a two-component medium, but rather have Gaussian-like  distributions (likely log-normal ones, see \citealt{owocki_2018} and \citealt{schultz_convectively_2020}) where the most probable density is quite close to the mean and the dispersion is large. 
Since similar results are suggested also by the 2D O-star simulations by \cite{debnath_2d_2024} and \cite{driesen_thesis}, the latter including effects of the (LDI, \citealt{owocki_rybicki_1984}) (see their Fig. 8.3), this may indicate issues also with the generic assumption applied here of an effective two-component medium consisting of clumps and an interclump medium. \blue{Additionally, the models by \citet{jiang_outbursts_2018} and \citet{debnath_2d_2024} that consider deep sub-surface layers clearly indicate that also the photosphere is very variable and structured, whereas in the clumping description applied here we assume an inner boundary for possible structure formation at the wind sonic point (see Fig. \ref{fig:clump_over_v}).}
In this respect, we note further that (as mentioned in Sect. 2) in our current approach the ionisation balance is only derived for an effective medium, and thus not for the interclump and clumped components separately. Overall, in view of these results a general re-calibration of methods used to account for wind clumping effects in spectroscopic studies may thus be needed (see \citealt{owocki_2018} for a first attempt that focuses on transfer effects arising from the turbulent density structures typically seen in multi-D simulations of hot star winds).      

%\subsection{Implications for mass loss in massive-star evolution models}
%
%\textbf{To be written. We could focus on some stuff from Robin's paper, and also perhaps have a look at Zsolt's paper with Jo on this. We haven't focused on the wind-momentum in the analysis though (stars close in L, so not so useful we think), so let's see. General point is that the bistability mass loss jump is HUGE (independent of luminosity and mass) in these evolution models.}

\section{Summary and outlook}
We have analysed the optical X-shooter and UV ULLYSES spectra of 15 OB supergiants in the LMC using model atmosphere code {\sc fastwind} and the GA fitting approach Kiwi-GA, resulting in 18 consistently determined stellar (7) stellar and (11) wind parameters.
Derived spectroscopic masses in our sample range from 15-75 $M_{\odot}$ and effective temperatures lie in the range 35-14kK with uncertainty margins of around $\pm 1500$K per star. 
Our sample of stars have been selected with the goal to determine if the so-called bistability jump, an upward jump in mass loss towards cooler photospheres within the observed temperature regime, is observable when the parameter degeneracy of wind clumping is broken by the availability of UV data. 
As such, we focused especially on determining accurate mass-loss rates. 
Our derived rates range from $-7.7$ to -6  $\log_{10}(M_{\odot}/$yr), taking into account inhomogeneities of the wind in the most detailed way current 1D atmosphere codes allow. 
%However, as we determined 18 total parameters along the way we were able to study the clumping and wind behaviour of these stars in depth.\\

By determining the mass-loss rates of all stars within $\pm 0.3$ dex for most of them, we can see that no sudden mass loss increase in this effective temperature regime is present in the empirical analyses. 
From comparisons to different theoretical predictions we see that the jump described by \citet{vink_mass-loss_2001} (which is the current standard recipe to include in applications like stellar evolution modelling) is drastically overestimating the mass-loss rates on the cool side of their predicted jump. 
For our sample we derive, similarly to how we derive the uncertainty margins on the cumulative clumping parameters in figure \ref{fig:gauss_fclump-fic-fvel}, an overestimation by about a factor $\sim 24$. 
By contrast, on the hotter side of the predicted jump these same predictions are typically rather well aligned with the empirically derived rates, on average the offset is $\sim 0.8$ times the empirically derived rates.
Comparing to the alternative theoretical rates by \citet{bjorklund_new_2023}, which decreases monotonically with temperature and thus do not find any mass loss jump (see discussions in previous sections). 
We notice that the trend with $T_{\rm eff}$ of these predictions is rather well aligned with the empirical data, showing a relatively constant underestimation and an offset of $\sim 0.5$ times the empirically derived rates. 
If a jump were present in the observed sample it would not be possible to find such a constant offset with a prescription that is only decreasing with decreasing $T_{\rm eff}$. 
Note that for the computation of these ratios we did not take into account the uncertainty margins of the prescriptions, which are always present as they are, on purpose, simple fits of a grid of models resulting in some scatter between the fits and the actual model results; as such, the quoted error-margins are likely to be slightly underestimated. The \citet{krticka_new_2024} prescriptions further also align overall well with the observed mass-loss rates, but our current data-set is not optimal to test the weaker mass-loss bump present at cooler temperatures, as it is also very localised( decreasing aggressively when moving toward even cooler temperatures, see Fig. \ref{fig:ULLYSES_bi}). 
As such we can here neither confirm nor exclude a presence of the modest bump seen in the models by \citet{krticka_new_2024}.  
The localised behaviour in temperature and comparatively weaker strength of the bump would make this feature relatively hard to verify in general, but impossible with our sample due to its temperature position.

Our derived terminal wind speeds are very much in line with the findings of \cite{hawcroft_x-shooting_2024}.
The terminal wind speeds show a very linear dependence on the effective temperature even when masses and luminosities vary heavily. 
When trying to derive $\varv_\infty$ from the C\,{\sc iv}1550 doublet, which typically is the main diagnostic for terminal wind speeds, we noticed that stars which have optical spectra that point clearly towards cool B stars sometimes still had strong C\,{\sc iv} lines, although they are too cool to ionise a sufficient amount of carbon atoms to C\,{\sc iv} in order to have strong C\,{\sc iv} lines. To solve this we included ionisation due to additional X-rays in the wind also for stars down to $T_{\rm eff} \approx 15$ kK. 
This produced a sufficient amount of C\,{\sc iv} atoms to model the C\,{\sc iv}1550 doublet, 
whereas, interestingly enough, it did neither change the Si\,{\sc iv} wind lines nor have a noticeable change on other fit parameters such as mass loss or clumping parameters. \blue{Such effects from X-ray ionisation for BSGs mimic the well known so-called wind 'super-ionisation' for O-stars (often discussed in the context of OVI, \citealt{cassinelli_effects_1979, macfarlane_x-ray--induced_1994, krticka_nlte_2009}). However, even in B-stars this effect has been noted although less commonly mentioned \citep{macfarlane_x-ray--induced_1994}.}

We derived constraints on a collection of clumping parameters which describe inhomogeneities in the wind ($f_{\rm cl}, f_{\rm ic}, f_{\rm vcl,start}, f_{\rm vcl,max}, f_{\rm vel}$).
For $f_{\rm cl}$ we obtained a sample average $18 \pm 10$, where the large error margin primarily is due to the large spread of best-fit clumping factors. 
Perhaps most strikingly in our empirical study of wind clumping are the high values of $f_{\rm ic}$. 
Here we obtain a sample average $0.38 \pm 0.23$, where again the large uncertainty primarily is due to the large spread of the best-fit values; indeed, for individual stars we find values up to $0.92 _{-0.22}^{+0.02}$. 
These high interclump densities may be viewed as surprising in the sense that a widely used assumption in 1D atmospheric and spectroscopic modelling has been that all wind mass is contained in the clumps; by contrast, here we find that typically about half of the wind mass is actually within the interclump regions. 
On the other hand, in view of recent multi-dimensional radiation-hydrodynamical simulations of hot star atmospheres with winds \citep{moens_first_2022, debnath_2d_2024} this may not be so surprising after all, since these tend to display Gaussian-like density distributions centred quite close to the mean rather than bimodal distributions with dominant over-dense clumps (which had been the prevailing thought based largely on 1D LDI simulations, e.g.,  \citealt{sundqvist_2013, driessen_theoretical_2019}). In the simulations by \citet{moens_first_2022} the accumulative density distributions indeed show that almost half of the wind mass is contained in parcels that have densities lower than the mean density of the wind. 
Since similar results seem to be indicated also by multi-D simulations including the LDI \citep{driesen_thesis}, 
but has yet to be further quantified, our results here indicate a need to rethink how the radiation-driven wind is described in 1D codes used for spectroscopic modelling, also as this very basic assumption greatly influences not only the density structure of the wind but also its ionisation structure. 

\begin{figure}
    \centering
    \includegraphics[width=0.45\textwidth]{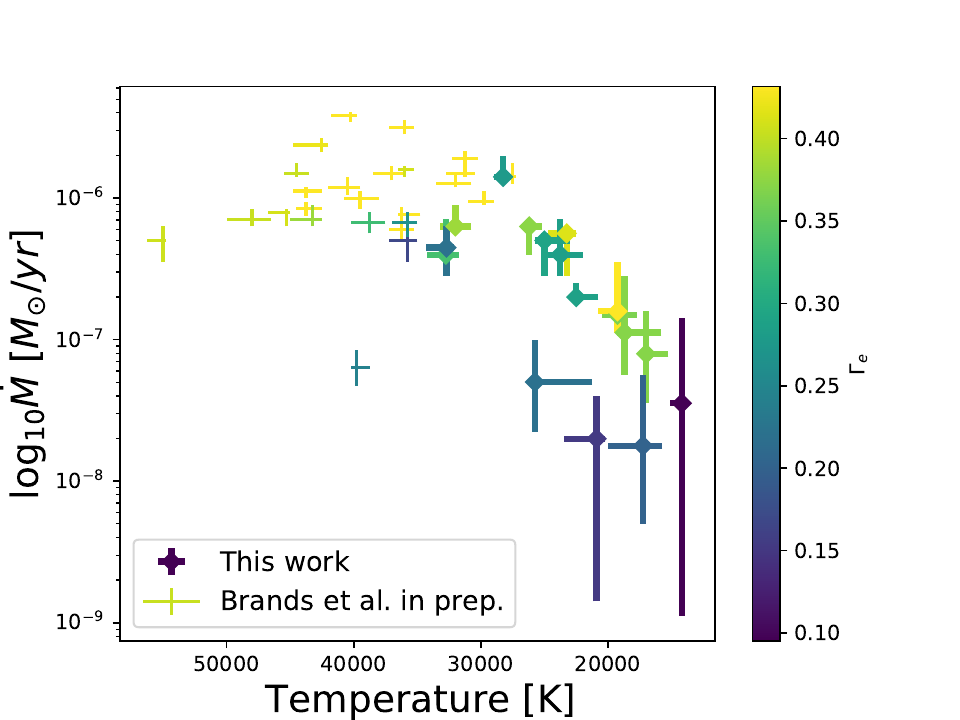}
    \caption{Expanding the present sample with study by Brands et al. (in prep.).}
    \label{fig:sarah_mass_loss}
\end{figure}

An interesting follow up to these results will be the study of similar objects in the SMC, for which we have equivalent data thanks to the ULLYSES program \citep{roman-duval_ultraviolet_2020} and the X-Shooting ULLYSES \citep{vink_x-shooting_2023, sana_x-shooting_2024}. 
The goal will be to find if these empirical mass loss and clumping trends continue to lower metallicity. Additionally, with very comparable methodology for SMC and LMC samples it should be possible to also study the mass-loss behaviour scaling with metallicity. 
To this end we would be able to compare to other SMC studies on the same data in other temperature regimes \citep{parsons_optically_2024, Backs_inprep, bernini-peron_x-shooting_2024} and other LMC samples studying the O-star regime \citep{brands_r136_2022, hawcroft_empirical_2024} (Brands et al. (in prep.)) and comparing to studies in the Milky Way \citep{hawcroft_empirical_2021, bernini-peron_clumping_2023, de_burgos_iacob_2024}.
\blue{A first look at how the sample range can easily be expanded by studies using similar fitting routines and methodology is shown in figure \ref{fig:sarah_mass_loss}.
Brands et al. (in prep.) studied O-stars in the LMC and therefore complements our sample to the hotter side. Figure \ref{fig:sarah_mass_loss} shows our sample again but the thinner crosses show the  mass-loss rates and effective temperatures of the study by Brands et al. (in prep.). We can see that the trend of increasing mass-loss rates with increasing temperatures and $\Gamma_e$ is still present in this larger sample. As expected the error margins in this high temperature regime are smaller, as we already saw in our sample, where the hotter stars have lower error in mass-loss rate.
Object sk$-68^{\circ}155$ is present in both samples and have very comparable fit-parameters in both studies (with the exception of the carbon abundance).} 

Finally, another interesting step would be to further study the effects of not including a strong bistability jump in evolution calculations. The lack of a prominent jump in observations means that stellar evolution codes currently being used as standards to describe single massive star evolution are also likely to significantly overestimate the loss of mass and angular momentum in this temperature and luminosity regime, perhaps then even influencing the viability of single stripped star formation (see, e.g., discussion in \citealt{bjorklund_new_2023}) or changing surface abundances \citep{josiek_impact_2024}, and rotation \cite{keszthelyi_modeling_2017, britavskiy_tracing_2024}. 
However, in this respect we note that for even higher luminosity to mass ratios than those examined in this paper, multi-dimensional models in this temperature regime suggest that strong winds can be ignited already from hot and optically thick sub-surface layers \citep{jiang_outbursts_2018}, which may enhance mass-loss rates beyond the standard luminosity scalings of the stars with line-driven winds studied in this paper. And indeed, empirically this then approaches the region where we find the so-called Luminous Blue Variables (LBVs), for which it is well known that they undergo vigorous and variable mass loss (see overview in \citealt{vink_review_2022}). 
We would argue that it should be a key focus for future (both empirical and theoretical) studies to try and better constrain the mass loss properties of this (very) high-luminosity region in the HR diagram. 

\section*{Data availability}
The fits of all stars discussed here and the distribution of the GA models are available at \url{https://zenodo.org/records/13948998}
\begin{acknowledgements}
The resources and services used in this work were provided by the VSC (Flemish Supercomputer Center), funded by the Research Foundation - Flanders (FWO) and the Flemish Government.
We would also like to thank professor Leen Decin for her contribution to this work.  
O.V. and J.S. acknowledge the support of 
the Belgian Research Foundation Flanders (FWO) Odysseus program under grant number G0H9218N and FWO grant G077822N and, KU Leuven C1 grant MAESTRO C16/17/007.
O.V. also acknowledges the FWO travel grant under grant number K210124N.
J.S., F.B., and P.S. further acknowledge the support of the European Research Council (ERC) Horizon Europe under grant agreement number 101044048. 
B.K. acknowledges the support from the Grant Agency of the Czech
Republic (GA\v{C}R 22-34467S) and RVO:67985815.
AACS and MBP are supported by the Deutsche Forschungsgemeinschaft (DFG - 
German Research Foundation) in the form of an Emmy Noether Research 
Group -- Project-ID 445674056 (SA4064/1-1, PI Sander). AACS and MBP 
further acknowledge funding from the Federal Ministry of Education and 
Research (BMBF) and the Baden-W{\"u}rttemberg Ministry of Science as 
part of the Excellence Strategy of the German Federal and State Governments.
RK acknowledges financial support via the Heisenberg Research Grant funded by the Deutsche Forschungsgemeinschaft (DFG, German Research Foundation) under grant no.~KU 2849/9, project no.~445783058.
\end{acknowledgements}
%the extend to which these changes will occur still needs to be studied in more detail. 


\addcontentsline{toc}{section}{Bibliography}
\bibliography{Bibliography/aanda.bib}
\clearpage
\begin{appendix}
\section{Remarks on the results}\label{app:CNO}
\begin{figure}[h!]
    \centering
    \includegraphics[width=0.45\textwidth]{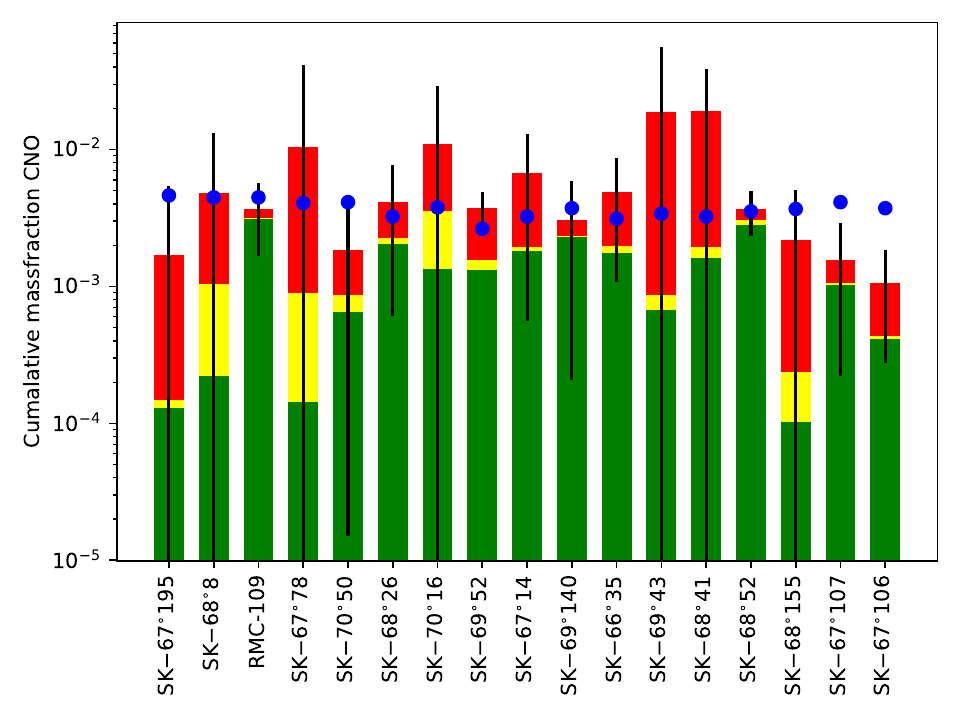}
    \caption{Cumulative CNO mass fraction; Here the CNO abundance is plotted for each star, (C in green, N in yellow, and O in red) in the sample with in black-lines the uncertainty on the sum of the CNO abundance. The blue dots indicates the expected cumulative CNO massfraction from \cite{vink_x-shooting_2023}}
    \label{fig:cummulative CNO}
\end{figure}

The goal of this paper was not to focus on the CNO abundances in particular.
However, as they are needed fit parameters to derive the other parameters we shall briefly highlight what we can learn from them.
The CNO abundance as a whole is expected to stay constant during the evolution of the star as one element gets converted into the other. 
To this effect to check the validity of our CNO abundances we have made figure \ref{fig:cummulative CNO} showing the sum of CNO massfractions and propagating the uncertainties for each of them.
As expected, the sum of CNO abundances with the uncertainty stays close to the base sum.
%\section{Notes on the micro turbulence}
\begin{figure}[h!]
    \centering
    \includegraphics[width=0.45\textwidth]{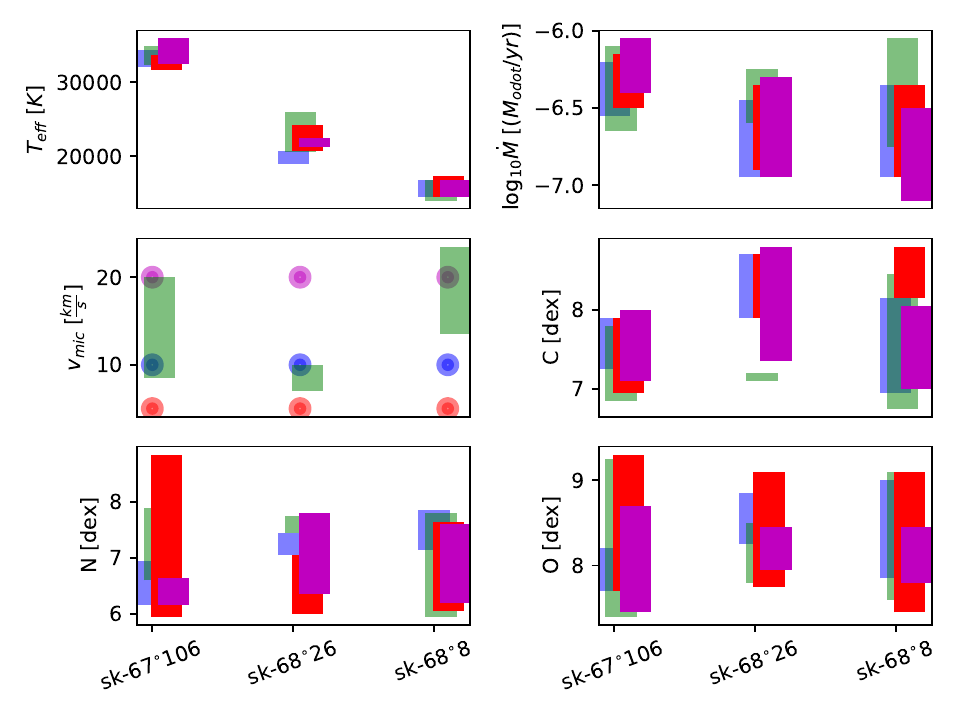}
    \caption{Results of the micro-turbulence variation tests. We show for three stars, four different fits. These 4 fits are very similar except for the photospheric micro turbulence, which is changed from a fixed 10 km/s we used in the paper (blue), to 5 km/s (red), to 20 km/s (purple), and left as a fit parameter (green). We plot the 1-$\sigma$ uncertainty of 5 parameters besides the $\varv_{\rm mic}$. These are: effective temperature, mass loss rate, and the CNO abundance. The three different stars each get a column in the 6 different panels and the 4 different fits are slightly overlapping in this column. }
    \label{fig:micro_test}
\end{figure}

%\section{Comparison of velocity filling factor}
\begin{figure}[h!]
    \centering
    \includegraphics[width=0.45\textwidth]{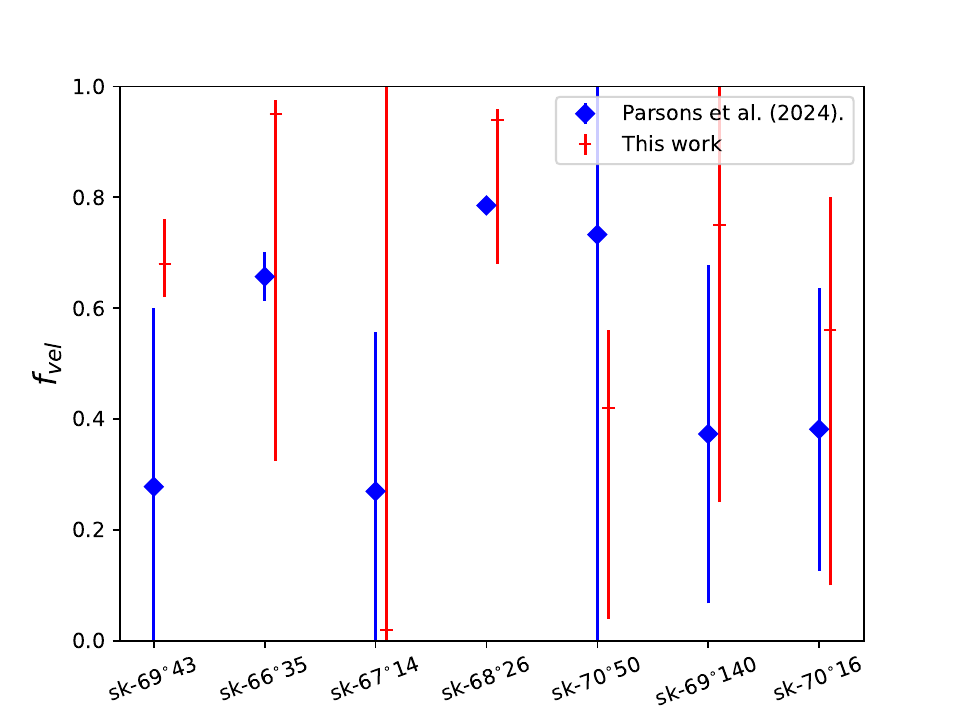}
    \caption{Velocity filling factor in this study compared to \citet{parsons_optically_2024}. This figure shows a comparison between the velocity filling factor we derived here, and the velocity filling factor from \citet{parsons_optically_2024}. We compute the velocity filling factor from optical depths of the red and blue absorption dips obtained by \citet{parsons_optically_2024} using equation 23 from \citet{sundqvist_2014}. When the ratio of the red and blue absorption dips reached values above 2 we excluded it here as this results in non-physical $f_{\rm vel}$. }
    \label{fig:parson_Test}
\end{figure}

\section{Line list}
\onecolumn
% \longtab{
% \begin{longtable}{lllrrr}
% \caption{\label{kstars} Sample stars with absolute magnitude}\\
% \hline\hline
% Catalogue& $M_{V}$ & Spectral & Distance & Mode & Count Rate \\
% \hline
% \endfirsthead
% \caption{continued.}\\
% \hline\hline
% Catalogue& $M_{V}$ & Spectral & Distance & Mode & Count Rate \\
% \hline
% \endhead
% \hline
% \endfoot
% %%
% Gl 33    & 6.37 & K2 V & 7.46 & S & 0.043170\\
% Gl 66AB  & 6.26 & K2 V & 8.15 & S & 0.260478\\
% Gl 68    & 5.87 & K1 V & 7.47 & P & 0.026610\\
%          &      &      &      & H & 0.008686\\
% Gl 86 
% \footnote{Source not included in the HRI catalog. See Sect.~5.4.2 for details.}
%          & 5.92 & K0 V & 10.91& S & 0.058230\\
% \end{longtable}}
% example for Table A.3:
\longtab[3]{
\begin{longtable}{lrcrrrrrrrrl}
\caption{Short caption of Table A.3.}\\
\hline
\hline
Def & mol & Ion & $\lambda$ & $\chi$ & $\log gf$ & N & e & rad & $\delta$ & $\delta$ red & References \\
\hline
\endfirsthead
\caption{continued}\\
\hline
Def & mol & Ion & $\lambda$ & $\chi$ & $\log gf$ & B & C & rad & $\delta$ & $\delta$ red & References \\
\hline
\endhead
\hline
\endfoot
\hline
\endlastfoot
A & CH & 1 &3638 & 0.002 & $-$2.551 & & & & $-$150 & 150 & Jorgensen et al. (1996) \\
\end{longtable}
}% End longtab 

\begin{longtable}{p{2.cm}p{2.cm}p{2.cm}}
    \caption{\normalsize{Detailed list of all spectral lines which are fitted. }}\label{tab:linelist}\\
        \hline
        Ion& Wavelength $[\AA]$&Line window\\
        \hline
        C\,{\sc iv} & 1168.9, 1169.0 & C\,{\sc iv}\,1196b\\
        C\,{\sc iii} & 1174.9, 1175.3, 1175.6  & C\,{\sc iv}\,1169b\\
         & 1175.7, 1176.0, 1176.4 & \\
        Si\,{\sc iv} & 1393.8, 1402.8 & Si\,{\sc iv}\,1400\\
        C\,{\sc iv} & 1548.2, 1550.8& C\,{\sc iv}\,1550\\
        \hline
        O\,{\sc iii} & 3961.6 & H$\epsilon$\\
        He\,{\sc i} &  3964.7 &H$\epsilon$\\
        H\,{\sc i} & 3970.1 & H$\epsilon$\\
        He\,{\sc ii} & 4025.4 & He\,{\sc i} 4026\\
        He\,{\sc i} & 4026.2 & He\,{\sc i} 4026\\
        C\,{\sc iii} & 4068.9, 4070.3 & C\,{\sc iii} 4070\\
        O\,{\sc ii} & 4069.6, 4069.9, & C\,{\sc iii} 4070 \\
            & 4072.16, 4075.86 & \\
        S\,{\sc iv} & 4088.9, 4116.1 & H$\delta$ \\
        N\,{\sc iii} & 4097.4, 4103.4 & H$\delta$ \\
        H\,{\sc i} & 4101.7 & H$\delta$ \\
        Si\,{\sc ii} & 4128.1, 4130.9 & Si\,{\sc ii} 4128\\
        N\,{\sc iii} & 4195.8, 4200.1, 4215.77 & He\,{\sc ii} 4200\\ 
        He\,{\sc ii} & 4199.6 & He\,{\sc ii} 4200\\
        C\,{\sc ii} & 4267.0, 4267.3 & C\,{\sc ii} 4267\\
        He\,{\sc ii} & 4338.7 & H$\gamma$\\
        H\,{\sc i} & 4340.5 & H$\gamma$\\
        O\,{\sc ii} & 4317.1,4319.6, 4366.9 & H$\gamma$\\
        N\,{\sc iii} & 4345.7, 4332.91 &  H$\gamma$\\
        N\,{\sc iii} & 4379.0, 4379.2 & He\,{\sc i} 4387 \\
        He\,{\sc i} & 4387.9 & HeI4387 \\
        O\,{\sc ii} & 4414.9, 4417.0 & O\,{\sc ii} 4416\\
        He\,{\sc i} & 4471.5 & He\,{\sc i} 4471\\
        N\,{\sc ii} & 4613.9, 4621.4, 4630.5 & N\,{\sc ii} 4601\\
         & 4601.5, 4607.2, 4643.1 & \\
        N\,{\sc iii} & 4534.6 & He\,{\sc ii} 4541\\
        He\,{\sc ii} & 4541.4 & He\,{\sc ii} 4541\\
        Si\,{\sc iii} &4552.6, 4567.8, 4574.8 & Si\,{\sc iii} 4552\\
        C\,{\sc iii} & 4647.4, 4650.2, 4651.5 & C\,{\sc iii} N\,{\sc iii} COLD/HOT\\
        N\,{\sc iii} & 4634.1, 4640.6, 4641.9 & C\,{\sc iii} N\,{\sc iii} COLD/HOT \\
        O\,{\sc ii} & 4638.9, 4641.8, & C\,{\sc iii} N\,{\sc iii} COLD\\
            &  4661.6, 4676.2& \\
        He\,{\sc ii} & 4685.6 & He\,{\sc ii} 4686 \\
        N\,{\sc iii} & 4858.7, 4859.0, 4861.3,  & H$\beta$\\
        & 4867.1, 4867.2, 4873.6 &\\
        He\,{\sc ii} & 4859.1 & H$\beta$ \\
        H\,{\sc i} & 4861.4 & H$\beta$\\
        He\,{\sc i} & 4921.9 & He\,{\sc i} 4922\\
        He\,{\sc ii} & 5411.3 & He\,{\sc ii} 5411\\
        O\,{\sc iii} & 5592.3 & O\,{\sc iii} 5592\\
        C\,{\sc iii} & 5695.9 & C\,{\sc iii} 5695 \\
        He\,{\sc i} & 5875.6 & He\,{\sc i} 5875 \\ 
        He\,{\sc ii} & 6527.1 & He\,{\sc ii} 6527\\
        He\,{\sc ii} & 6559.8 & H\,${\alpha}$\\
        H\,{\sc i} & 6562.8 & H\,${\alpha}$\\
        C\,{\sc ii} & 6578.1, 6582.9 & C\,{\sc ii} 6578\\
        He\,{\sc i} & 6678.2 & He\,{\sc ii} 6683\\
        He\,{\sc ii} & 6682.8 & He\,{\sc ii} 6683\\
        He\,{\sc i} & 7065.2 & He\,{\sc i} 7065\\
        \hline
        \normalsize
\end{longtable}
        \begin{tablenotes}
            \item \textbf{Note:} The first column shows the atom and its ionisation stage which is responsible for the transition. The second column shows the corresponding wavelength with possible multiplets. The third column shows where to find this line in the fit summary which is available on Zenodo.
        \end{tablenotes}

\twocolumn

\section{Fit summaries}\label{sec:fitsummary}
\begin{figure}[htp!]
    \centering
    \subfigure{\includegraphics[width=0.55\textwidth]{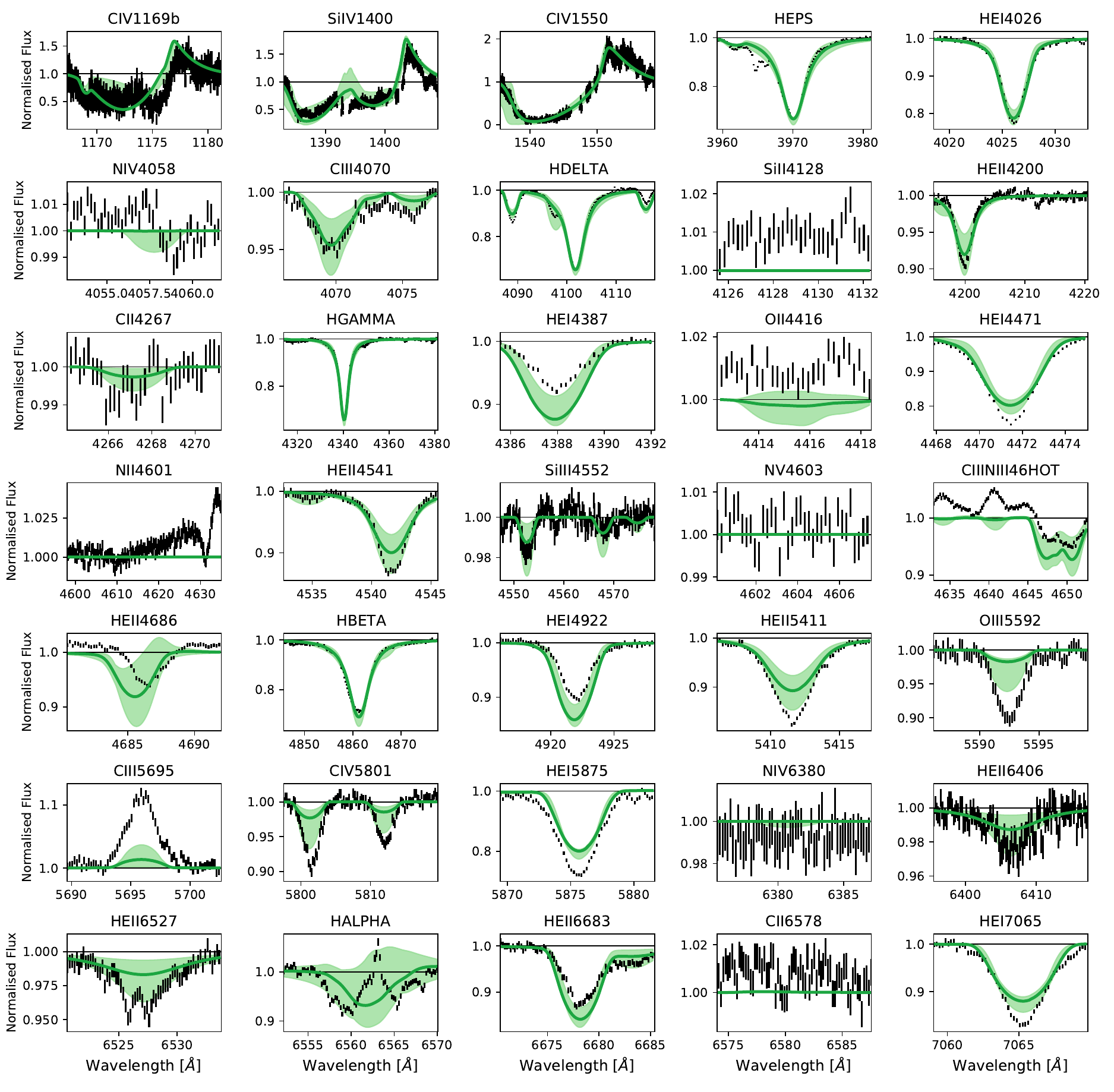}}
    \subfigure{\includegraphics[width=0.55\textwidth]{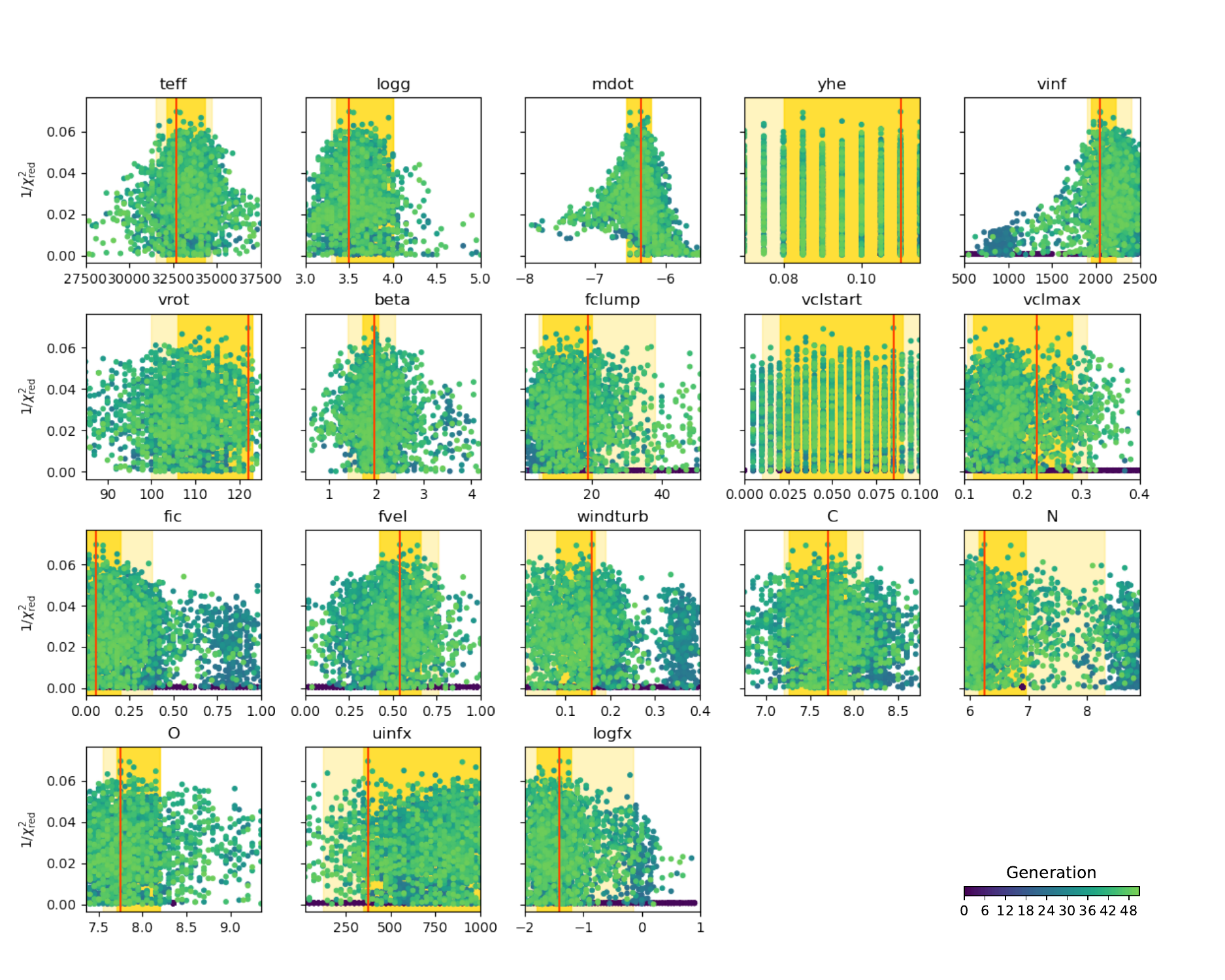}}
    \caption{Fit summary of object Sk$-67^{\circ}106$. The top of this figure shows the fitted line windows with in solid green line the best model fit and in the light green area the $1-\sigma$ uncertainty area of the fit. The data is shown in black.
    In the bottom part we show the resulting probability distribution of these fits in the top. Each model produced while running the GA is one dot in this plot with the colour showing the generation of the model. The better the model fits the higher the $1/\chi^2_{\rm red}$. The best model is noted by the red line and the 1 and 2 $\sigma$ uncertainties are highlighted by the yellow regions. }\label{fig:full_fits}
\end{figure}
\begin{figure*}
    \centering
    \subfigure{\includegraphics[width=0.7\textwidth]{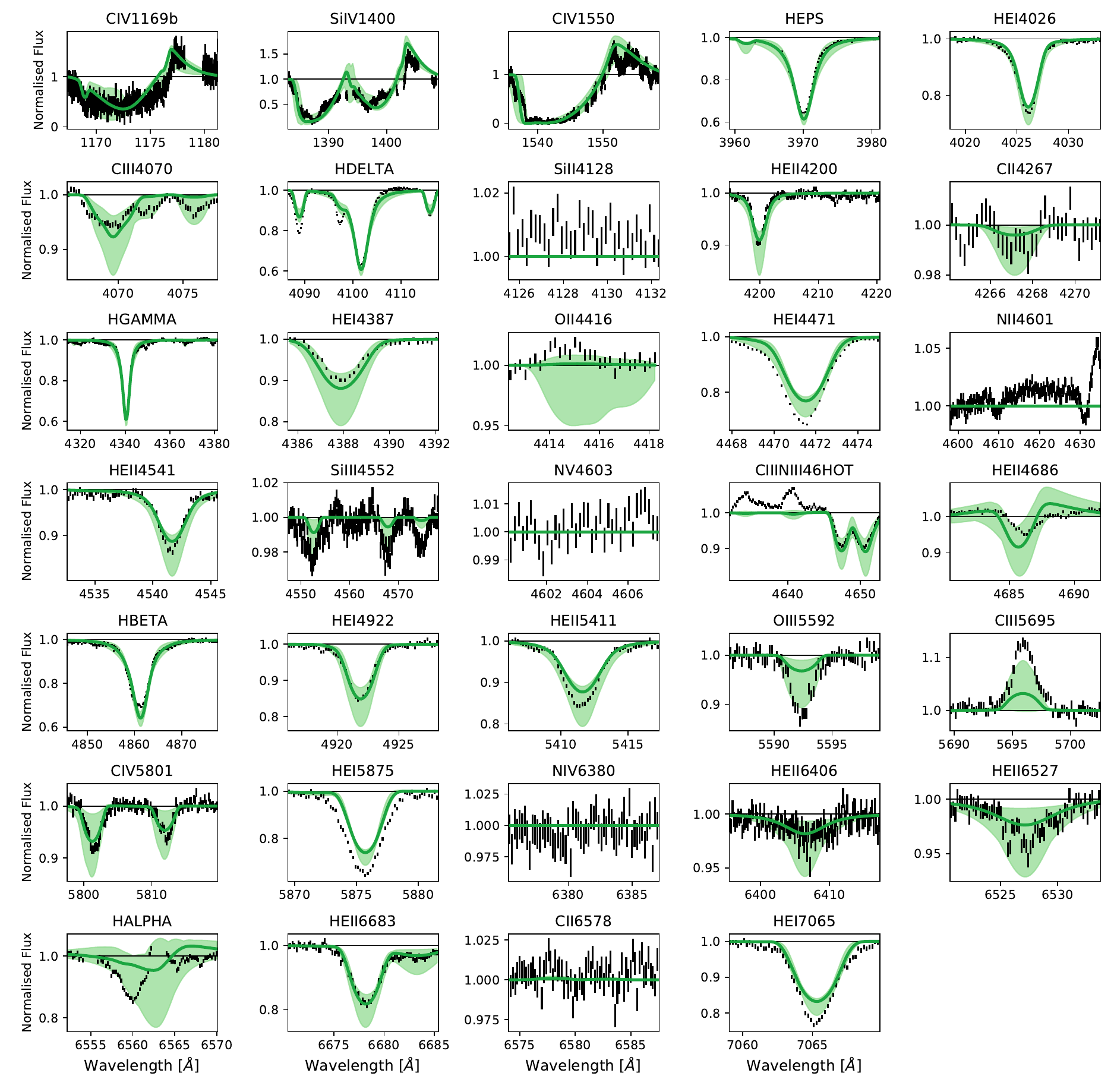}}
    \subfigure{\includegraphics[width=0.7\textwidth]{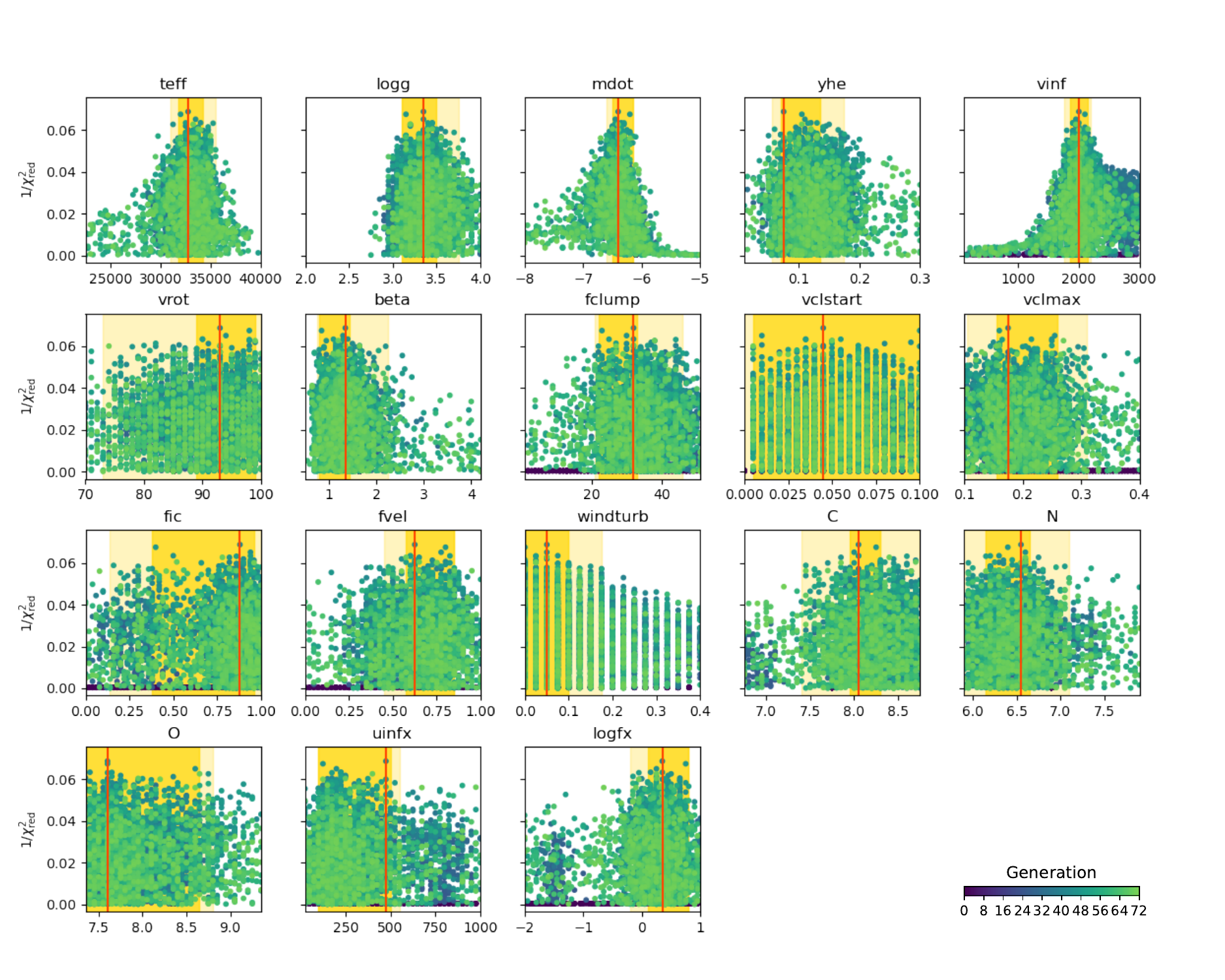}}
    \caption{Same as figure \ref{fig:full_fits} but for object Sk$-67^{\circ}107$.}
\end{figure*}

\begin{figure*}
    \centering
    \subfigure{\includegraphics[width=0.7\textwidth]{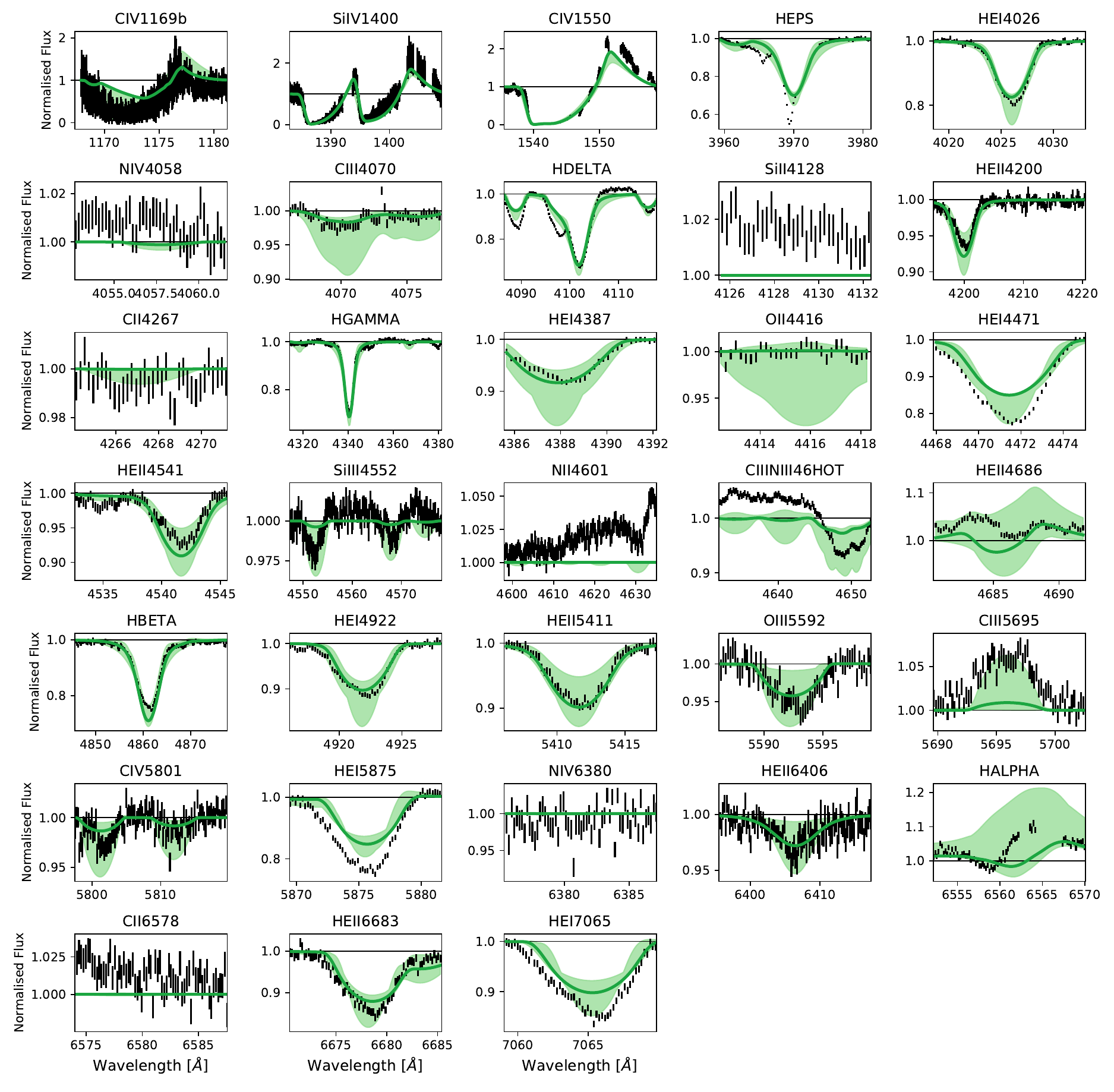}}
    \subfigure{\includegraphics[width=0.7\textwidth]{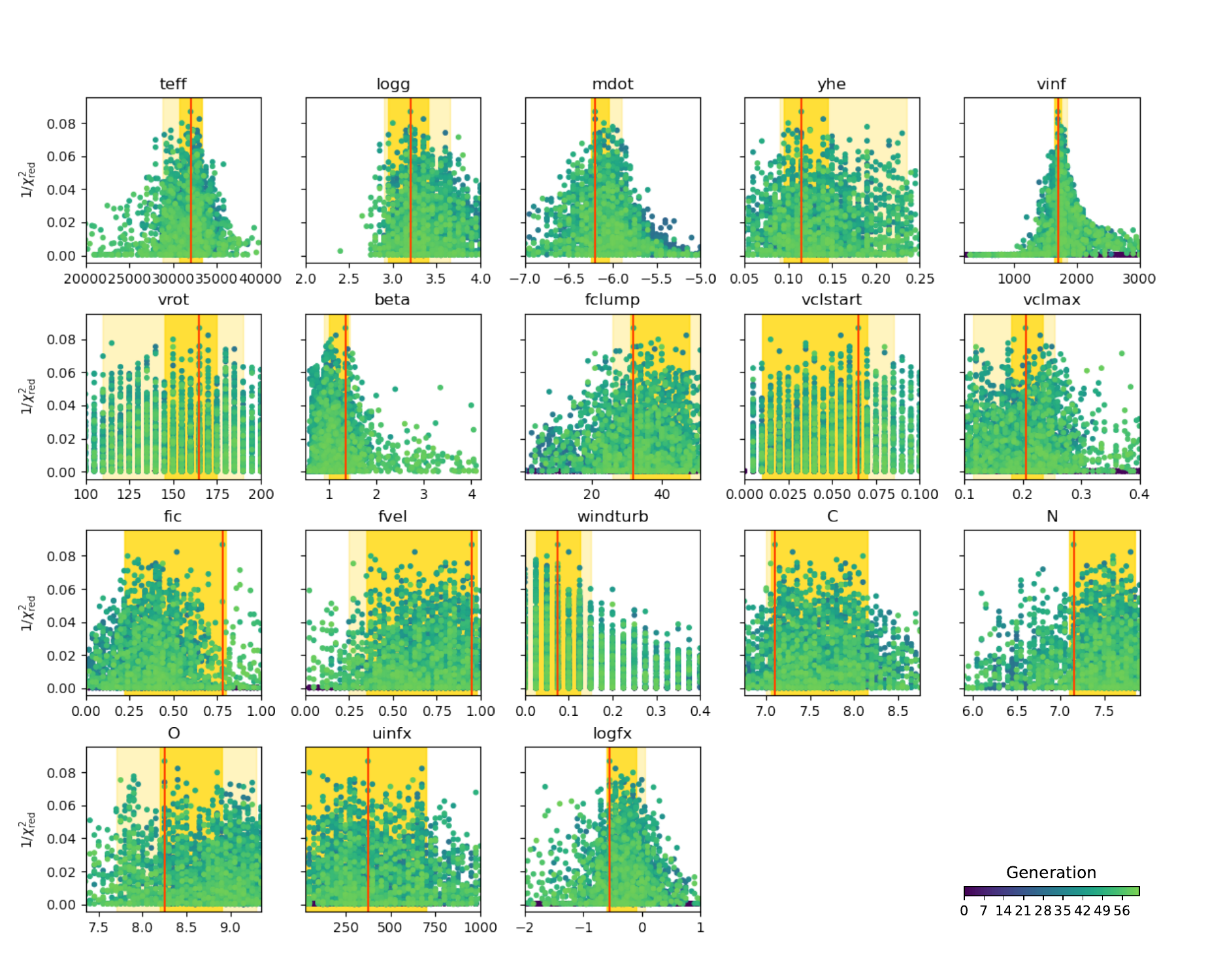}}
    \caption{Same as figure \ref{fig:full_fits} but for object Sk$-68^{\circ}155$.}
\end{figure*}

\begin{figure*}
    \centering
    \subfigure{\includegraphics[width=0.7\textwidth]{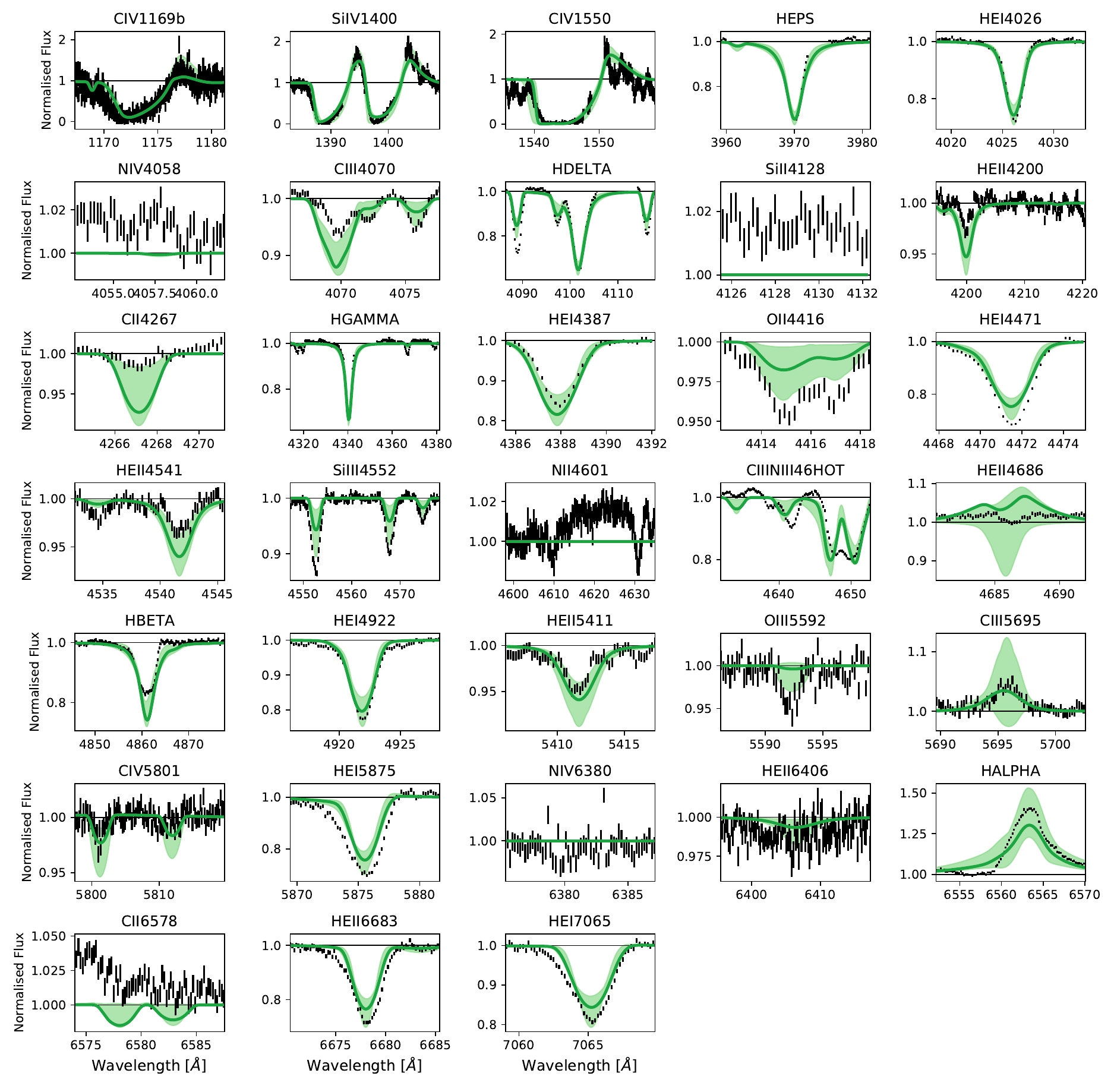}}
    \subfigure{\includegraphics[width=0.7\textwidth]{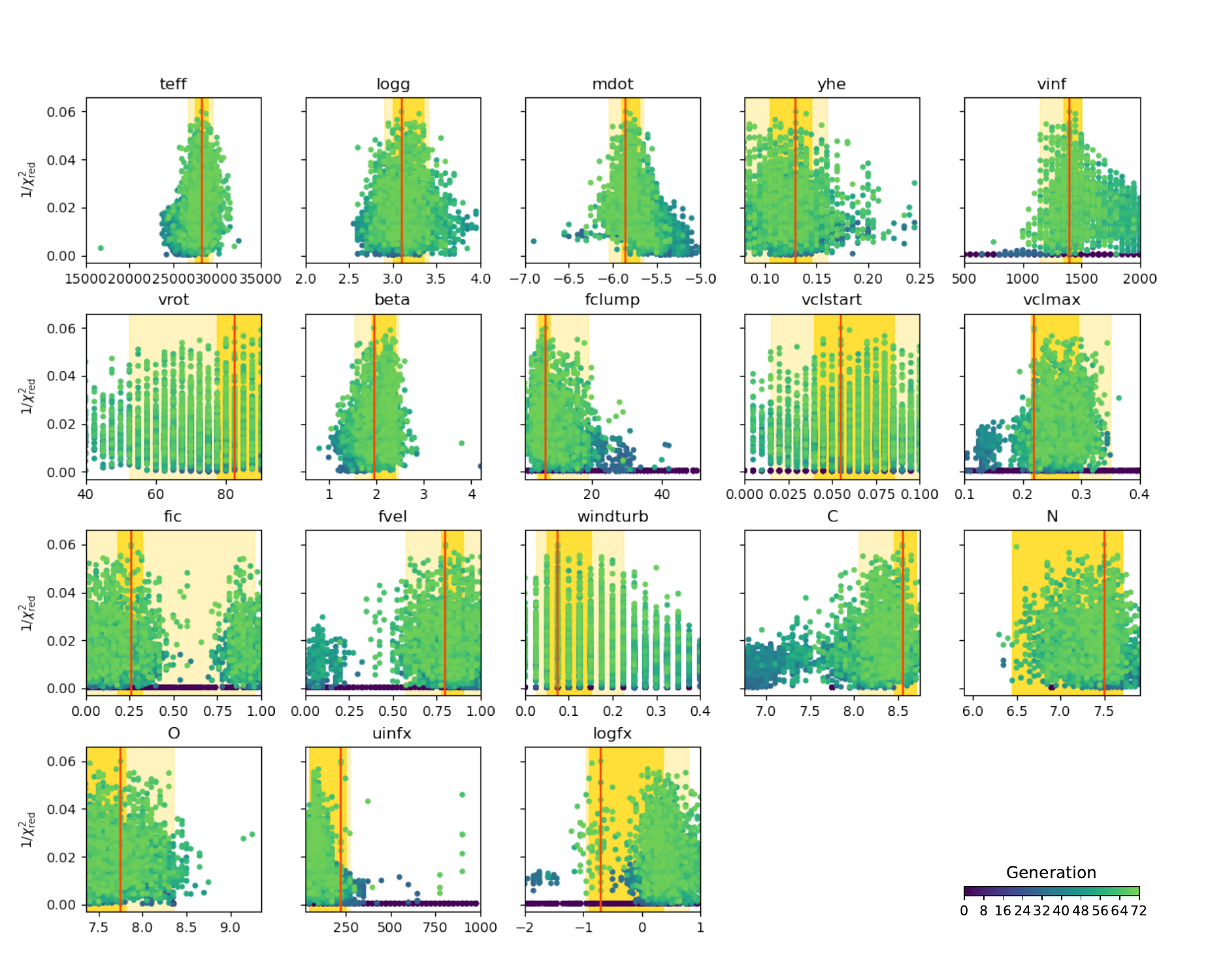}}
    \caption{Same as figure \ref{fig:full_fits} but for object Sk$-68^{\circ}52$.}
\end{figure*}

\begin{figure*}
    \centering
    \subfigure{\includegraphics[width=0.7\textwidth]{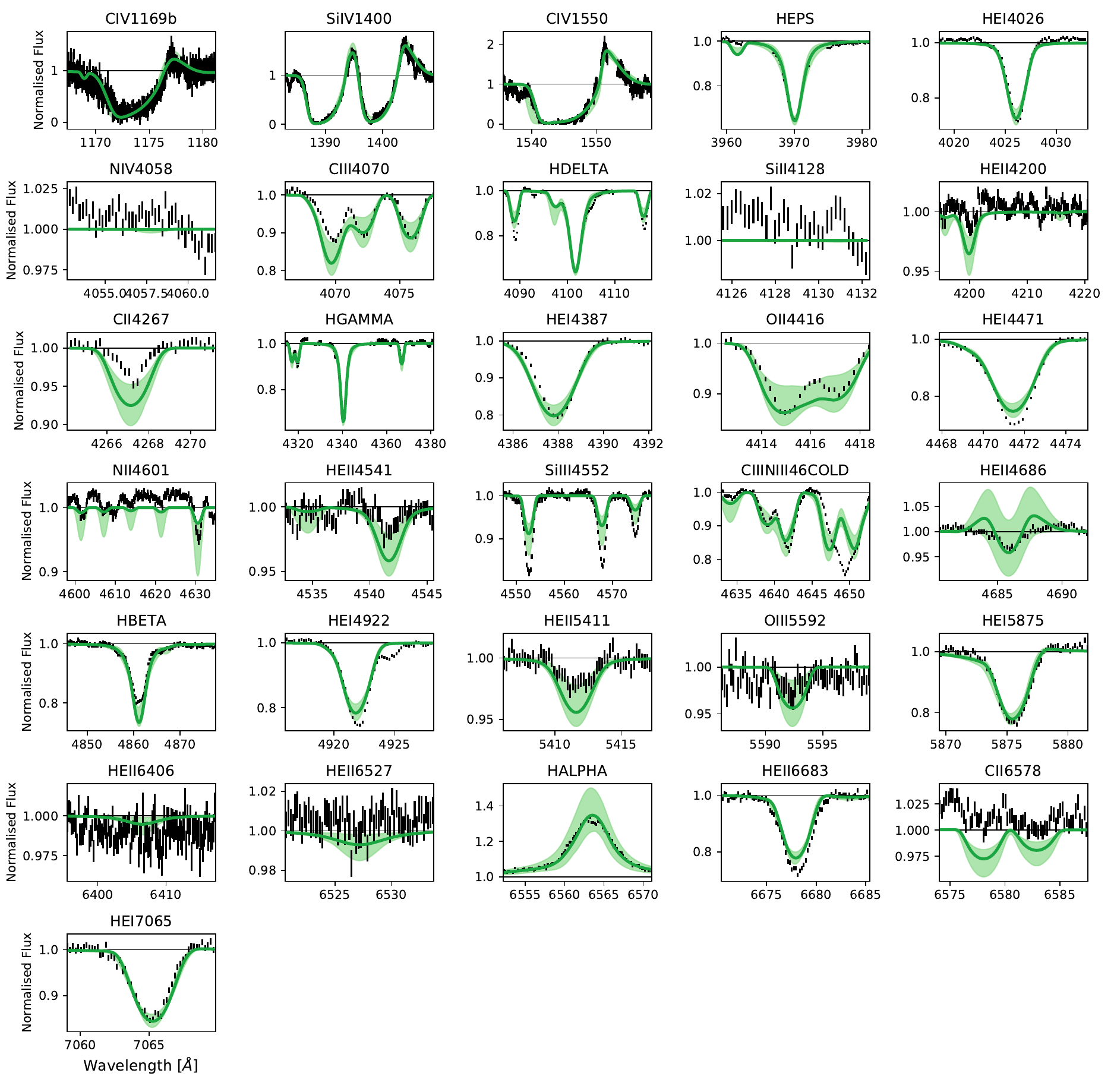}}
    \subfigure{\includegraphics[width=0.7\textwidth]{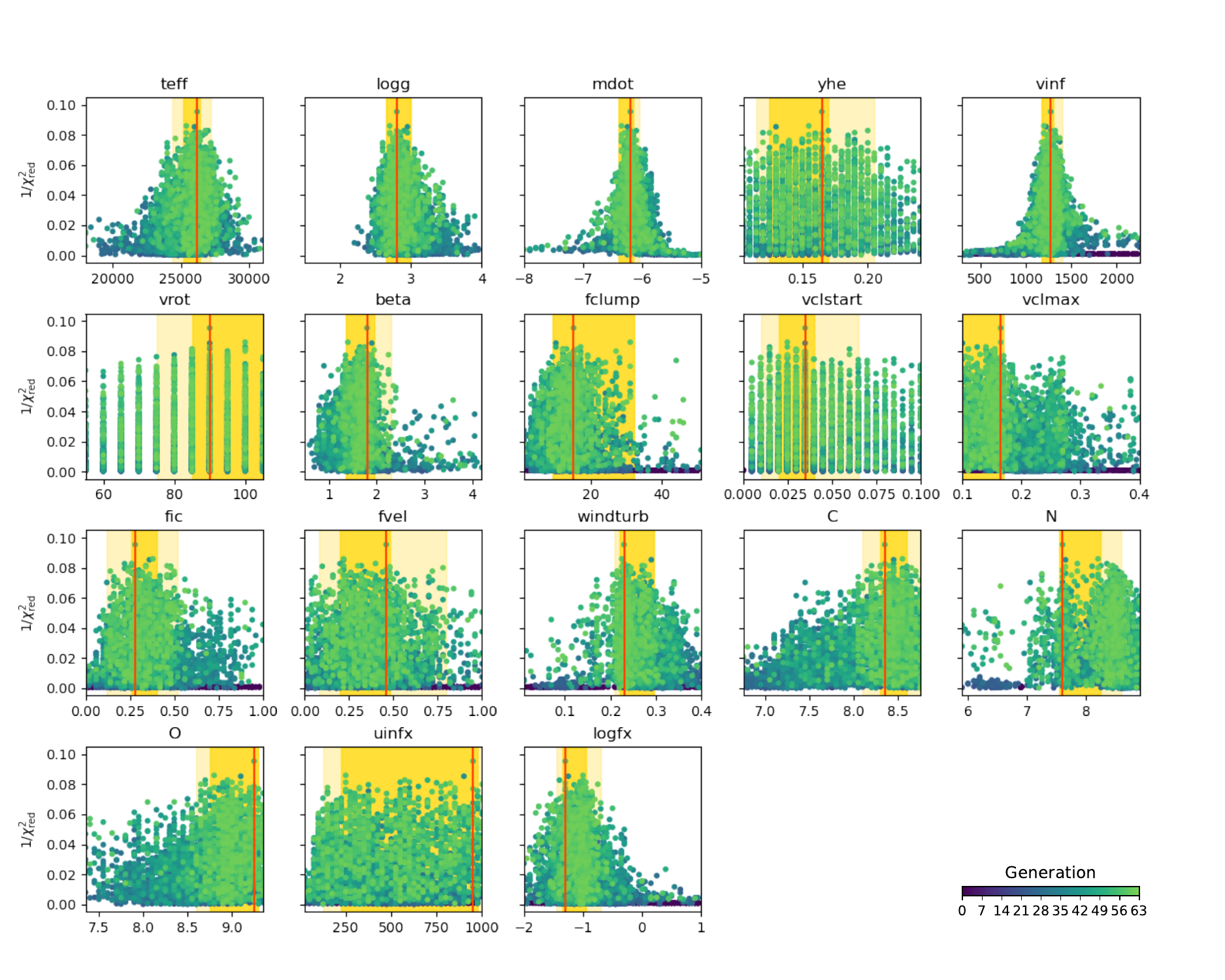}}
    \caption{Same as figure \ref{fig:full_fits} but for object Sk$-68^{\circ}41$.}
\end{figure*}

\begin{figure*}
    \centering
    \subfigure{\includegraphics[width=0.7\textwidth]{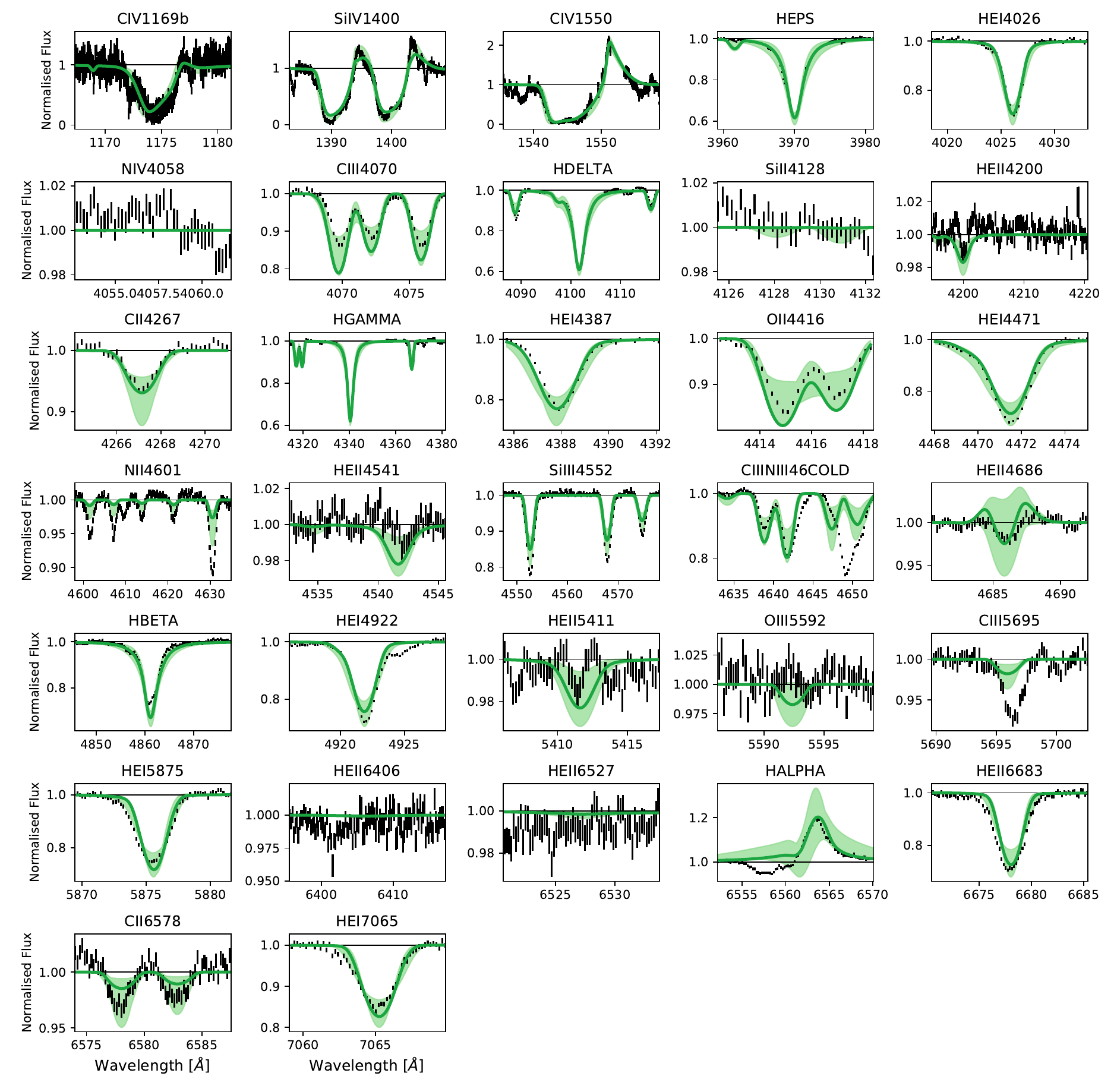}}
    \subfigure{\includegraphics[width=0.7\textwidth]{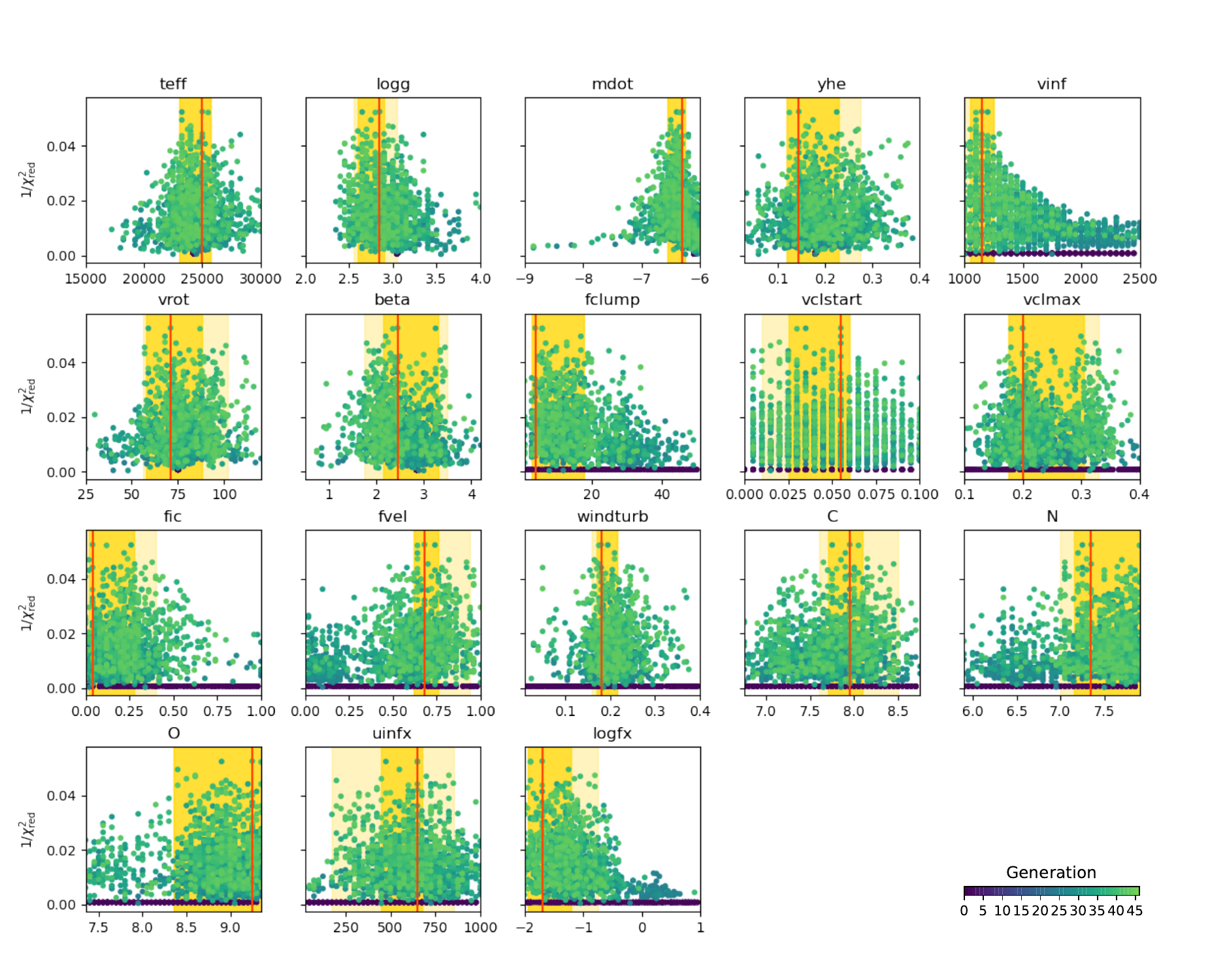}}
    \caption{Same as figure \ref{fig:full_fits} but for object Sk$-69^{\circ}43$.}
\end{figure*}

\begin{figure*}
    \centering
    \subfigure{\includegraphics[width=0.7\textwidth]{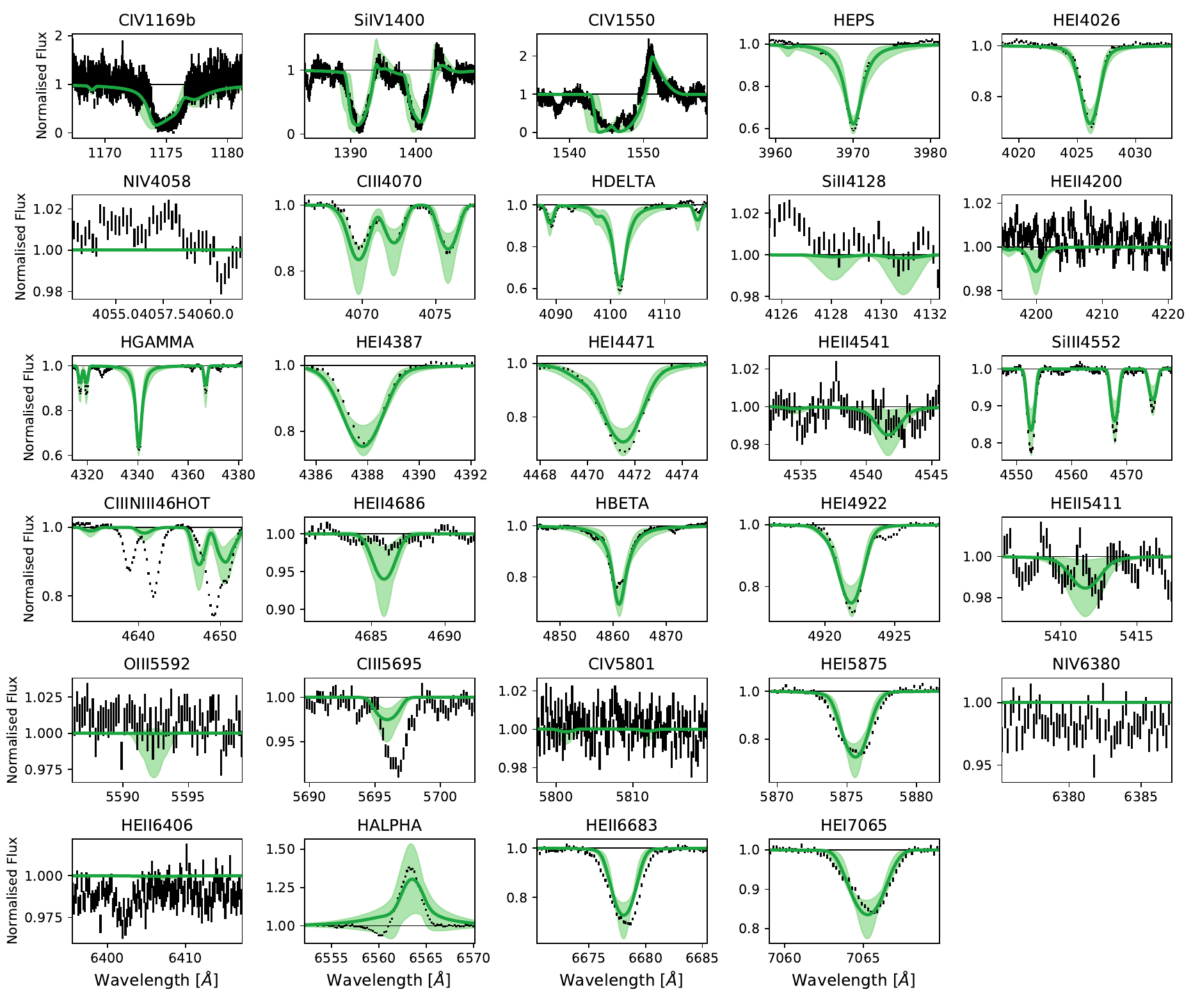}}
    \subfigure{\includegraphics[width=0.7\textwidth]{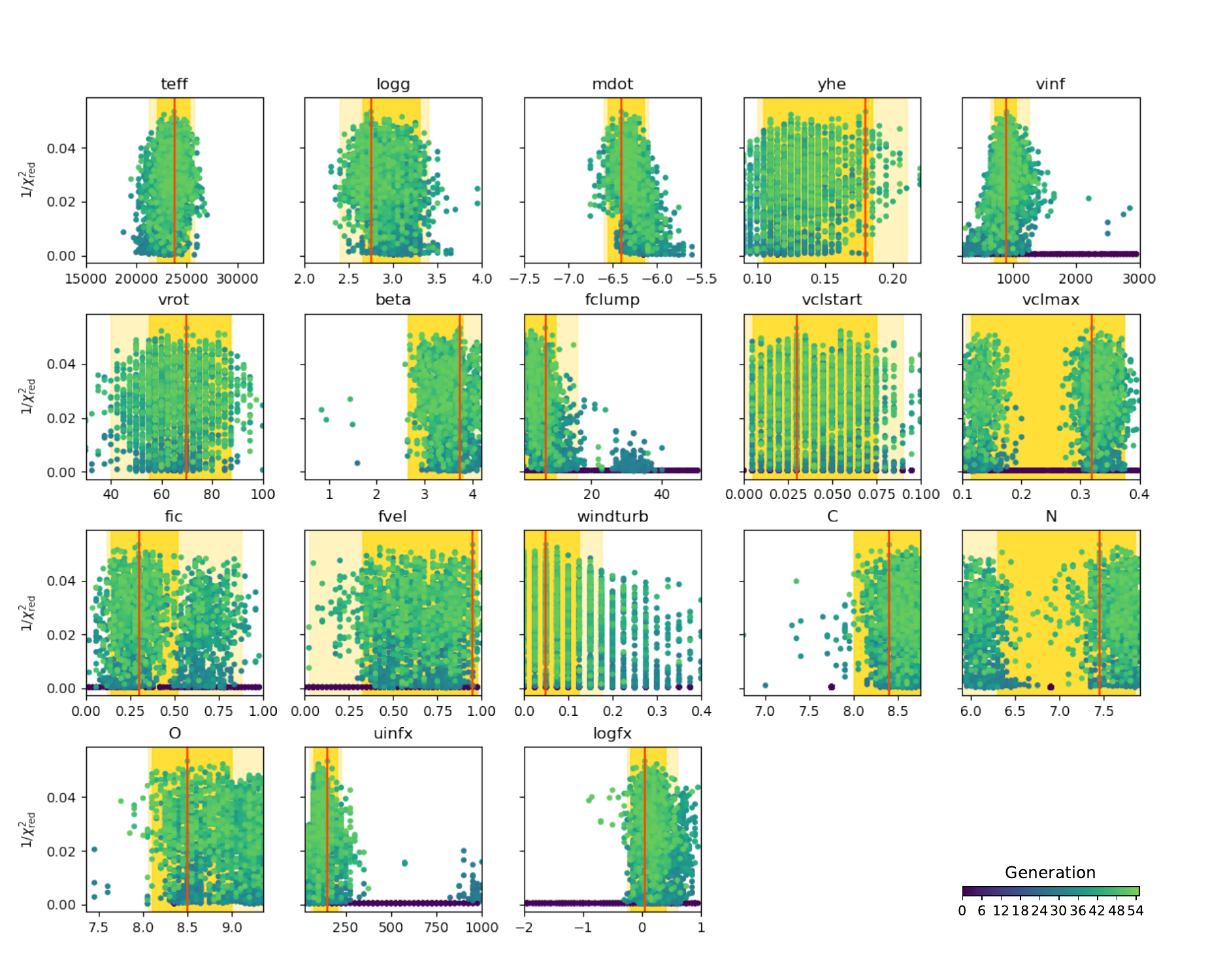}}
    \caption{Same as figure \ref{fig:full_fits} but for object Sk$-66^{\circ}35$.}
\end{figure*}

\begin{figure*}
    \centering
    \subfigure{\includegraphics[width=0.7\textwidth]{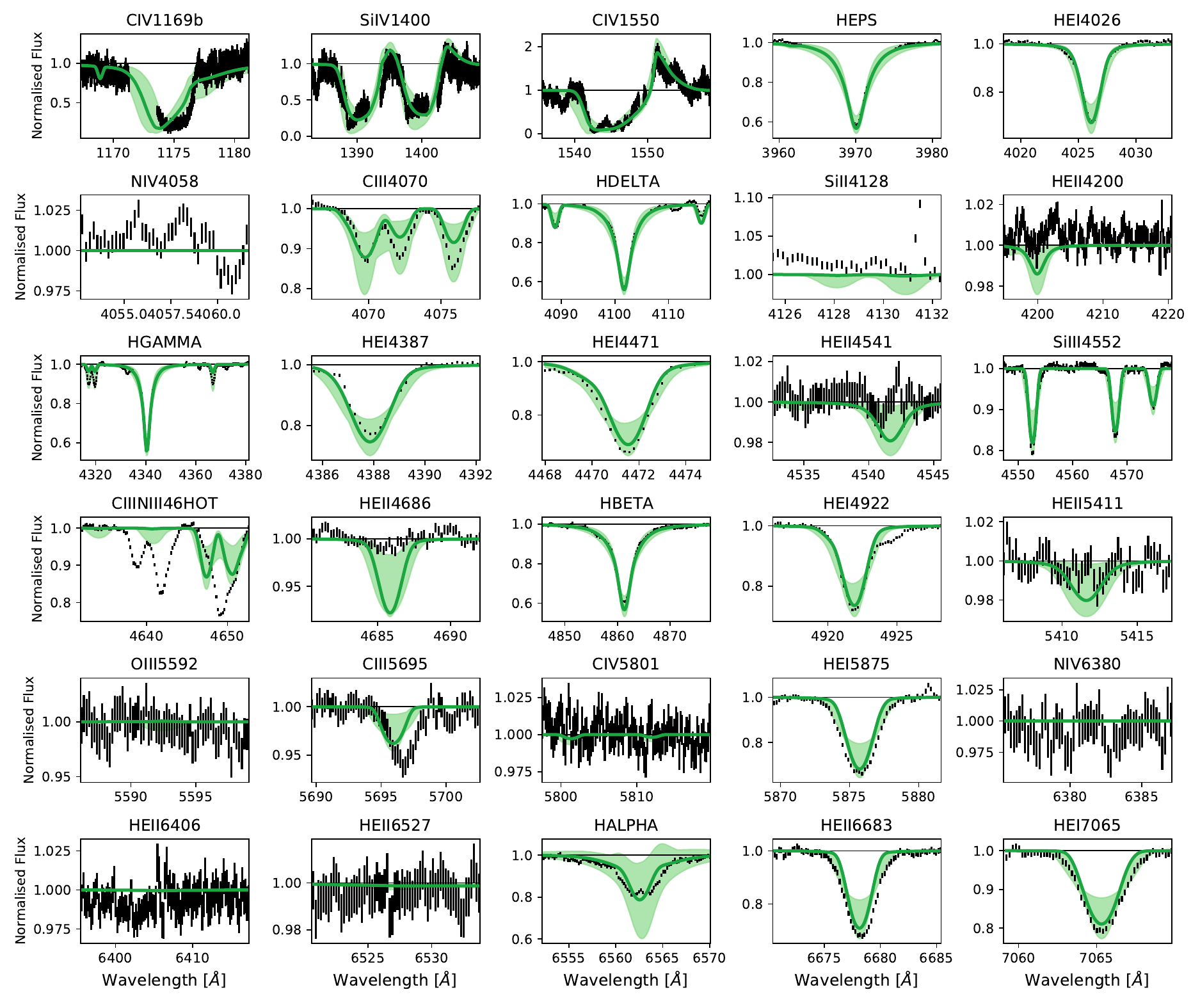}}
    \subfigure{\includegraphics[width=0.7\textwidth]{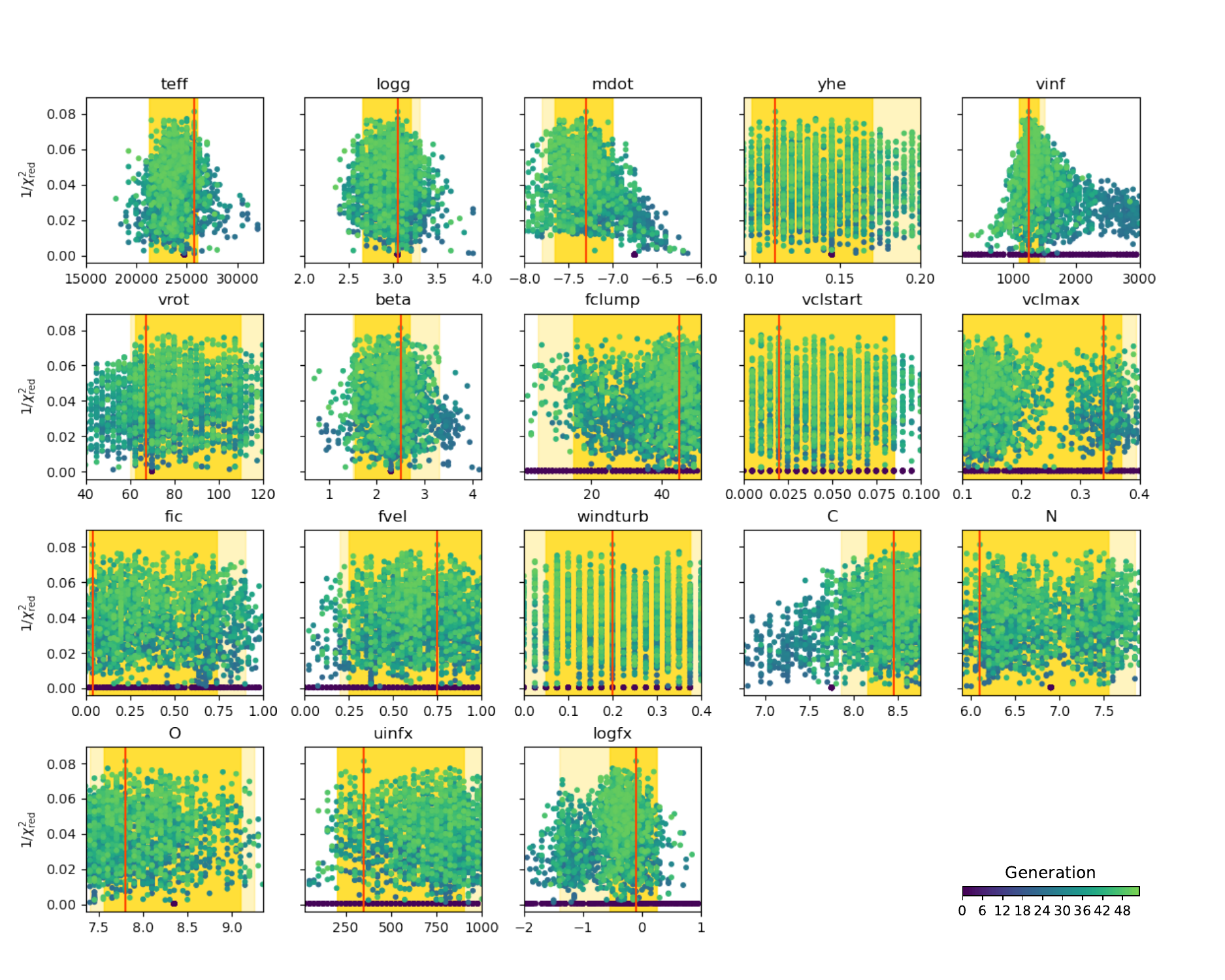}}
    \caption{Same as figure \ref{fig:full_fits} but for object Sk-$69^{\circ}140$.}
\end{figure*}

\begin{figure*}
    \centering
    \subfigure{\includegraphics[width=0.7\textwidth]{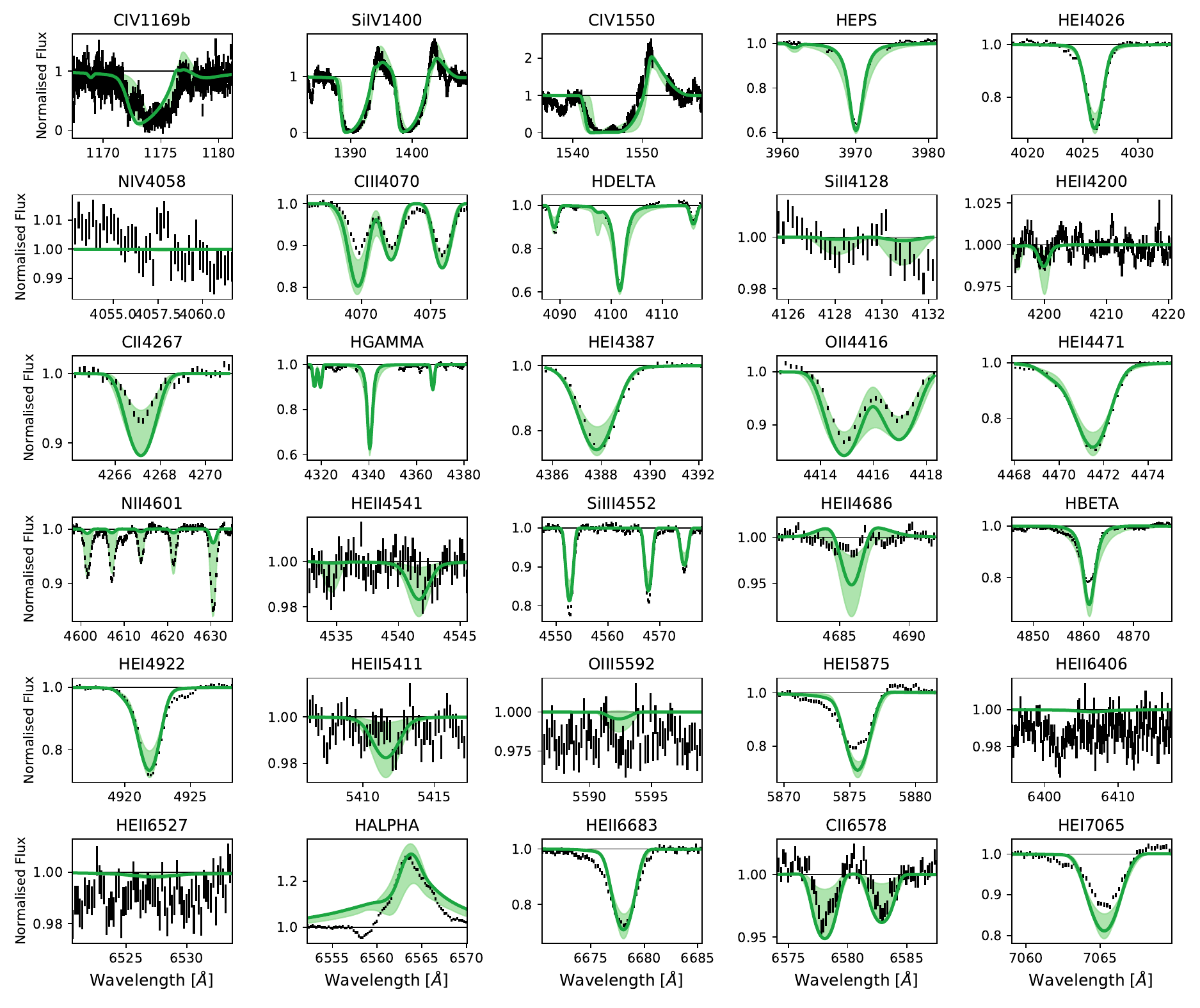}}
    \subfigure{\includegraphics[width=0.7\textwidth]{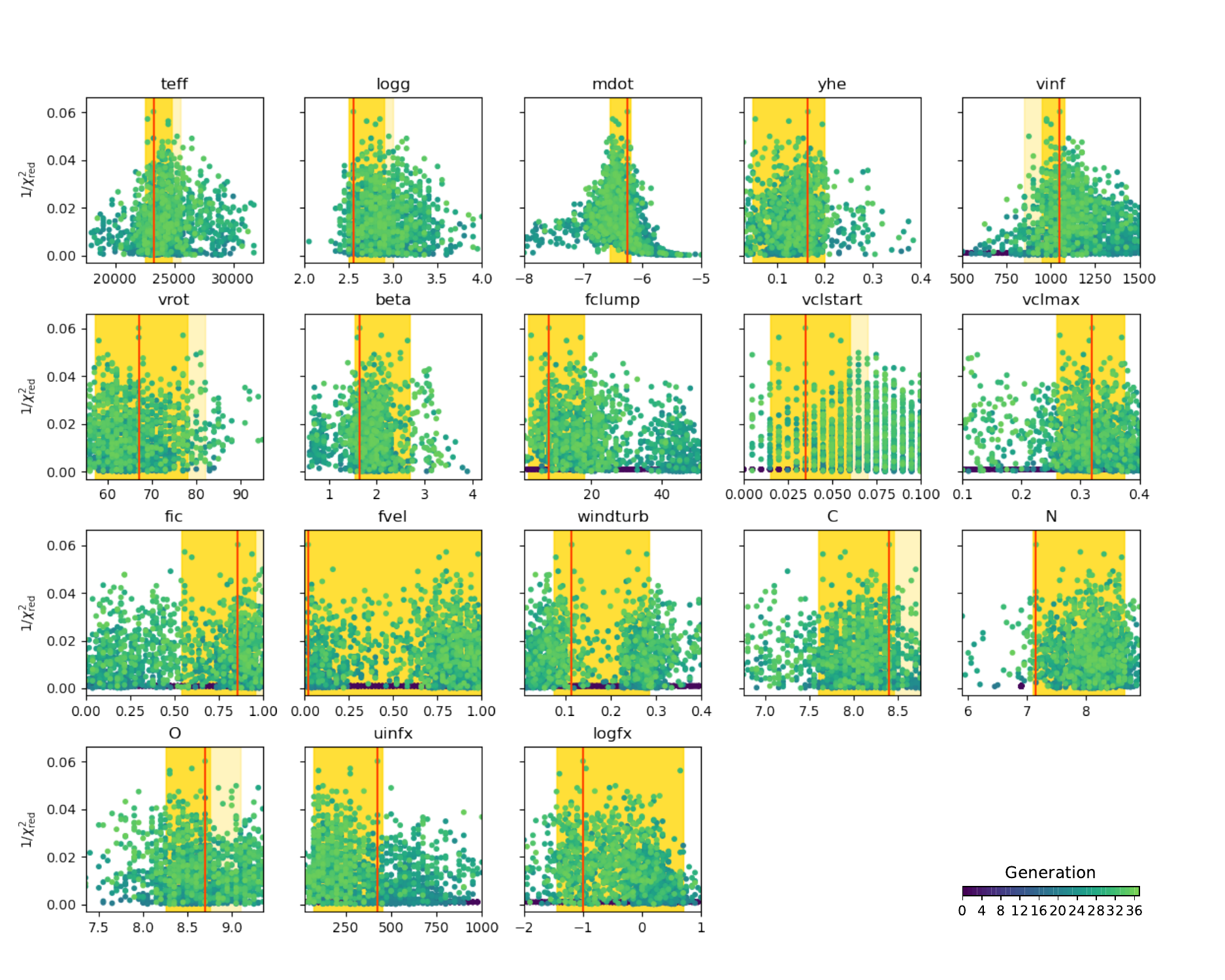}}
    \caption{Same as figure \ref{fig:full_fits} but for object Sk$-67^{\circ}14$.}
\end{figure*}

\begin{figure*}
    \centering
    \subfigure{\includegraphics[width=0.7\textwidth]{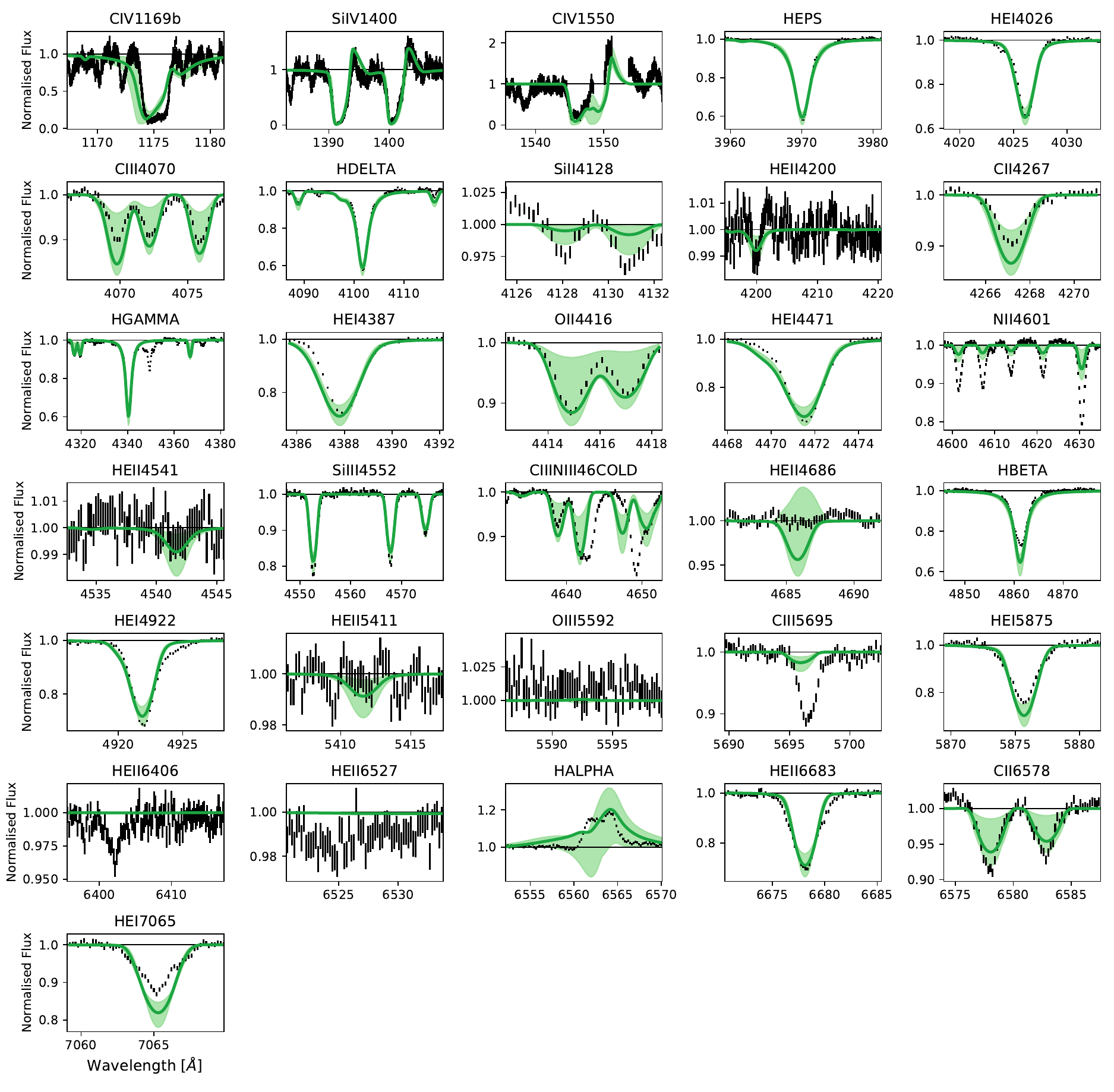}}
    \subfigure{\includegraphics[width=0.7\textwidth]{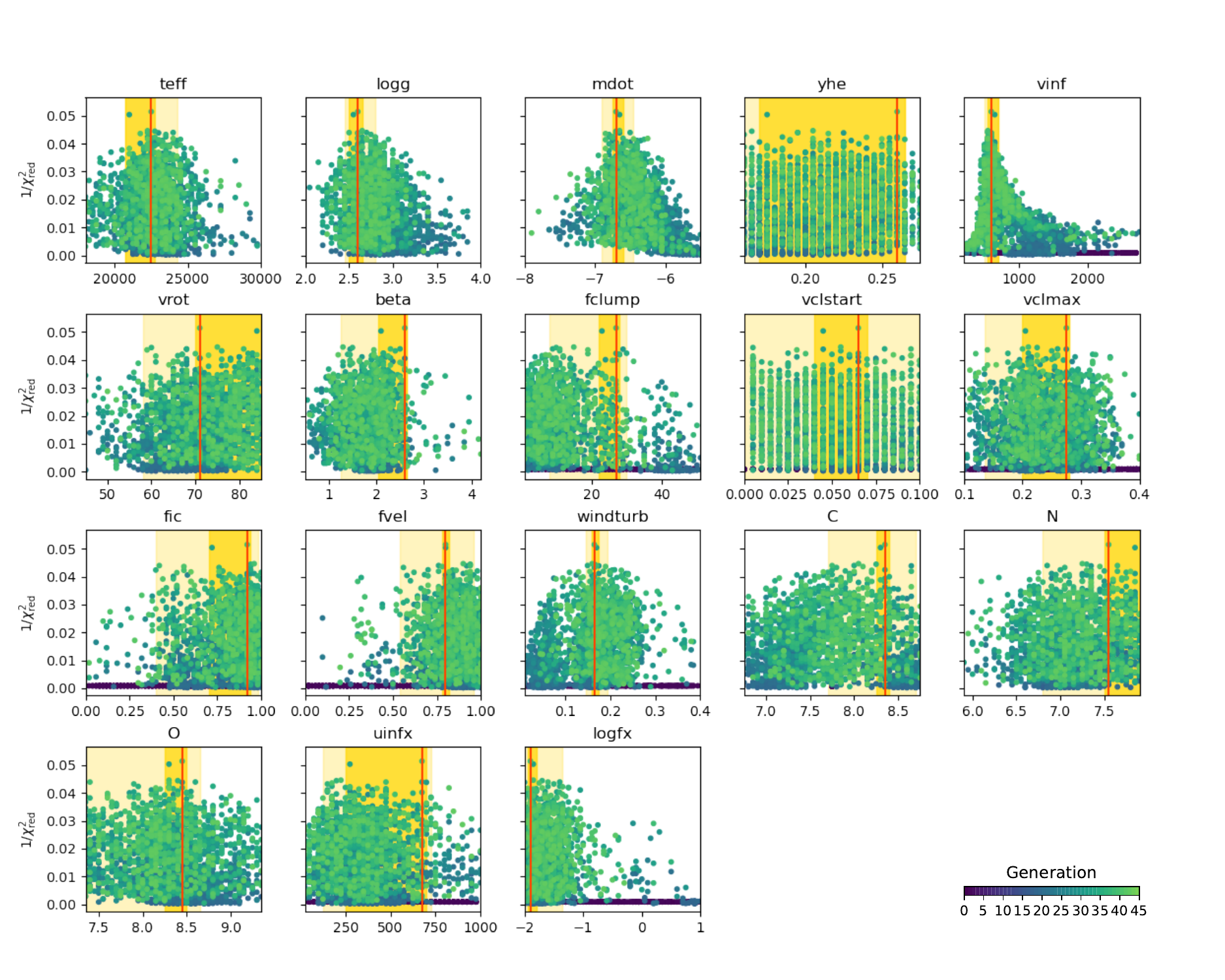}}
    \caption{Same as figure \ref{fig:full_fits} but for object Sk$-69^{\circ}52$.}
\end{figure*}

\begin{figure*}
    \centering
    \subfigure{\includegraphics[width=0.7\textwidth]{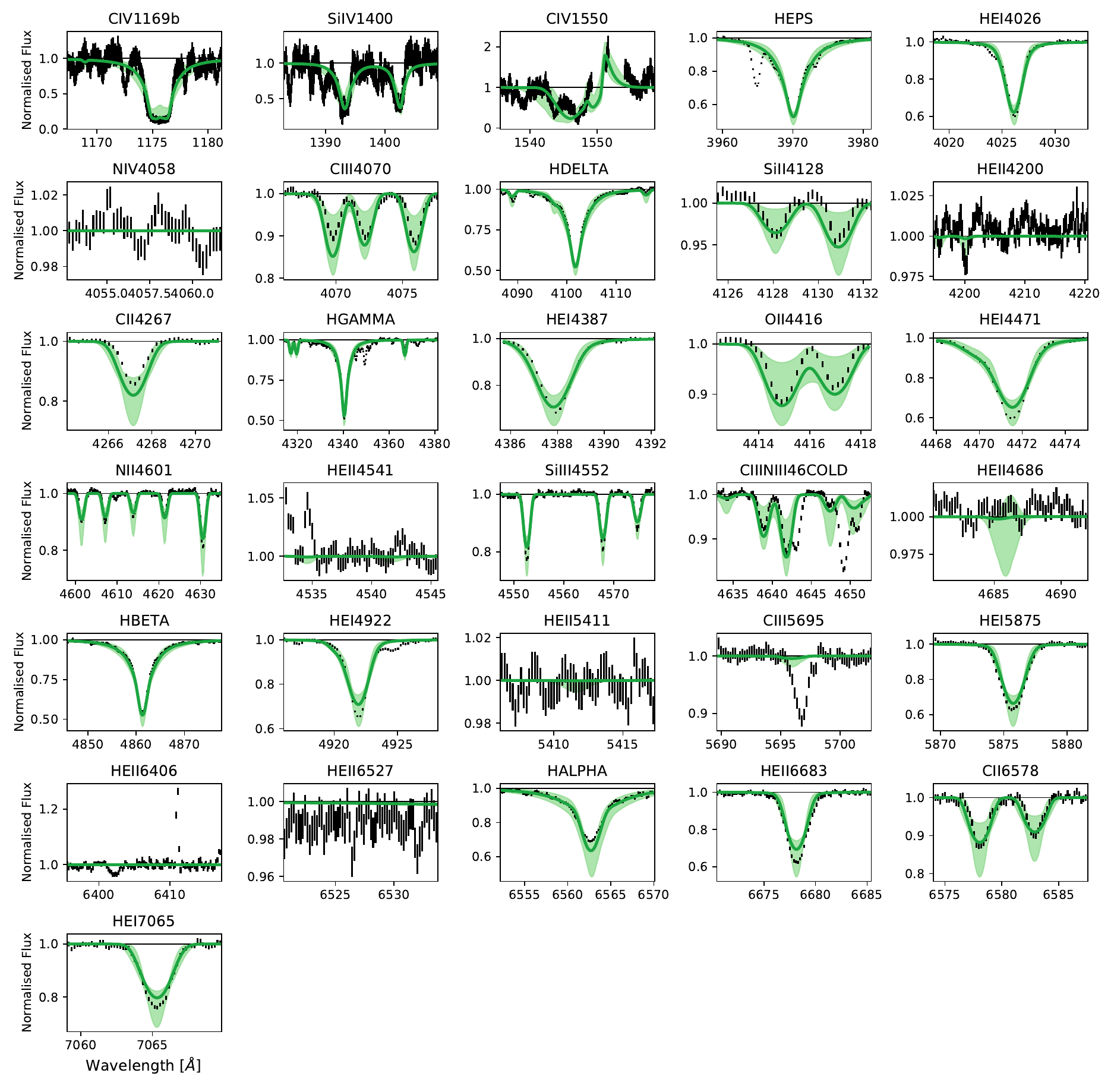}}
    \subfigure{\includegraphics[width=0.7\textwidth]{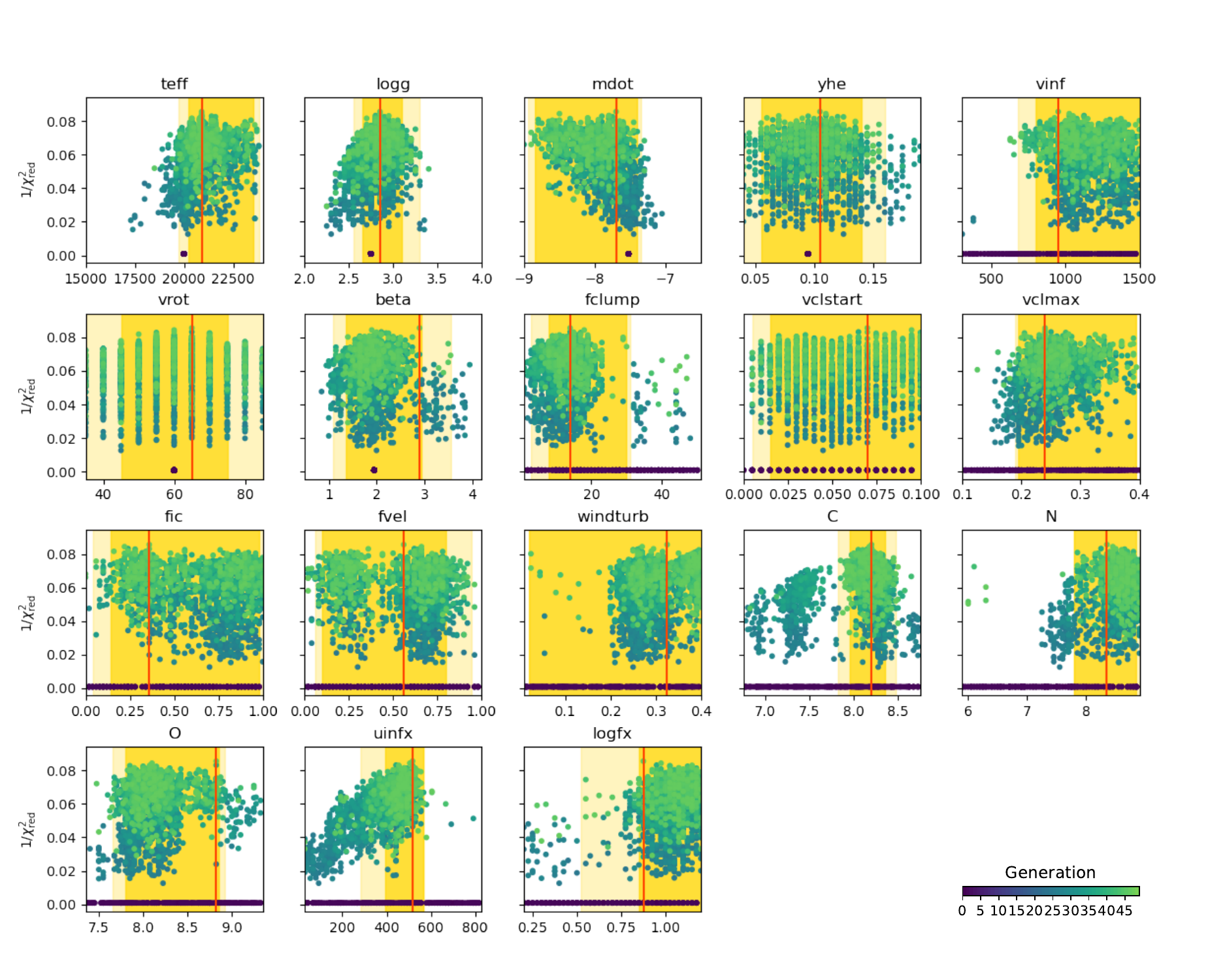}}
    \caption{Same as figure \ref{fig:full_fits} but for object Sk$-70^{\circ}16$.}
\end{figure*}

\begin{figure*}
    \centering
    \subfigure{\includegraphics[width=0.7\textwidth]{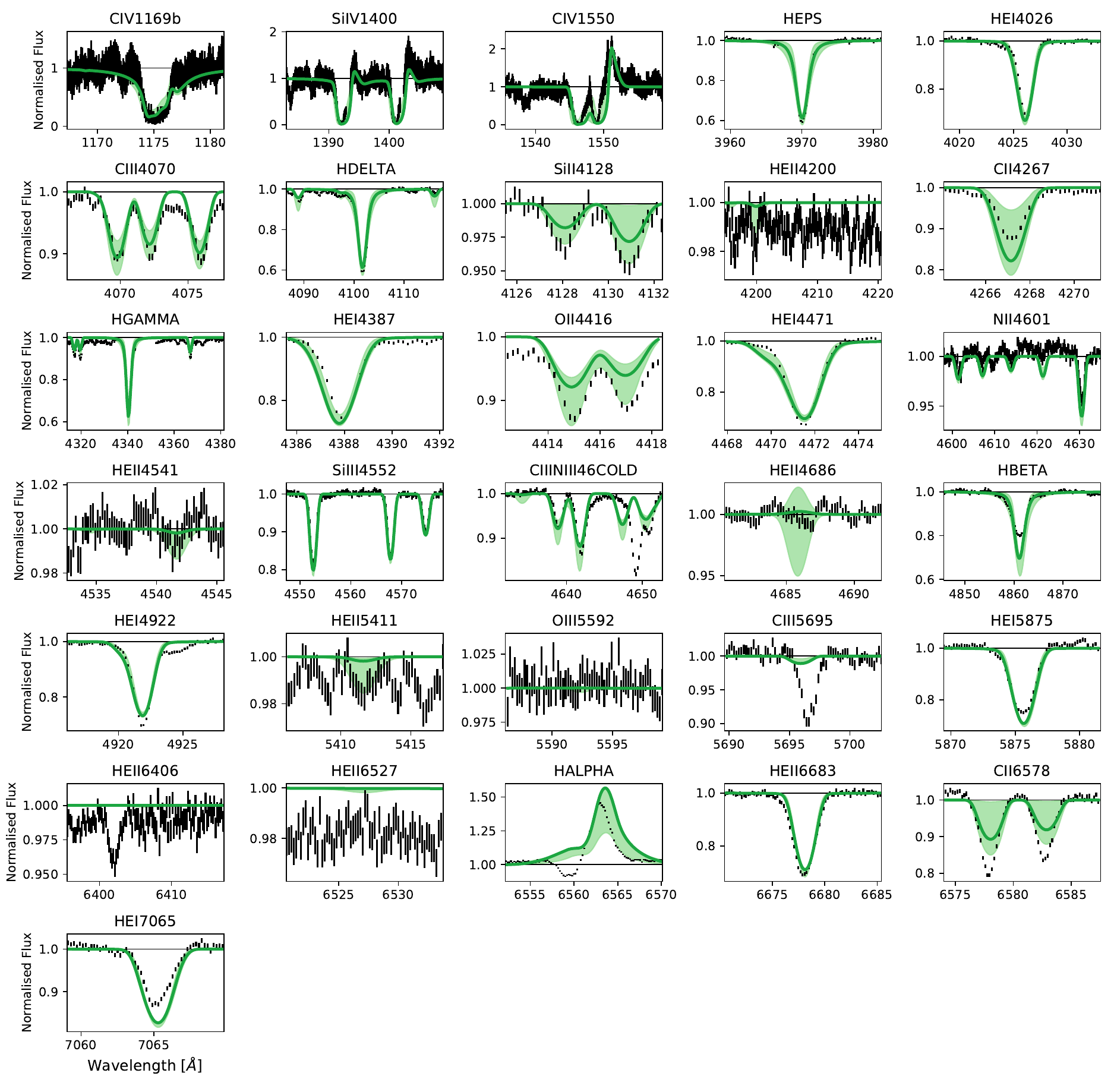}}
    \subfigure{\includegraphics[width=0.7\textwidth]{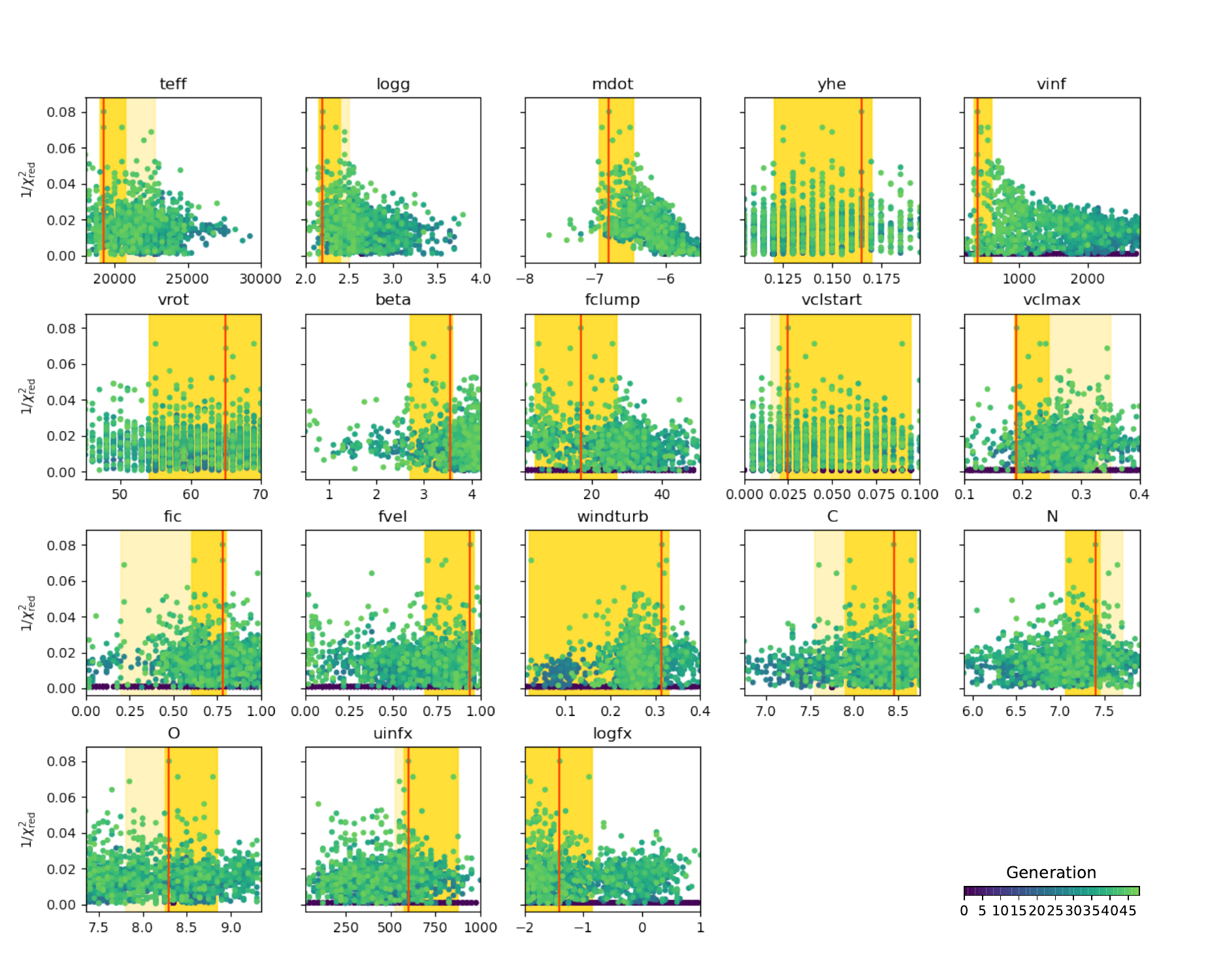}}
    \caption{Same as figure \ref{fig:full_fits} but for object Sk$-68^{\circ}26$.}
\end{figure*}

\begin{figure*}
    \centering
    \subfigure{\includegraphics[width=0.7\textwidth]{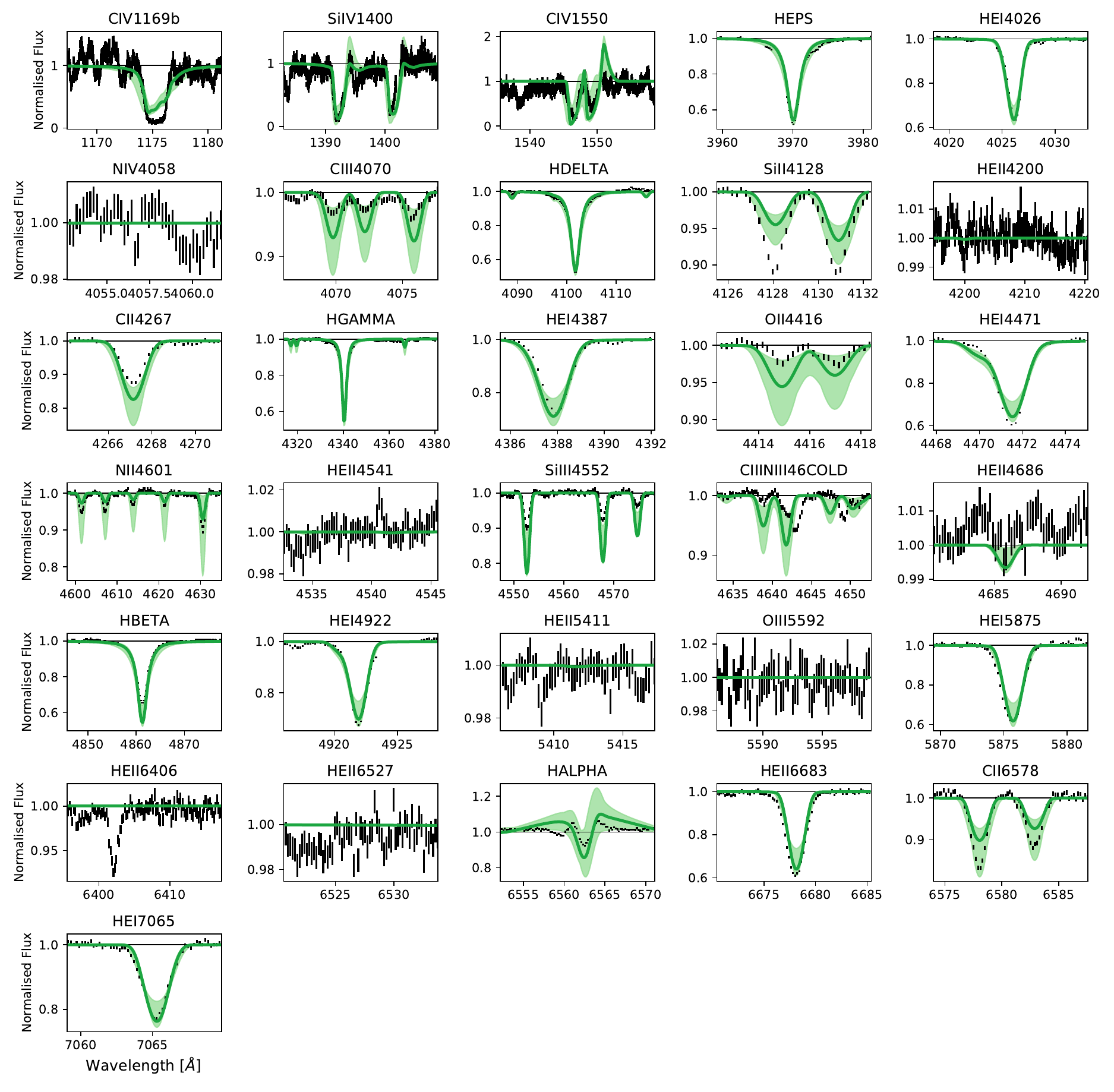}}
    \subfigure{\includegraphics[width=0.7\textwidth]{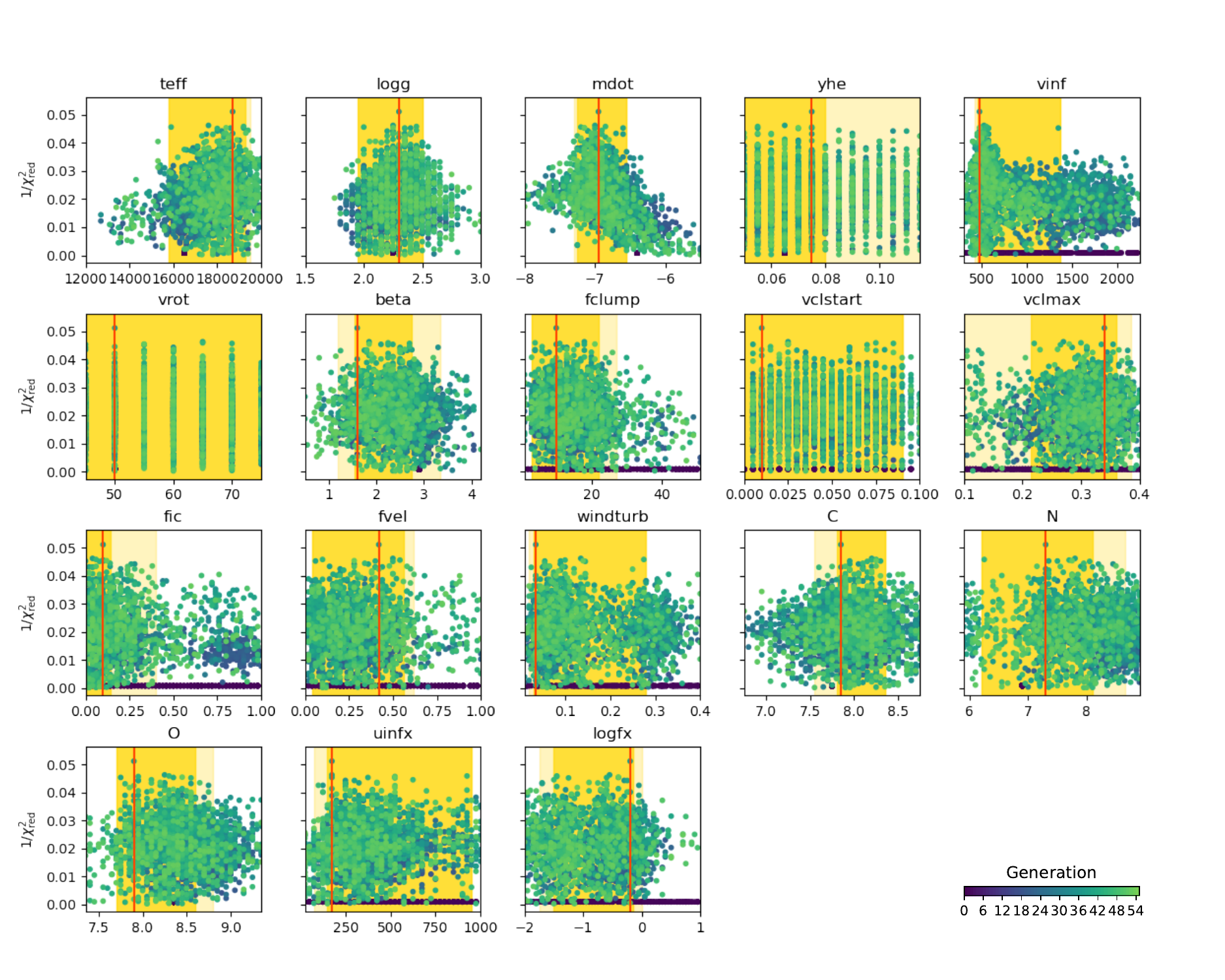}}
    \caption{Same as figure \ref{fig:full_fits} but for object Sk$70^{\circ}50$.}
\end{figure*}

\begin{figure*}
    \centering
    \subfigure{\includegraphics[width=0.7\textwidth]{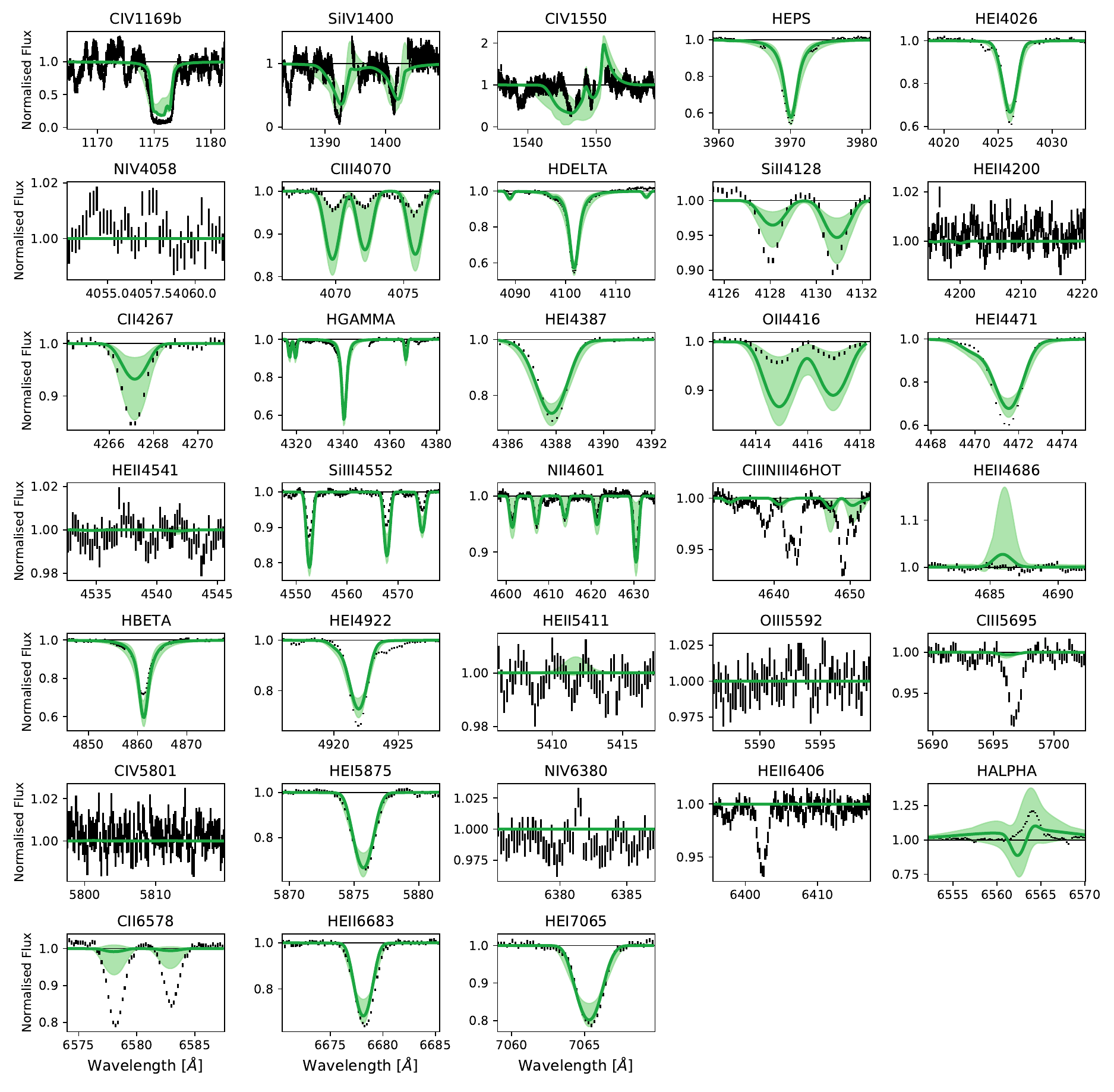}}
    \subfigure{\includegraphics[width=0.7\textwidth]{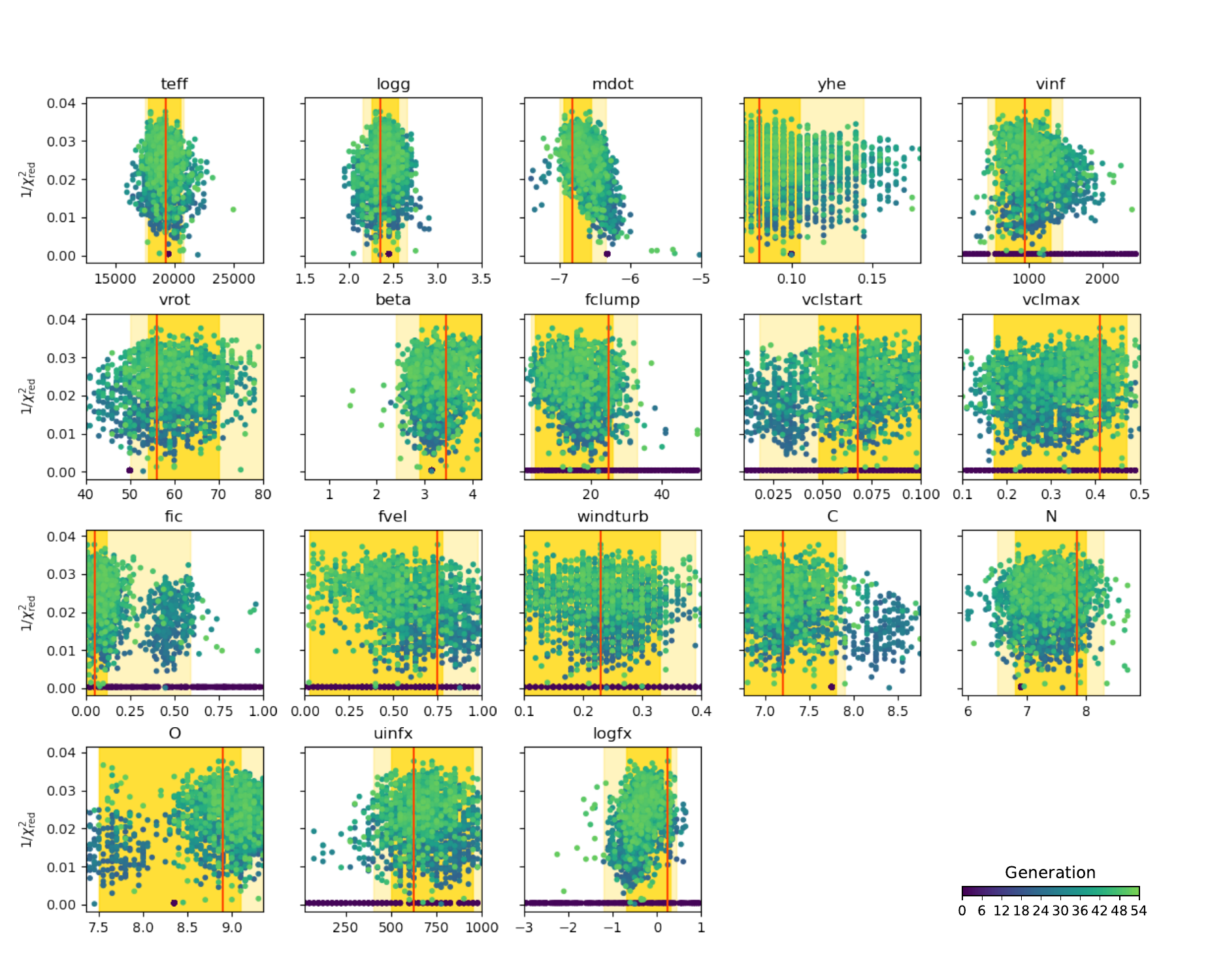}}
    \caption{Same as figure \ref{fig:full_fits} but for object Sk$-67^{\circ}78$.}
\end{figure*}

\begin{figure*}
    \centering
    \subfigure{\includegraphics[width=0.7\textwidth]{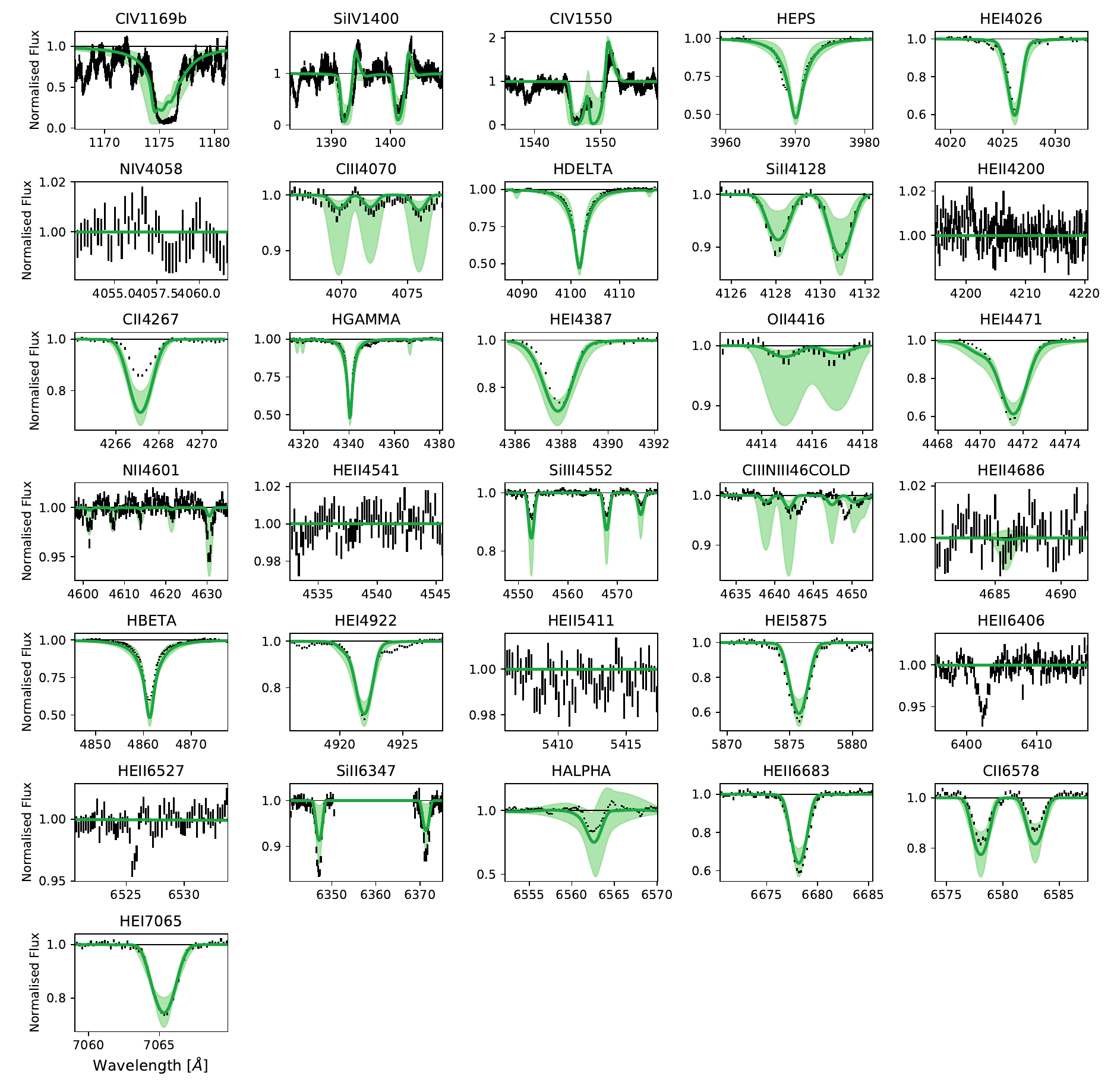}}
    \subfigure{\includegraphics[width=0.7\textwidth]{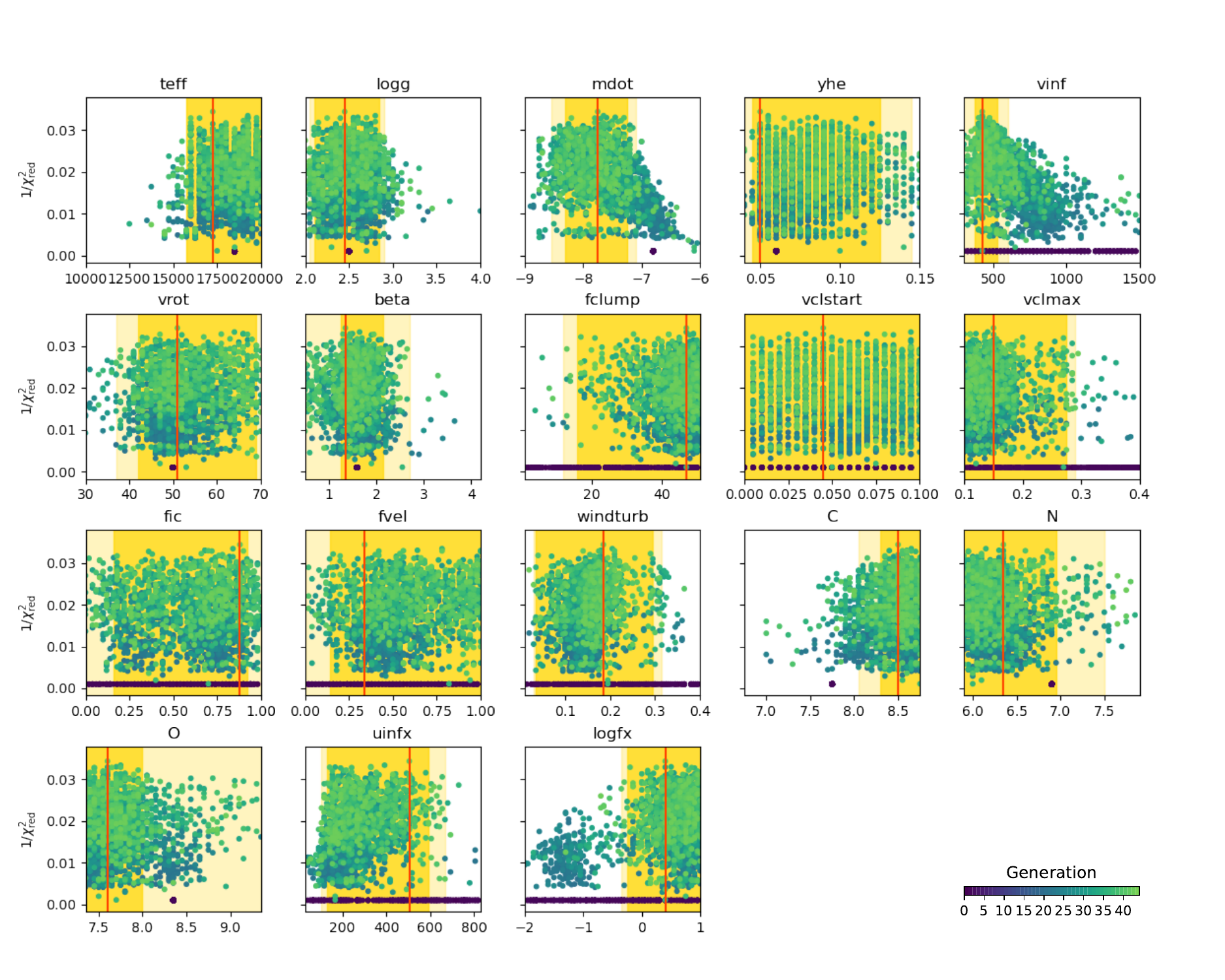}}
    \caption{Same as figure \ref{fig:full_fits} but for object RMC-109.}
\end{figure*}

\begin{figure*}
    \centering
    \subfigure{\includegraphics[width=0.7\textwidth]{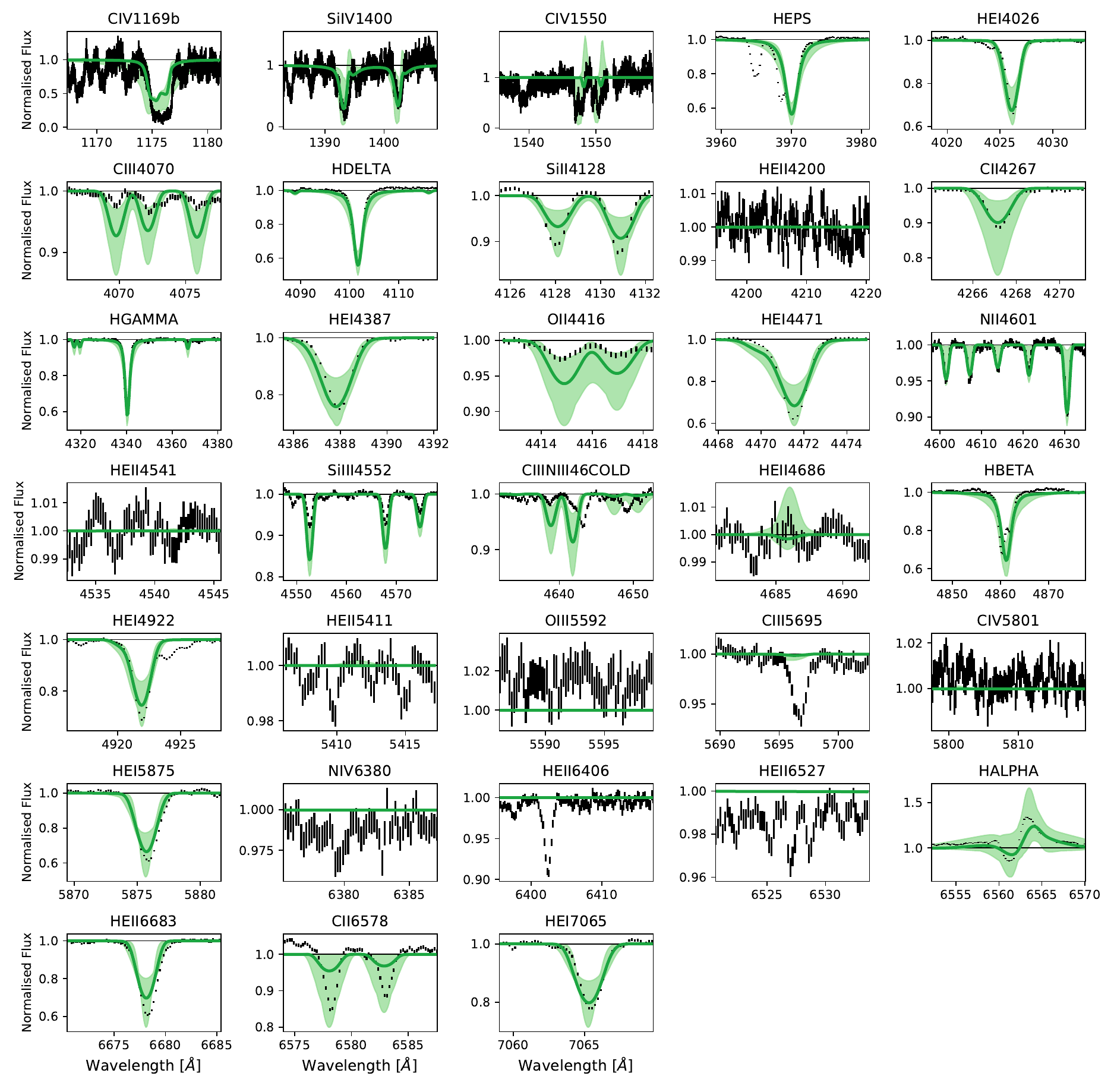}}
    \subfigure{\includegraphics[width=0.7\textwidth]{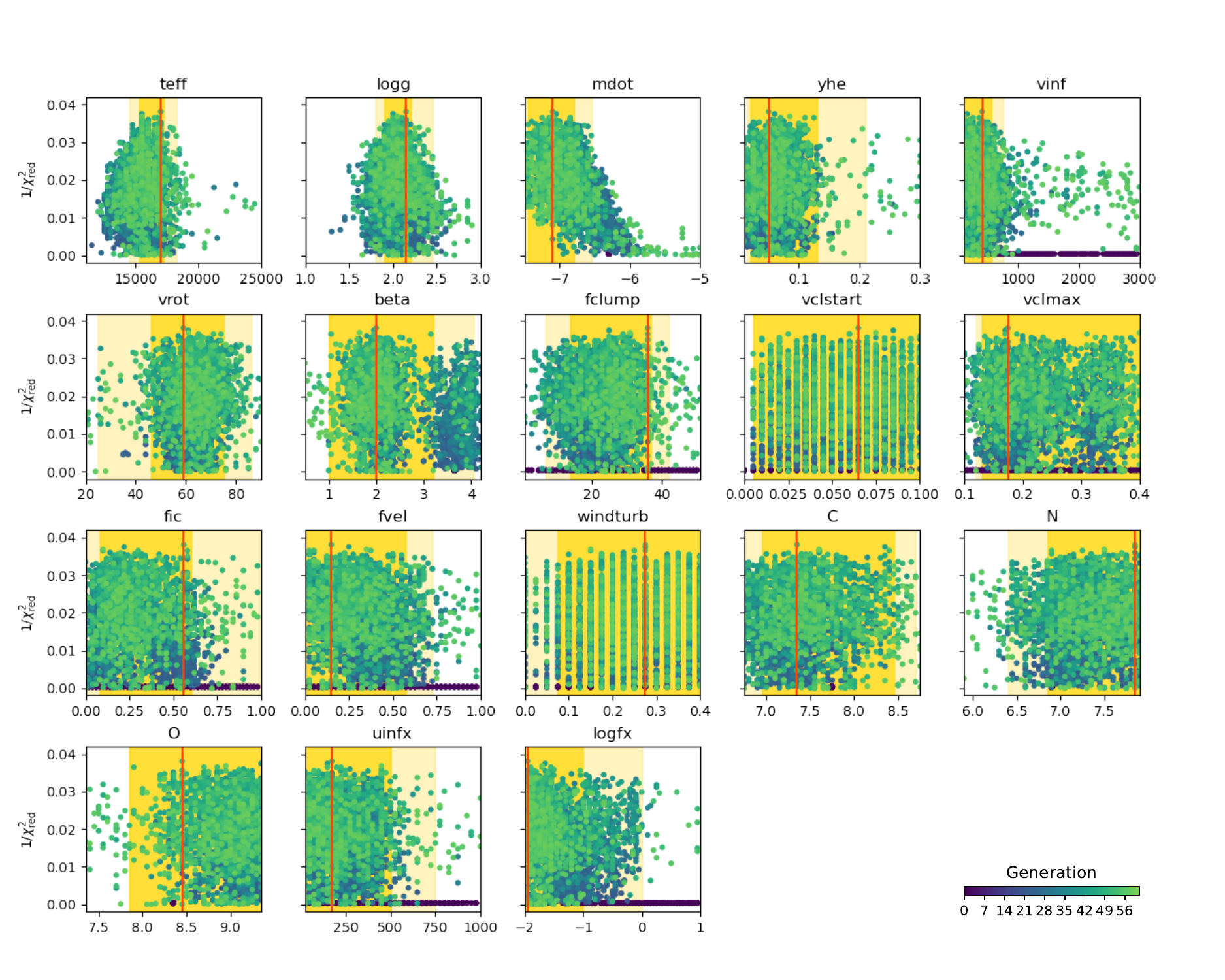}}
    \caption{Same as figure \ref{fig:full_fits} but for object Sk$-68^{\circ}8$.}
\end{figure*}

\begin{figure*}
    \centering
    \subfigure{\includegraphics[width=0.7\textwidth]{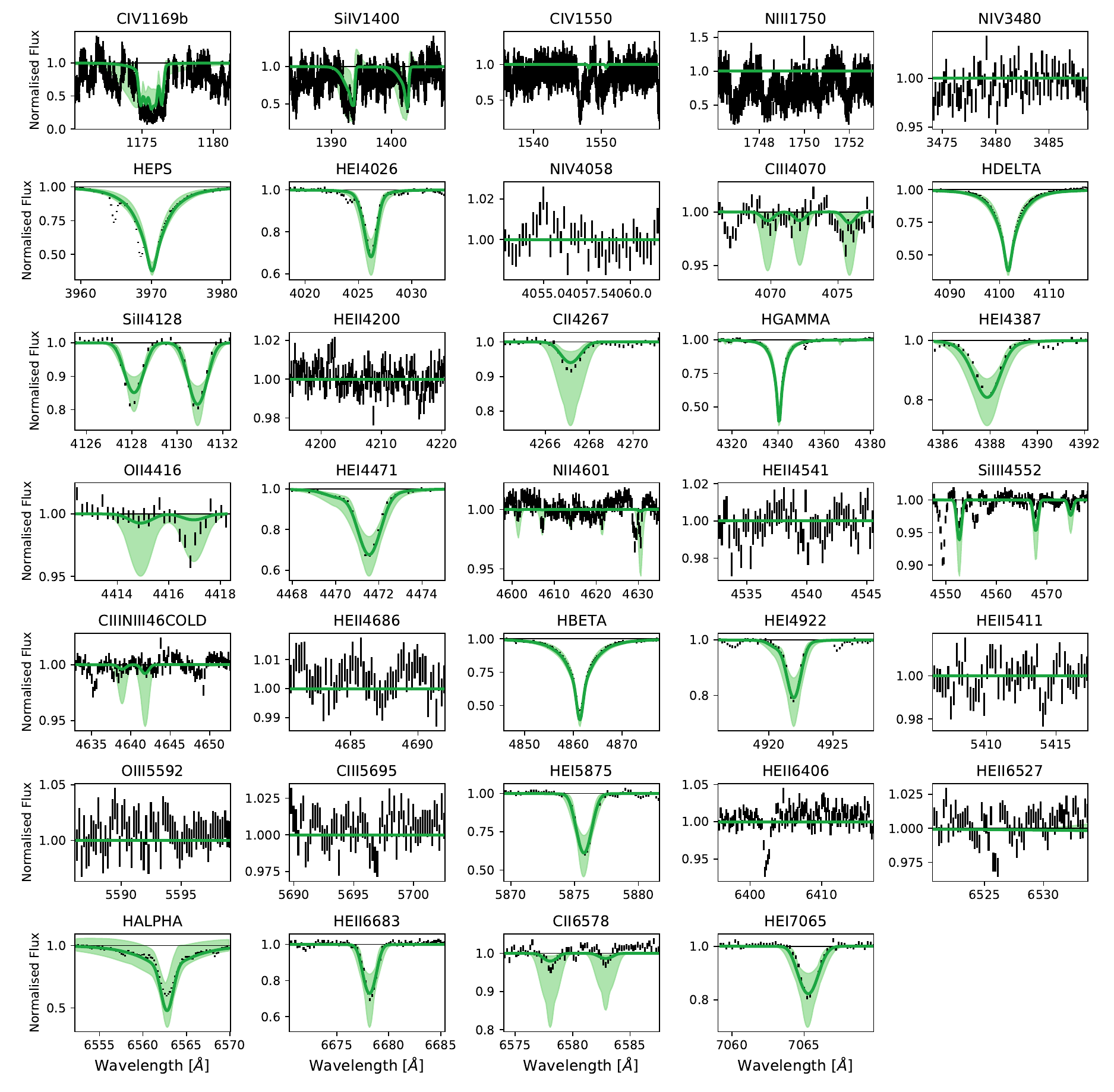}}
    \subfigure{\includegraphics[width=0.7\textwidth]{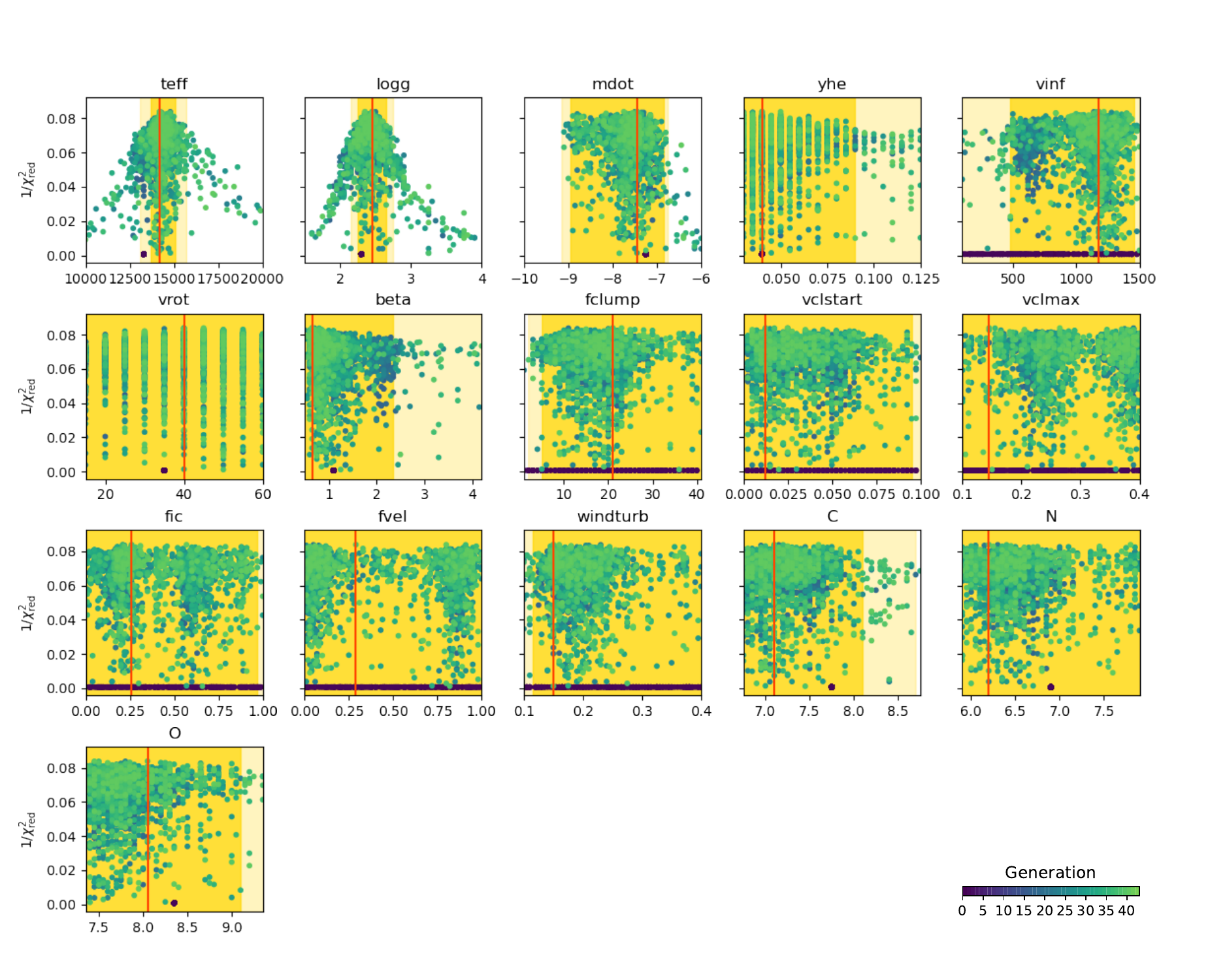}}
    \caption{Same as figure \ref{fig:full_fits} but for object Sk$-67^{\circ}195$.}
    \label{fig:app_sk67-195}
\end{figure*}
\end{appendix}

\end{document}